\documentclass{aa}

\usepackage{amsmath}
\usepackage{longtable} 
\usepackage{lscape}

\usepackage{natbib}
\bibpunct{(}{)}{;}{a}{}{,} 

\usepackage{graphicx}
\usepackage{txfonts}
\usepackage{hhline}

\usepackage{subfigure}
\usepackage{color}

\DeclareMathOperator\erf{erf}
\newcommand{\be}{\begin{equation}}
\newcommand{\ee}{\end{equation}}
\newcommand{\barr}{\begin{array}}
\newcommand{\earr}{\end{array}}

\begin{document}


    \title{Multiplicity and clustering in Taurus star-forming region.}
\subtitle{I. Unexpected ultra-wide pairs of high-order multiplicity in Taurus}

\author{Isabelle Joncour\inst{1,2} 
 \thanks{\emph{\tiny{Email:}} {\scriptsize isabelle.joncour@univ-grenoble-alpes.fr; joncouri@gaia.astro.umd.edu}} , Gaspard Duch\^ene\inst{1,3} 
      \and Estelle Moraux\inst{1}   } 

\titlerunning{I. Unexpected ultra-wide pairs of high-order multiplicity in Taurus}
\authorrunning{Isabelle Joncour et al.}

\institute{Univ. Grenoble Alpes, IPAG, F-38000 Grenoble, France; CNRS, IPAG, F-38000 Grenoble, France 
  \and Department of Astronomy, University of Maryland, College Park, MD 20742, USA
  \and Astronomy Department, University of California, Berkeley, CA 94720-3411, USA}

\date{Received date: 26 July 2016 / Accepted date: 29 November 2016}
    \abstract
 {}
   {This work analyses the spatial distribution of stars in Taurus with a specific focus on multiple stars  and wide pairs  in order to derive new constraints on star formation and early dynamical evolution scenarios.
   }
   {We collected the multiplicity data  of stars in Taurus to build an up-to-date stellar/multiplicity catalog. We first present a general study of nearest-neighbor statistics on spatial random distribution, comparing its analytical distribution and moments to those obtained from Monte Carlo samplings. We introduce the one-point correlation $\Psi$ function  to complement  the pair correlation function and define the spatial regimes  departing from randomness in Taurus.  We then perform a set of statistical studies  to characterize the binary regime that prevails in Taurus. 
   }
   {The $\Psi$ function in Taurus has a scale-free trend with a similar exponent as the correlation function at small scale. It extends almost 3 decades up to $\sim 60$ kAU showing a potential extended wide binary regime. This was hidden in the correlation function due to the clustering pattern blending. Distinguishing two stellar populations, single stars  versus multiple systems (separation  $ \leq 1$ kAU),  within Class II/III stars observed at high angular resolution,  we highlight  a major spatial neighborhood difference between the two populations using nearest-neighbor statistics. The multiple systems are three times more likely to have a distant companion within 10 kAU when compared to single stars.   We show that this is due to the presence of most probable physical ultra-wide pairs (UWPs, defined as such from their mutual nearest neighbor property), that are themselves generally
composed of multiple systems containing up to five stars altogether. More generally, our work highlights; 1)  a new  large population of candidate UWPs in Taurus  within the range 1-60 kAU in Taurus and 2) the major local structural role they play up to $60$  kAU. There are three different types of UWPs;  either composed of two tight and comparatively massive stars (MM), by  one single and one multiple (SM), or by two distant low-mass singles (SS) stars. 
  These UWPs are biased towards high multiplicity and higher-stellar-mass components at shorter separations. The multiplicity fraction per ultra-wide pair with separation less than 10 kAU may  be as high as 83.5 $\pm$ 19.6\%.  }
{We suggest that these young pre-main sequence UWPs  may be pristine imprints of their spatial configuration at  birth resulting from a cascade fragmentation scenario of the natal molecular core. They could be  the older counterparts, at least for those separated by less than 10 kAU,  to the $\le$ 0.5 Myr prestellar cores/Class 0 multiple objects observed at radio/millimeter wavelengths.
 }

   {}

\keywords{Methods: statistical -- binaries: visual -- Stars: formation -- Stars: pre-main sequence -- Stars: statistics -- Galaxy: open clusters and associations: individual: Taurus}

   \maketitle
%
%

\section{Introduction}

 The formation of stars results fundamentally from 
the  runaway unbalance between overwhelming  inwards  self-gravity and outwards pressure-gradient forces (thermal, turbulence,  magnetic, radiation, etc.) 
inside a perturbed gas cloud. 
Star formation has been known to be a complex process since at least the first model proposed almost three decades ago  of a single isolated low-mass star formed out of one dense gas core  \citep{ShuEtAl1987}. It also involves radiative processes (heating and cooling), generation and decay of turbulence and magnetic fields, chemical evolution of molecules, as well as dust and feedback mechanisms of newly formed stars in their natal environment. A realistic picture must be even more complicated since the stars are rarely born in isolation \citep{LadaEtAl2003} and, moreover, they are generally not single but part of multiple systems \citep{Mathieu1994,DucheneKraus2013}. Overall, the whole story of star formation spans over ten  orders of magnitude length-scale range. Giant gas clouds, having typical sizes  from 10 pc to 100 pc \citep{DobbsEtAl2014}, fragment, cool, and collapse to form dynamic clusters of young multiple objets whose sizes are approximately a  solar radius. Since newly born stellar objects are closely  linked to their gaseous progenitors \citep{Hartmann2002}, their spatial distribution and their multiplicity properties offer  important tests on star formation models.

Giant molecular clouds appear to be anything but uniform. They are highly substructured into a hierarchy of gas clumps and  into highly intertwined filamentary networks, as strikingly revealed 
by the recent Herschel  Gould  Belt  Survey \citep{AndreEtAl2014}. In turn these filaments shelter single or chains of molecular cores  \citep{TafallaHacar2015}. Whether such  heterogeneous patterns remain as the relics of the star formation processes  in young stellar populations or are totally and quickly  erased by subsequent dynamical evolution is an important question.

To seek such imprints of star formation, the giant Taurus molecular cloud is an ideal nursery. Its  proximity (137 pc, 
\citeauthor{TorresEtAl2007} \citeyear{TorresEtAl2007}), the  youth  (1--10 Myr) of its stellar population 
 \citep[]{BertoutEtAl2007, AndrewsEtAl2013}, and  its low stellar density  bring a complete census of approximately 350 young low-mass stars down to 0.02 $M_\odot$ \citep{LuhmanEtAl2010, RebullEtAl2010}. As there is  a small spatial dispersion of (proto-)stars within the gaseous filament \citep{Hartmann2002,CoveyEtAl2006}, these stars are  most probably born where they are now observed.

 The current paradigm of star formation attributes a seemingly fundamental  role to dense molecular cores thought to be the very cradles of stars   \citep{Ward-ThompsonEtAl2007}. The long-standing scenario proposed so far to form one low-mass star from one collapsing single core \citep{ShuEtAl1987} predicts that the dense cores set the final mass of young stars. This is supported  by the correspondence  of the shapes of the core mass function (CMF)   and  the initial mass function (IMF) of stars, although the peak of the IMF is shifted by a  factor  2-3 towards lower mass.
This shift may be due to a partial 30-40\% efficiency gas to star conversion \citep{AlvesEtAl2007,LadaEtAl2008,MarshEtAl2016,KonyvesEtAl2015}, although this estimate is higher than the star-formation efficiency (1--10\%) estimated for the whole cloud. Alternatively, such a  shift  between the core mass function  and the initial mass function may also result from the multiplicity of stars born within one individual core that has undergone multiple fragmentation \citep{GoodwinEtAl2008}. This may lead to a core-to-stars efficiency as high as 100\%  (\citeauthor{HolmanEtAl2013} \citeyear{HolmanEtAl2013}; see for a review \citeauthor{OffnerEtAl2014} \citeyear{OffnerEtAl2014}).

The multiplicity appears to be a key feature in star-forming regions. The companion frequency of young stars in Taurus is generally twice that of field stars \citep{DucheneKraus2013}, 
and the formation of coeval  and wide binaries up to 5 kAU  is a common outcome of the star formation process \citep{KrausHillenbrand2009b}. 
Furthermore, since multiplicity appears to decrease with age (see for a review \citeauthor{DucheneKraus2013} \citeyear{DucheneKraus2013}) due to dynamical evolution,  observing multiplicity at this young age means that the multiple system-core picture of star formation is the rule rather than the exception. In this picture, the maximal size of multiple systems  cannot exceed the size of their progenitor cores, typically approximately $20$ kAU, that is, $0.1\,$pc \citep{Ward-ThompsonEtAl2007}. If the dense cores are {\it de facto} the smallest bricks to build a young stellar population, we then expect that spacing between stars as well as binary semi-major axes be related to parental molecular core size and spacing, provided that dynamical evolution has not erased early imprints. Furthermore, if the formation of multiple systems and single stars share a common scenario, we do not  {\it  a priori}  expect  any difference in spatial distribution between the two.

Several spatial studies have been performed in Taurus. Some were based on the stellar  two-point correlation function and  the related mean surface density of companions  to examine the scale-free behavior of clustering modes \citep{GomezEtAl1993, Larson1995,Simon1997,GladwinEtAl1999,Hartmann2002,KrausHillenbrand2008}. Other studies based on first-nearest-neighbor separation (1-NNS) statistics analysis aimed either at comparing the spatial distribution of stars in Taurus to a random distribution \citep{GomezEtAl1993} or at studying their distribution as a function of  their mass and spectral energy distribution (SED) \citep{Luhman2006,LuhmanEtAl2010}.

In this work, after a description of the stellar data catalog we have built to perform our studies (section 2), we apply these two  statistical tools (1-NNS distribution and two-point correlation function)  to analyse the spatial distribution of stars in Taurus with a focus on multiples with respect to single stars, while introducing the one-point correlation function $\Psi$ (section 3). We then assess the statistical properties of  1-NNS couples while defining ultra-wide pairs (UWP)  as mutual nearest neighbors in the range of 1--100 kAU and we show the crucial role they play in spatial clustering features (section 4). We make a synthesis of results and open a discussion (section 5) before a summary to conclude the paper (section 6).

\section{Data}
\label{Sect_Data}
\subsection{Input catalog}

\begin{figure*}
\includegraphics[width=2\columnwidth]{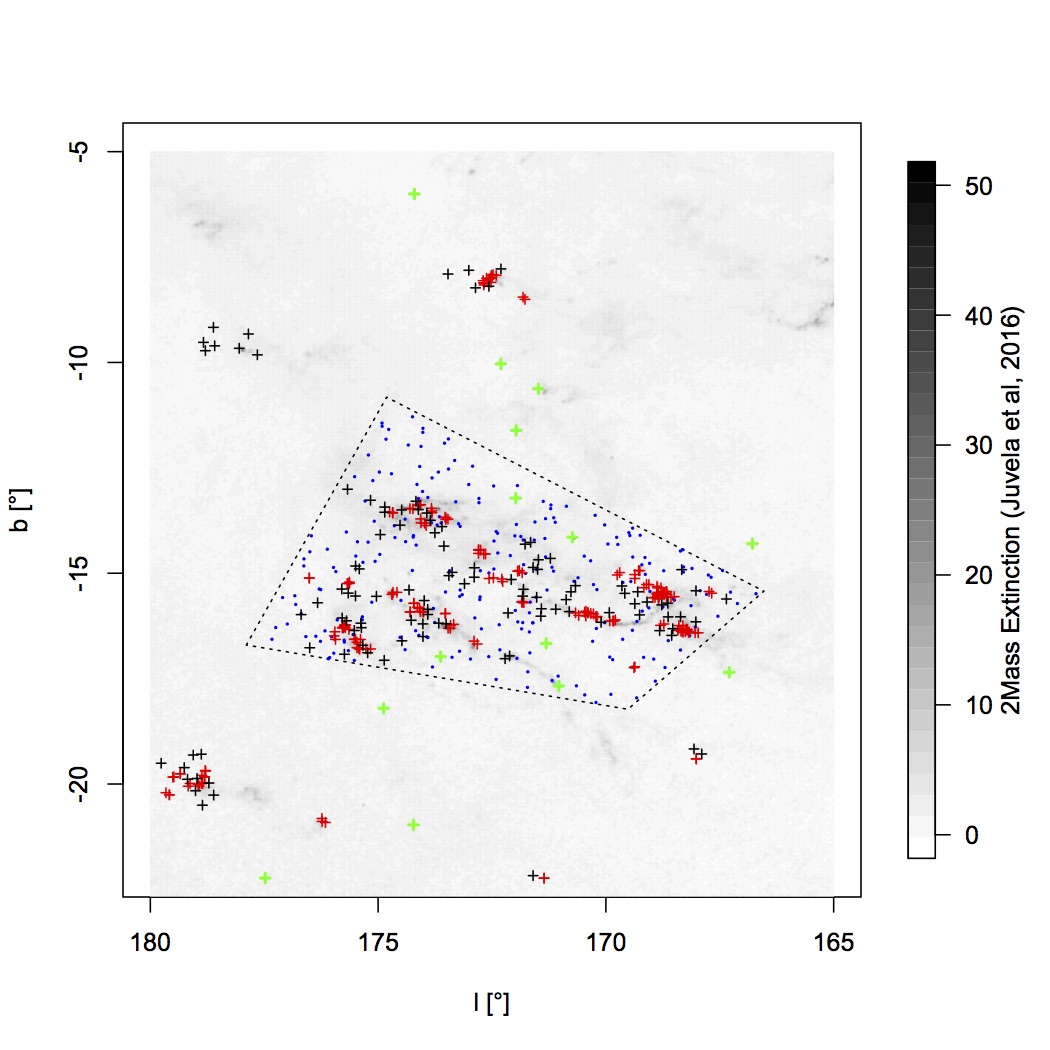}
\caption{Spatial distribution of stars within the Taurus region superimposed on near-infrared extinction (unit mag, gray scale) from \cite{JuvelaEtAl2016}. Restricted spatial window $W_{in}$ (dotted line) in Taurus complex; blue filled dots: random sampling. Color-coding for Taurus stars:  red plus marks: clustered stars [1-NNS $\le$0.1$^\circ$]; black plus marks: <<inhibited>> regime stars  [0.1$^\circ$ < 1-NNS $\le$0.55$^\circ$];  green plus marks: isolated stars [1-NNS beyond 0.55$^\circ$] (see Sect. \ref{Sec:NNS}).
}
 \label{Fig:TauG_Dobashi2005_DHT2001}
\end{figure*}

\longtab{
\onecolumn
 \begin{landscape}
{\scriptsize
\begin{longtable}{rllrrlrlrlllll}
  \caption{\label{Tab:stars}  Taurus stars catalog}\\
  \hline
  \# & 2MASS & Name &  RA$\_{J2000}$ &  DEC$\_{J2000}$&SpT   & M [$M_\odot$] & Class & n$_{*}$ &  Sep ["] & HAR & HAR ref. \\  \hline
  \hline
  \endfirsthead
\caption{continued.}\\
\hline\hline
  \# & 2MASS & Name &  RA$\_{J2000}$ &  DEC$\_{J2000}$&SpT   & M [$M_\odot$] & Class & n$_{*}$ &  Sep ["] & HAR & HAR ref. \\  \hline
\endhead
\hline
  1 & J04034930+2610520 & HBC358A+B+C & 60.955 & 26.18 & M3.5 & 0.335 & III & 4 & 0.15,1.58,3.15 & Y & L93,D99 \\ 
  2 & J04034997+2620382 & XEST06-006 & 60.958 & 26.34 & M5.25 & 0.157 & III & 1 & - & N & \\ 
  3 & J04035084+2610531 & HBC359 & 60.962 & 26.18 & M2 & 0.575 & III & 1 & - & N & \\ 
  4 & J04043936+2158186 & HBC360 & 61.164 & 21.97 & M3.5 & 0.335 & III & 1 & - & Y & L93 \\ 
  5 & J04043984+2158215 & HBC361 & 61.166 & 21.97 & M3 & 0.398 & III & 1 & - & Y & L93 \\ 
  6 & J04044307+2618563 & IRAS04016+2610 & 61.179 & 26.32 & K3 & 1.796 & I & 1 & - & Y & P99,C08 \\ 
  7 & J04053087+2151106 & HBC362 & 61.379 & 21.85 & M2 & 0.575 & III & 1 & - & N & \\ 
  8 & J04080782+2807280 & - & 62.033 & 28.12 & M3.75 & 0.303 & III & 2 & 0.052 & Y & K11,K12\\
\hline
\end{longtable} }
\tablefoot{This catalog was built from \cite{LuhmanEtAl2010} and \cite{KirkMyers2011}. All the multiple systems with a separation less than 1 kAU is set to be one entry (see Sect. \ref{Sect_Data}). The following sources are listed in \cite{LuhmanEtAl2010} but not included here:  HBC351,  HBC352, HBC353, HBC355(+354), HBC356, V410TauAnon20, LH0422+15, CIDA13.
Columns 1--3: Star reference number (this work),  2MASS Point Source Catalog and common Name. Column 4--5: Ecliptic right ascension and Declination (Epoch J2000). Columns 6--8: Spectral type, (Primary) mass, Class. Columns 9--15: Total number of stars ($n_{*}$) within 1000AU (spectroscopic binaries not counted as membership), Separation of companions stars with the primary,  Flag (Y/N) to mention wether the stellar system/star has been observed at High Angular Resolution and reference papers associated to each star/stellar system.}
 \tablebib{B96:~\cite{burrows96}; C04:~\cite{chakraborty04}; C08:~\cite{connelley08}; C09:~\cite{connelley09}; C06:~\cite{correia06}; D15:~\cite{daemgen15}; D99:~\cite{duchene99}; D03:~\cite{duchene03}; D07:~\cite{duchene07}; D16:~Duch\^ene et al. (in prep.); F14:~\cite{difolco14}; G93:~\cite{ghez93}; G97:~\cite{ghez97}; I05:~\cite{itoh05}; I08:~\cite{ireland08}; K98:~\cite{koehler98}; K07:~\cite{konopacky07}; K00:~\cite{koresko00}; K06:~\cite{kraus06}; K11:~\cite{kraus11}; K12:~\cite{kraus12}; L93:~\cite{leinert93}; L97:~\cite{leinert97}; M08:~\cite{monnier08}; P99:~\cite{padgett99}; R93:~\cite{reipurth93}; S98:~\cite{sartoretti98}; S95:~\cite{simon95}; S99:~\cite{simon99}; S05:~\cite{smith05}; T10:~\cite{todorov10}; T14:~\cite{todorov14}; W01:~\cite{white01}; W16:~White et al. (in prep.)}
 \end{landscape}
\twocolumn
}

\addtocounter{table}{-1}

 \begin{table}[b]
 \centering
 \caption{Sub-samples of stars in Taurus}
\begin{tabular}{|r|r|r|r|r|}
  \hline
Sub-sample & Spatial region   & Class & HAR  & N\\
  \hline
$S1$ &whole&  All & No& 338\\
$S2$ &$W_{in}$ & All & No   &252\\
$S3$ &whole  & II, III & Yes  &179\\
     \hline
\end{tabular}
\tablefoot{Sub-samples of our star catalog depending on their spatial location  (whole region of Taurus or within $W_{in}$, see Fig.~\ref{Fig:TauG_Dobashi2005_DHT2001}), their Class, or whether or not they all have been observed at HAR. N: total numbers of stars within the sub-sample.}
\label{Tab:TauG_star_samples}
\end{table}

Although a recent catalog of Taurus members has been released including newly detected mid-infrared  WISE (Wide-field Infrared Survey Explorer)  sources  \citep{EsplinEtAl2014}, we adopted the catalog containing  352 Taurus members that offers  a full census of members down to $0.02\, M_{\odot}$ \citep{LuhmanEtAl2010, RebullEtAl2010}, which we supplemented with  stellar multiplicity data.
The newly identified WISE members are mainly due to a wider coverage of the Taurus region compared to Spitzer's, similar to the new candidates recently identified in Taurus using the  Sloan Digital Sky Survey \citep{LuhmanEtAl2016Arxiv}. These new members belong primarily to the dispersed stellar population and have typically not been observed at high angular resolution (HAR), hence multiplicity information is not complete for these sources. Therefore, setting this population aside should not significantly affect our conclusions surrounding local pairing statistical properties. Moreover, the   \cite{LuhmanEtAl2010}  catalog provides
a full and coherent analysis of SEDs  as obtained from the Spitzer Infrared
Array Camera  and Multiband Imaging Photometer, 
leading to a system of four Classes of SED (Class 0, I, II, and III objects) that we will use in our statistical studies. This  classification  is widely used to assess the evolutionary state of the  material surrounding the star  \citep{LadaWilking1984, Lada1987,AndreEtAl1993,GreeneEtAl1994}. 
Further studies have revealed that pure SED analyses can be misleading \citep{CarneyEtAl2016} for individual objects. However, although uncertainties remain on the individual classification, there is a clear statistical sequence of ages as a function of SED classification when the population is taken as an ensemble.

Taurus is known to be the archetype of low-mass star formation in loose groups \citep{JonesHerbig1979,GomezEtAl1993,KirkMyers2011}. The star mass estimates  were taken from the latter work exploiting the same catalog of stars from \cite{LuhmanEtAl2010}. We note that the stellar masses  have been estimated from their spectral type while assuming a constant age of 1--2 Myr for all stars of the complex and using
an ad-hoc isochrone at 1--2 Myr composed of a sequence of several theoretical evolutionary tracks depending on the mass range (see \citeauthor{KirkMyers2011}  \citeyear{KirkMyers2011} for details). Since a recent careful analysis based on optical spectroscopic study of young stars in star-forming regions, including Taurus, 
has shown that spectral types measured over the past decade are mostly consistent with the new evaluation \citep{HerczegHillenbrand2014}, 
the main source of uncertainties surrounding mass estimation 
comes from the hypothesis of a unique age for all the members (1--2 Myr). 

The mass was estimated for each star using  the effective temperature converted from its spectral type and an ad-hoc isochrone at 1--2 Myr composed of a sequence of several theoretical evolutionary tracks depending on the mass range (see \citeauthor{KirkMyers2011}  \citeyear{KirkMyers2011} for details).  For a given spectral type, assuming a younger age for a star than reality will tend to underestimate  its mass. But, even if the absolute value of the mass is subject to great uncertainties (30 to 50\%), yet we do not expect a systematic differential bias in  mass estimation. Besides, there is no indication of a mixture of components in the age distribution that would suggest  the presence of distinct episodic generation of stars, nor any evidence of a systemic dynamical evolution that would divide the stars into distinct age-based populations. New generation of stars and dynamical evolution appear to be continuous processes that globally lead to a smooth dispersion of age. For instance, multiple systems and single stars have a same mean age and display the same age dispersion \citep{white01}.  
Thus, assuming a single age for all the stars will bias the absolute value of each derived mass, but the relative difference between them, which is used in our analysis, should be much less affected.

\subsection{Multiplicity}
 To construct a census of stellar systems  in Taurus, we began by compiling  a list of all known multiple members of the region from almost three decades of increasingly higher-angular-resolution observations (see references in Table~\ref{Tab:stars}). We also assigned a flag to each star when it has been observed at HAR. 

In order to separately study the populations of single stars and multiple systems and to distinguish clustering from multiplicity, we grouped together all stars that are separated by less than $7''$ ($0.92 \mathrm{kAU} \sim 1$ kAU) to form a single entity.  Throughout this paper we refer to them simply as <<multiple systems>>. This threshold is, to a certain extent, arbitrary. However, two reasons motivate this choice. First of all, it is  the lower threshold that defines wide binaries (separation > 1 kAU) and secondly (the main reason), this threshold is close to the  $6\arcsec$ beam of the Spitzer-MIPS data. 
Thus,  we ensure that the census of neighbor stars beyond $1$ kAU is completely independent of the Class of the Taurus members.

 We assigned  the total number of stars within a 1 kAU  radius $n_*$ (for a single star $n_*=1$, for a binary  $n_*=2$, \ldots) to each entry of the catalog, and we note that this parameter is only reliable for Class II and III type stars that have been observed at HAR, 
 using techniques such as speckle interferometry imaging and adaptive optics  with spatial resolution between $0.05\arcsec$ to $2\arcsec$. 
These techniques probe the immediate neighborhood of a star, at least beyond $\sim 10 $ AU at the distance of Taurus (i.e., $5-20$ AU depending on the technique). 
Very close binaries (i.e., $\lesssim 10$ AU) most probably originate from a distinct scenario from other binaries, as they are probably not formed by fragmentation in situ \citep{BateEtAl2003}. 
We thus consider spectroscopic binaries and the closest visual binaries (below $\sim 20$ AU) as single stars to  ensure the consistency and completeness of our visual binary analysis, since no exhaustive spectroscopic binaries survey has been performed till now.

 The resulting catalog (see Table~\ref{Tab:stars}) contains 338 sources. Of these, 250 have been observed at HAR and amongst the latter 94 are multiple systems, resulting in a mean multiplicity fraction (MF)  of $40 \%,$ which could truly be as high as  54\%, if  all the remaining stars that have not been observed at HAR are multiple systems. A similar multiplicity fraction is obtained when considering only Class II/III stars (ie., ${\rm MF} = 40\% \pm 10\% $). The  mean companion frequency for Class II/III stars observed at HAR is 48\%.  We note that a  chi-squared test shows that the proportion of Class I, II, and III in the star sample observed at HAR is  statistically the same  when considering either  single stars or multiple stars. The disk fraction (the number of Class II stars over the number  of Class II and III stars) is also statistically the same in the two populations. Using Class as a proxy for the age, we obtain, as already stated by \cite{white01} that uses theoretical stellar evolution models, that the two populations share the same age and the same range of age dispersion.

\subsection{Samples}
The sources in Taurus, as was first pointed out by \cite{GomezEtAl1993}, are not randomly distributed on the sky (see Fig.~ \ref{Fig:TauG_Dobashi2005_DHT2001}). Some sets of stars appear grouped at a global scale in two structures located North (in L1507 molecular cloud) and South-West (in L1543 molecular cloud) with respect to the three main filaments, while some of them are tightly subgrouped at small scales within the main central filaments. Yet, other stars appear more spatially dispersed between the main structures. One of the goals of this work is to more precisely quantify their spatial distribution based on nearest-neighbor analysis, now that the census of star population in Taurus has more than doubled since the seminal work of \cite{GomezEtAl1993}.

From our catalog we extracted three distinct samples (defined in Table~\ref{Tab:TauG_star_samples}) that we use to test different hypotheses throughout this work, notably for multiplicity testing.

\section{Local spatial analysis}
\label{Sec:NNS}

In this section,  we perform the spatial analysis of stars in the entire Taurus region and we assess the difference in the spatial distribution between multiple systems and single stars based on their respective first nearest neighbor separation (1-NNS) statistics. Nearest-neighbor statistics  have  been used for decades in Taurus
to; (1) demonstrate departures from random uniform distributions 
\citep{GomezEtAl1993}, (2) show that the spatial distribution of brown dwarfs and stars are undistinguishable \citep{Luhman2004b}, and (3) study the spatial distribution of stars as a function of their Class  \citep{LuhmanEtAl2010}.
We start with some preliminary work on  1-NNS statistics 
 to advocate the use of the theoretical 1-NNS distribution derived for a random distribution in an infinite medium as a reliable proxy for a random distribution enclosed in a  finite and irregularly shaped window provided that the window is large and well-populated.  
We then define the one-point correlation function $\Psi$, which complements the two-point correlation function, to study the binary regime. This function allows us to quantify the departure of a population from a random uniform one and to identify distinct spatial regimes (clustering, inhibition, and dispersion). 
 
 Most of our work was performed using the free R software environment for statistical computing and graphics \citep{R-ManualR2015} using a set of packages  such as {\it Spatstat, Hmisc, fields, FITSio, calibrate, xtable, astrolab, deming, mixtools,} and {\it magicaxis}    \citep{R-Spatstat2005,R-Hmisc2015, R-fields2015,R-FITSio2013,R-calibrate2013,R-xtable2014, R-astrolab2014, R-deming,R-mixtools2009, R-magicaxis}.

\subsection{1-NNS study in Taurus}

\begin{table*}[ht]
\centering
\caption{Moments  of the 1-NNS distribution in Taurus}
\begin{tabular}{|r||r|rrrrr|}
  \hline
Spatial window & Stars sample &  $\overline{r}$ [pc] &  $ r_{0.5}$ [pc] & $\sigma^2$ [ pc$^2$] & $ \sigma$ [pc] & $\gamma$  \\ 
  \hline
    \hline
whole region  Taurus& $S_1$ & 0.38 $\pm$ 0.04 & 0.17$\pm$0.06 & 0.55$\pm$ 0.05  & 0.74 $\pm$ 0.03& 5.91 $\pm$ 0.32 \\ 
     \hline
$W_{{\rm in}}$ Taurus  & $S_2$  & 0.28 $\pm$ 0.02& 0.16 $\pm$0.03& 0.14$\pm$ 0.01 & 0.37 $\pm$ 0.02& 3.18 $\pm$ 0.19\\ 
     \hline
     \hline
  Infinite  &Rand. Theo  & 0.54 & 0.50 & 0.08 & 0.28 & 0.63 \\ 
     \hline
$W_{{\rm in}}$  &Rand. MC   & 0.55 $\pm 0.02$ & 0.51 $\pm 0.02$ & 0.09 $\pm 0.01$& 0.30 $\pm 0.01$& 0.84 $\pm 0.05$ \\ 
     \hline
\end{tabular}
\tablefoot{The moments   of the 1-NNS distribution are given for:
all stars in the whole Taurus region (sample $S_1$, see Table~\ref{Tab:TauG_star_samples}),  stars within the $W_{{\rm in}}$ window (sample $S_2$, see Table~\ref{Tab:TauG_star_samples} and Fig.~\ref{Fig:TauG_Dobashi2005_DHT2001}), the random  theoretical 1-NNS distribution <<Rand. Theo >> in an infinite medium  $w_R(  r )=  2 \pi \rho r \exp(- \pi \rho r^2)$ (equation \ref{Eq:P_nnd_k1}), and  the 1-NNS distribution  obtained from the 10\,000
random Monte Carlo  samplings <<Rand. MC>>  within the $W_{{\rm in}}$ window. The two random distributions are computed using the same intensity process $\rho=5 \,\deg^{-2}$ (see equation \ref{Eq:P_meanRho_wind_taurus}). The uncertainties are evaluated from equations \ref{Eq:NND_k1_pdf_MC_resume}, with N=252.}
\label{Tab:TauRand_1NND_CharVal}
\end{table*}

This subsection aims at analyzing  the spatial distribution in Taurus based on a 1-NNS analysis. The 1-NNS of a star is, by definition, the distance to its nearest neighbor star.  Throughout this paper, we deal with the projected 1-NNS, computed for each star as the minimum value of projected angular distance between each pair of stars onto the celestial sphere (see Appendix \ref{Appendix_SpheGeo}).

After estimating the mean surface density of stars, we first compare the Taurus 1-NNS distribution in the three main filaments to that associated to the entire region. We then compare the theoretical random 1-NNS distribution obtained for an infinite random spatial process to the 1-NNS  distribution derived from Monte-Carlo samplings in a finite irregularly shaped window, and finally we compare Taurus 1-NNS distribution to the  1-NNS random distribution.

 \subsubsection{Mean stellar surface density in Taurus}
\label{Subsubsect:MeanRho}

\begin{figure}
\includegraphics[width=\columnwidth]{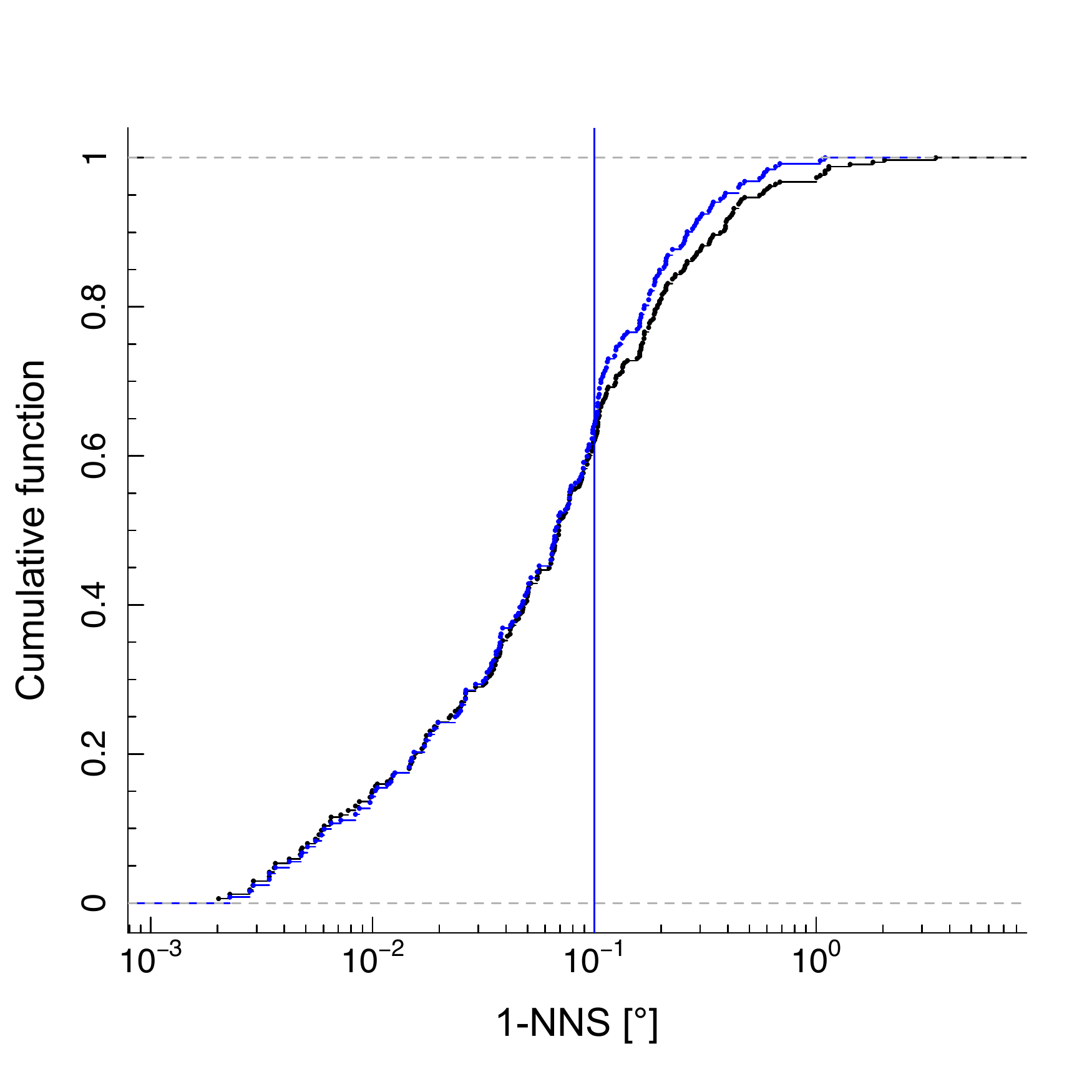}
    \caption{Empirical cumulative distribution of the 1-NNS distribution for the whole region (black) and for the central part within $W_{{\rm in}}$ region (blue). The blue vertical line represents the clustering length, $r_c=0.1^\circ$.}
   \label{Fig:TauG_ECDF_Whole_testowin}
\end{figure}

To derive an estimate of the mean surface density $\rho_w$ of the spatial process in Taurus, we focus on its central part  as the least heterogeneous area and define 
 the spatial polygonal window  $W_{{\rm in}}$ that encloses the central three main filaments  (see Fig.~\ref{Fig:TauG_Dobashi2005_DHT2001}), to get:
\begin{equation}
\rho_w=N_w/A_w \sim 5 \,\deg^{-2} \sim 1 \, {\rm pc}^{-2},
\label{Eq:P_meanRho_wind_taurus}
\end{equation}
where $A_w= 48.5\deg^2$ (i.e. $ 289.57 \,{\rm pc}^2$) is the projected area of the window $W_{{\rm in}}$ that encloses $N_w= 252$ stars. This window defines our star sample $S_2$ (see Table~\ref{Tab:TauG_star_samples}). 

In the following paragraph, we estimate the uncertainty of the mean projected stellar surface density of Taurus.
A conservative estimate of the uncertainty on the surface density may be obtained when choosing for the window, the convex hull 
of the stars enclosed within  the polygonal window $W_{{\rm in}}$. The convex hull  of a set of points in the Euclidean space  is defined as  the smallest convex set of points that contains the whole set of points.The area of the convex hull, $A \sim 33\deg^2$,  being less than the polygonal window, the surface density estimate mechanically increases by a factor of $1.6$ ($N_w/A \sim 8 \,\deg^{-2}$). 
Conversely, taking the convex hull area of the whole Taurus region ($\sim 202\deg^2$) containing 338 objects lowers the mean surface density to $ \sim 2 \deg^{-2}$. Combined, these extreme cases yield a probable range for the projected stellar density of Taurus of
\begin{equation}
\rho \sim 5 \pm 3  \,{\rm stars /\deg^{-2}}.
\end{equation}

\subsubsection{1-NNS distribution across Taurus}

The 1-NNS median value evaluated within the full region of Taurus (see  Table~\ref{Tab:TauRand_1NND_CharVal}) does not differ from the 1-NNS median value evaluated within the three main filaments, 
and their 1-NNS cumulative distributions do not differ either for length scale below   $r_c\sim0.1^{\circ}$ (see Fig.~\ref{Fig:TauG_ECDF_Whole_testowin}). This indicates that the spatial properties of the spatial distribution of stars are similar across the whole region of Taurus below this angular scale. Above the $0.1^{\circ}$ scale,  the two cumulative distributions diverge due to the presence of highly dispersed stars between the three main filaments and other main stellar concentrations in Taurus. As a result of these, the 1-NNS distribution of the whole region has a more populated large-scale tail. Nonetheless, up to the third quantile, the  1-NNS quartiles are similar. As a matter of fact, due to its highly asymmetrical shape, the Taurus 1-NNS distribution is better evaluated using quantiles rather than the mean and standard deviation that are more conveniently suited for symmetrical functions.

\subsubsection{1-NNS distribution: theoretical random vs Monte-Carlo samplings }

\begin{figure}
\includegraphics[width=\columnwidth]{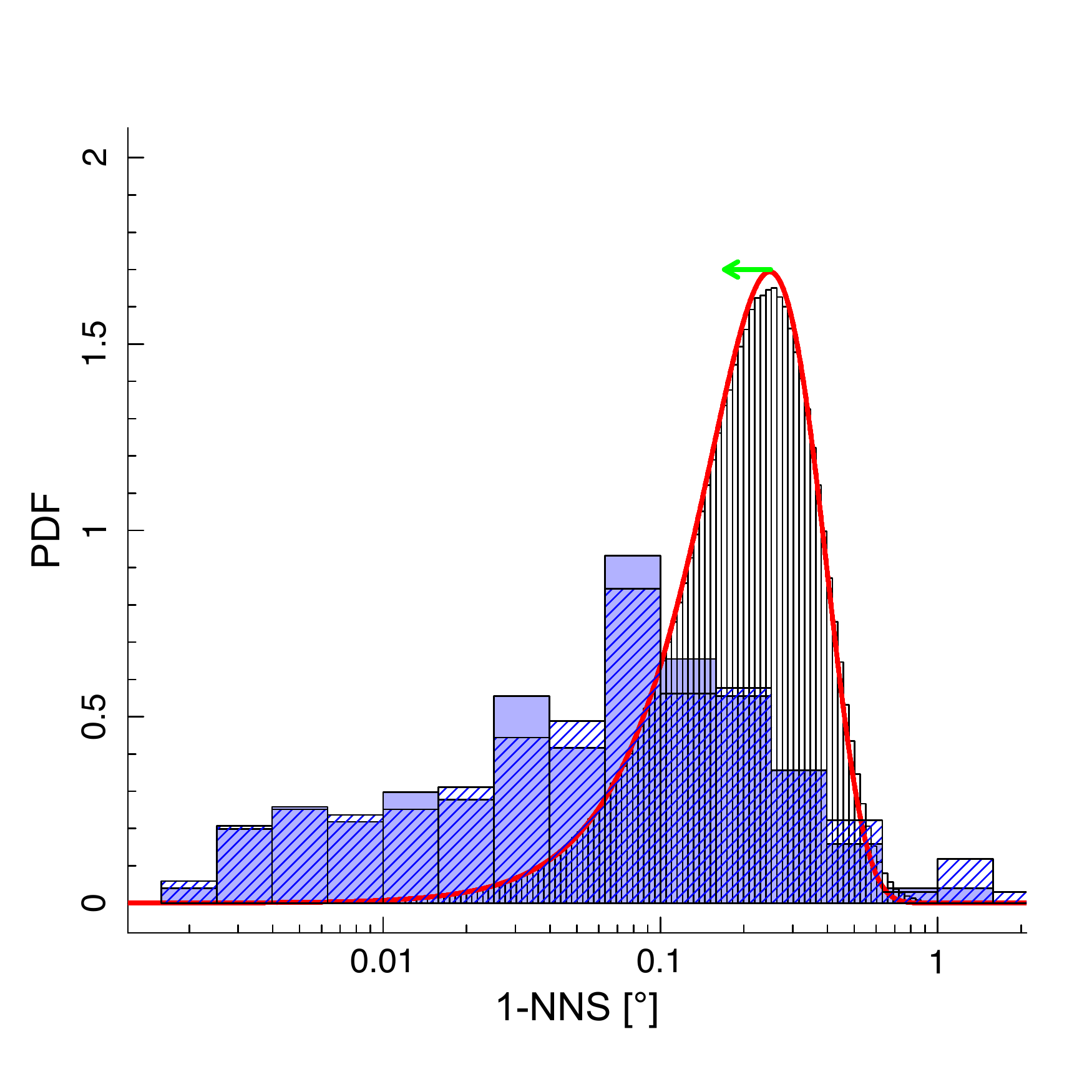}
\caption{Probability distribution function (PDF) of the 1-NNS distributions observed in Taurus (solid blue histogram: within the limited window $W_{{\rm in}}$, that is, the three main filaments; dashed blue histogram: the entire Taurus region) and predicted for a random distribution (black histogram: for a Monte Carlo sampling within the window $W_{{\rm in}}$; red curve: unlimited theoretical random 1-NNS distribution, see equation \ref{Eq:1-LogNNS}). Both random populations are drawn using the same mean surface density $\rho_w=5\deg^2$.  Increasing the mean density by a factor of two  leads to a shift  towards the left whose amplitude is represented by the green arrow. The Monte-Carlo sampling in a finite window and the unlimited theoretical random 1-NNS do not significantly differ. The  1-NNS distributions computed either in the whole Taurus or within $W_{{\rm in}}$ do not significantly differ either. However, the Taurus and random 1-NNS distributions are statistically highly inconsistent, with a p-value of $\sim 10^{-16}$ for the KS test. The spatial distribution in Taurus is clearly far from being random.
}
 \label{Fig:NND_k1_TauG_theo_MC_means}
\end{figure}
In this subsection, we compare the theoretical 1-NNS distribution for a spatially random process in an infinite media to the 1-NNS distribution obtained from Monte-Carlo samplings in a finite, irregularly-shaped window to establish that the former can be used as a reliable proxy for the latter.

The random theoretical  1-NNS distribution $w_R( \log r )$ derived for an infinite media (see Appendix \ref{Appendix_k-NND} for the full development) is given by: 
 \be
w_R( \log r ) = 2 \, \ln(10)  \, \pi  \,  \rho  \,  r^2  \,  \exp(- \pi \rho r^2),
\label{Eq:1-LogNNS}
\ee
 where $\rho$ is the mean intensity of the random process (the average surface density of stars) and $r$ is the 1-NNS. We want to compare this theoretical distribution to the 1-NNS   histogram obtained from   10\,000 random Monte-Carlo samplings within the finite and irregularly-shaped window $W_{{\rm in}}$ (see Fig.~\ref{Fig:NND_k1_TauG_theo_MC_means}). Their respective moments are similar (see Table~\ref{Tab:TauRand_1NND_CharVal}), with theoretical moments computed from the following expressions:

\begin{subequations}
\label{Eq:NND_k1_pdf_theo_resume}
  \begin{equation}
  \begin{array}{lll}
\bar{r} &\simeq &1/(2 \sqrt{\rho}),
\label{Eq:NND_k1_pdf_theo_resume:mean}
\end{array}
\end{equation}
  \begin{equation}
    \begin{array}{lll}
\sigma^2 & = & (4- \pi) (4 \pi \rho)^{-1},
\label{Eq:NND_k1_pdf_theo_resume:sig}
\end{array}
\end{equation}
  \begin{equation}
    \begin{array}{lll}
\gamma &=& 2 \sqrt{\pi} (\pi-3)/(4-\pi)^{3/2},
\label{Eq:NND_k1_pdf_theo_resume:gam}
\end{array} 
    \end{equation}
  \begin{equation}
    \begin{array}{lll}
    r_{0.5}&= & \sqrt{\ln 2/ (\pi \rho)},
    \label{Eq:NND_k1_pdf_theo_resume:med}
    \end{array}
    \end{equation}
\end{subequations}
where   $\overline{r}$, $r_{0.5}$,  $\sigma^2$,  $\sigma^2$ , and $\gamma$ are the mean, median, variance,  standard deviation, and skewness factor 
(see Appendix \ref{Appendix_k-NND} for their derivation).
The moments from random  Monte Carlo samplings 
were obtained from:
 \begin{equation}
\begin{array}{lll}
\overline{r}=\frac{1}{N-1} \sum_1^{N} r_{i} \pm \left[ N_i-1 \right]^{-1/2} \, \sigma_{Tau},\\
\sigma^2=\frac{1}{(N-1)} \sum_1^{N} (r_{i}-\overline{r})^2 \pm \sqrt{2} \left[  N-1 \right]^{-1/2} \sigma^2,\\
\sigma=\sqrt{\frac{1}{N-1} \sum_1^{N} (r_{i}-\overline{r})^2 } \pm \left[ 2\, (N-1) \right]^{-1/2} \sigma, \\
\gamma = \frac{N}{((N-1)(N-2))} \sum_1^{N} (r_{i}-\overline{r})^3 \pm \left (N-1 \right)^{-1/2} \gamma,\\
r_{0.5} = \Lambda_{0.5}  \pm  \sigma \sqrt{\pi/(2 (N-1))}
\end{array}
\label{Eq:NND_k1_pdf_MC_resume}
\end{equation}
where $\Lambda$ is  the ordered sequence of the 1-NNS from the smallest to the highest value, the standard errors of the moments and the standard  error  of the median  are taken respectively from \cite{AhnEtAl2003} and \cite{KendallStuart1977}.

We thus conclude that finite size and edge window effects   do not significantly alter the  distribution of 1-NNS from that obtained in an infinite medium at the level of our observational uncertainties. To prove this statement quantitatively, we use a bootstrapping technique, comparing 10\,000 realizations of a) a sample of $N \sim N_w=252 $ 1-NNS from standard rejection sampling technique using analytical random 1-NNS distribution (equation \ref{Eq:1-LogNNS}) to b) the 1-NNS of a sample of 252 random 2D spatial observations  directly enclosed within $W_{{\rm in}}$. 
The   two-sample Kolmogorov-Smirnov (KS) and Anderson-Darling (AD) statistical tests indicate that the two 1-NNS distributions cannot be statistically distinguished with a  mean  p-value  and standard deviation respectively of $0.36\pm0.28$ for the KS test and  $0.34\pm0.27$ for the AD test. 

We have therefore shown that the random theoretical 1-NNS  probability density function (equation \ref{Eq:1-LogNNS})  may be used  as a reliable proxy to describe a 1-NNS random distribution even in the case of a random population enclosed in  a finite and irregularly shaped window, such as  $W_{{\rm in}}$.  We mention, as an asset, that  using random theoretical 1-NNS distribution proxy avoids large  Monte Carlo sampling computations required to populate highly improbable bins of very small spacings associated with tightly clustered regions.

\subsubsection{Spatial distribution in Taurus}

The comparison between the Taurus and random 1-NNS distributions clearly reveals  an excess of short spacings in Taurus  (see Fig.~\ref{Fig:NND_k1_TauG_theo_MC_means}), for which the 1-NNS distribution has smaller central values (median and mean) and a wider dispersion  (see Table \ref{Tab:TauRand_1NND_CharVal}). The 2-sample KS
and AD tests, 
give a p-value less than $\le  2.2 \times 10^{-16} $ and   $\le 10^{-40} $ , respectively, that the two samples are mutually consistent. Thus, we can firmly reject the hypothesis that the  spatial distribution of stars within Taurus is random.

This result is not surprising, since the same conclusion was already reached by \cite{GomezEtAl1993} based on a sample of 139 stars. But the expansion of the stellar census by nearly 200 new stars in Taurus 
strengthens this conclusion. The median value of the 1-NNS distribution we derived is $r_{0.5}\sim 0.067^\circ$ ($\sim$ 0.16 pc, ie 33 kAU); almost half the value obtained by these authors. This is what we expect when  the spatial distribution of additional stars follows a similar spatial pattern from the original population, since from equation     \ref{Eq:NND_k1_pdf_theo_resume:med}, the median value is reduced by a factor of $\sim \sqrt{338/139}\sim 1.5$. 

This also underlines the modest usefulness of the absolute value of the median nearest-neighbor distance to determine clustering property when the stellar census is incomplete. In the following section, we introduce an alternative  criterion based on the one-point correlation function to quantitatively assess local departures from pure randomness and to determine characteristic clustering scales.

\subsection{One and two-point correlation statistics}
\label{Sect:OneTwoCorelFunct}

\begin{figure}
\includegraphics[width=\columnwidth]{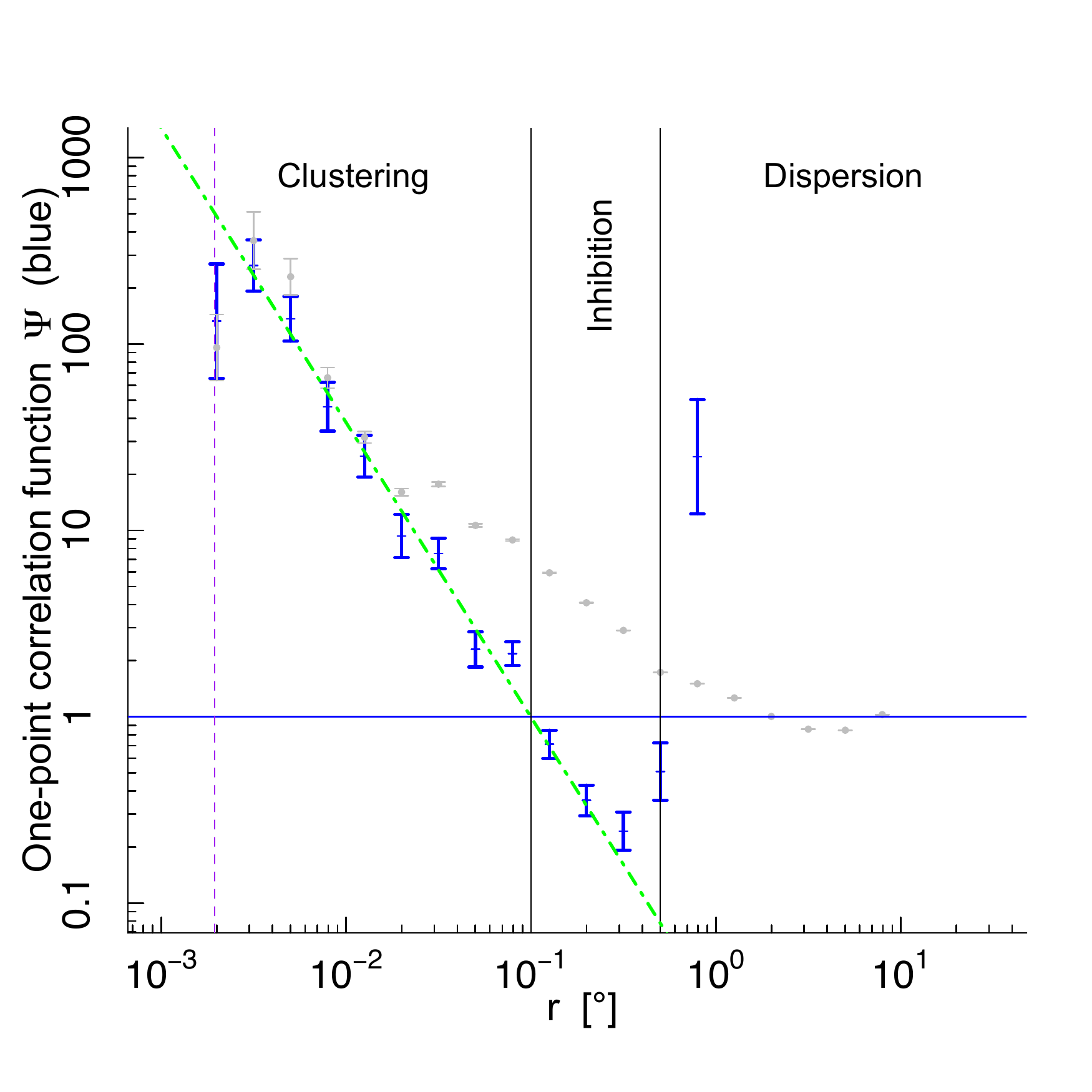} 
    \caption{
   One-point correlation function $\Psi$ (equation \ref{Eq:Psi}, blue symbols) and estimated pair correlation function  $g$ (gray symbols) using the sample of stars within $W_{{\rm in}}$.  The pair correlation function has been estimated  using the estimator given in equation \ref{Eq:pcf} while performing 10\,000 random Monte Carlo samplings within $W_{{\rm in}}$.  The $1\sigma$  vertical errorbars of $\Psi$ and $g$ are estimated from Poisson statistics.
    The solid horizontal blue line represents both the constant $\Psi=1$ and $\hat{g}=1$ functions, that is, the one-point and pair correlation function expected for a spatial random distribution. Vertical black solid lines delimit  three spatial regimes,  clustering, inhibition, and dispersion, derived from the crossing of the Taurus $\Psi$ function  with the $\Psi=1$ horizontal line associated to random spatial distribution.
The vertical dashed purple line indicates the 7" spacing threshold used to group stars into a unique multiple system. The dotted-dashed green line is the $\Psi \propto r^{-1.5}$ function resulting from a  linear regression fit taking into account  vertical errorbars 
 over ten 1-NNS logarithm bins ($\Delta \log r=0.2$). 
    The $\Psi$ function may reveal an extended binary regime  from $1.6$~kAU to $100$~kAU in Taurus, which 
has remained unidentified beyond approximately $5$~kAU in previous works that used the pair correlation function alone, since the latter reveals a power law break around that separation.
         }
   \label{Fig:NNDLog_k1Ratio_TauG_Theo}
\end{figure}

 In order to identify a  criterion that quantifies the departure of a spatial distribution of stars from  randomness, we introduce the one-point correlation function based on the 1-NNS distribution. This function is a new tool to assess and quantify the binary and spatial clustering regimes of stars and aims at complementing the  two-point correlation statistics.

 \subsubsection{The two-point and pair correlation functions}
The   two-point correlation function (TPCF) $\xi_V({\bf r})$ is defined for a 3D spatial process as a measure of the  probability   to get any excess of stellar pairs with a separation distance vector  ${\bf r}$ above random expectation: $dP = \rho_V^2 (1+\xi_V({\bf r})) {\rm d}V_1   {\rm d}V_2 $, 
 where $\rho_V$ is the mean volumic density of the uniform random process and dP is the infinitesimal joint probability to find a pair of stars respectively centered in the volume elements ${\rm d}V_1$ and ${\rm d}V_2$ and separated by {\bf r}   \citep{Peebles1980}. 
Similarly, for an   isotropic and stationary spatial process, the standard projected angular two-point correlation function 
 $\xi(r)$ is then defined as the  probability to find a pair of stars each separated by a projected angular distance $r$ above random expectation: $dP = \rho^2 (1+\xi(r)) {\rm d}\Omega_1 {\rm d}\Omega_2 $, where $\rho$ is the density of stars per steradian
and dP is the infinitesimal joint probability to find a pair of stars each within a unit solid angle ${\rm d}\Omega_1$, $ {\rm d}\Omega_2 $ and separated by a projected angular distance $r$.

Besides the two-point correlation function, the second order statistics may also be  evaluated from related entities:
 \begin{itemize}
\item  the main surface density of companions (MSDC) above random expectation: $\Sigma(r)=\rho \xi(r)$;
\item  the pair correlation function  $g(r)$ as a measure of the probability to have any pair of stars separated by $r$: $g(r) = 1+\xi(r)$.
 \end{itemize}
When the spatial distribution is random, all these functions are constant ($\xi(r)=0, g( r )= 1, \Sigma(r)=0$). Although different estimators may be used
to compute the  pair correlation function \citep{LandyEtAl1993,KerscherEtAl2000}, we use the simple and natural one defined as:
 \be
 \hat{g}(r) =\frac{N^{{\rm Pairs}}_{{\rm Tau}} (r)}{N^{{\rm Pairs}}_{\rm R}(r)},
 \label{Eq:pcf}
 \ee
where $N^{{\rm Pairs}}_{{\rm Tau}} (r)$ and $N^{{\rm Pairs}}_{\rm R}(r)$ are the  histogram of separations between stars within Taurus and in a randomly distributed catalogue, respectively.

Since the pair correlation function estimate is mainly sensitive to the   shape and extension of the window, no straightforward theoretical expression can be derived  for $N^{{\rm Pairs}}_{\rm R}(r)$ on general grounds, even in  the random spatial distribution case. 
We then compute the pair correlation function using equation \ref{Eq:pcf} while performing 100\,000 random Monte Carlo samplings within $W_{{\rm in}}$ (see Fig.~\ref{Fig:NNDLog_k1Ratio_TauG_Theo}). 

 The  slope and the shape of the resulting Taurus pair correlation fonction are comparable to those obtained in previous works studying  the two-point correlation function  or the main surface density of companions 
\citep{GomezEtAl1993,Larson1995,Simon1997,GladwinEtAl1999,Hartmann2002,KrausHillenbrand2008}. 
Together with our own, these studies  show that the second order statistics exhibit at least  two distinct power-laws: a steep  binary/multiple regime at small scale and a smoother clustering regime beyond a breaking point. \cite{Larson1995} suggested that; (1) the observed break at 0.04 pc  ($8.25$ kAU) can be the signature of the transition from a binary to a clustering regime, (2) it could correspond to  the Jeans length for the gas in cool dense molecular cores, (3)  the scale-free binary regime may reflect the scale-free fragmentation of collapsing clumps, and (4) the self-similar clustering regime may reflect the hierarchical, fractal, and spatial distribution of the gas from which the stars formed.

Later on, \cite{KrausHillenbrand2008} 
found an extended transition zone between  the binary regime and  the global clustering of stars in groups rather than a sharp break between the two regimes. This transition zone  was found to begin at 0.02 pc (4.25 kAU) and end at  0.2 pc (42.5 kAU).  They proposed that part of this transition zone, which is almost flat from 0.11 pc (23.52 kAU) to 0.2 pc (42.5 kAU), is due to the random motion of stars  that may smooth out  primordial stellar association and lead to a quasi-constant surface density of pairs.

However, some caution has to be taken when interpreting the two-point correlation function  beyond the binary regime, especially when the spatial distribution of stars deviates from  isotropy. Indeed  the standard form of the two-point correlation function  is obtained with the hypothesis of a homogeneous and isotropic distribution. However, the Taurus complex is not isotropic, as evidenced by its elongated gaseous structures in which stars are preferentially located. The break in the two-point correlation function between the binary and the clustering regimes   may thus be partly due to  the  filamentary geometrical anisotropy,  rather than being only a signature of a hierarchical/fractal gas structure as proposed by  \cite{Larson1995}. 

Indeed, the observed break in the two-point correlation function  could also be linked to the observed width of filaments, which was found to be relatively universal at approximately 0.1 pc \citep{AndreEtAl2014} even though  \cite{YsardEtAl2013} found that filament width may vary  by up to a factor of approximately four. This spread of filament widths 
 may also contribute to the transition elbow zone that is present between the binary and clustering regimes.  Thus, the elbow in the correlation function that extends up to 0.25 pc may be  related to  geometrical spatial stellar structures (see Joncour et al. in prep.).

  \subsubsection{The one-point correlation function}
 Similar to the pair correlation function, we define the one-point correlation function, $\Psi$, as the probability of having a closest star located at $r$ from any chosen star at random, which in turn defines an equivalent local stellar density function $\sigma$: ${\rm d}P= \sigma (r ) {\rm dV}$, with $\sigma(r ) = \rho \Psi(r ) $. The $\Psi$ function, which describes the spatial location of stars, gives first order variation trends of the spatial process relative to a random distribution. Instead, the pair correlation function is associated to second-order characteristics  describing the spatial co-location of stars.
One estimator of  the $\Psi$  function may be defined  from the ratio 
of the 1-NNS Taurus distribution $w_{Tau}( \log r ) $ to the 1-NNS random distribution $w_{R}( \log r )$ obtained either by Monte-Carlo simulations or from theoretical random probability density function (equation \ref{Eq:1-LogNNS}):
 \begin{equation}
\Psi (\log r )  = \frac{w_{Tau}( \log r )}{w_R( \log r )}
\label{Eq:Psi}
.\end{equation} 
We note that this function is usually called the nearest neighbor ratio when it is evaluated as the average over the whole range of 1-NNS values. Here we evaluate it bin per bin of spacing. For a random spatial distribution, the one-point correlation function reduces to unity ($\Psi( r )=1$).

The one-point correlation  $\Psi$ function can be fitted  from 1.6 kAU to $\sim 100$  kAU, by a power law using Pearson's chi-squared linear regression taking into account vertical errorbars (see Fig~\ref{Fig:NNDLog_k1Ratio_TauG_Theo}):
\be
\Psi = \frac{w_{Tau}}{w_R} \propto  r^{-\alpha}
\label{eq:PsiPowerLaw}
,\ee
with $\alpha=1.5^{+0.2}_{0.3}$ at the 95\% confidence level. 
This allows us to derive a fractal dimension $D_f$ associated to the binary regime in Taurus. Since  
the random distribution  in 2D increases as $w_{R} \propto r^2$  for $r \ll 1/\sqrt{\pi \rho}$ (see Appendix \ref{Appendix_k-NND}),  the 1-NNS distribution within Taurus  $w_{Tau}$ follows a shallow, rising power-law function with the following exponent: 
\be
D_f \sim 2- \alpha =0.5.
\label{FracDimNNS}
\ee
We note, that this scale-free behavior of the $\Psi$ function with an exponent lower than the projected 2D euclidian space dimension is equivalent  to a uniform distribution in a 2D projected fractal space  with the fractal dimension $D_f=0.5$ \citep{SakhrNieminen2006}.

 \subsubsection{Departure from randomness}
 
 The one point correlation  function $\Psi$ may be used to assess departure from a spatially random distribution.
From its definition, regimes where $\Psi$ is below unity indicate a deficit of stars (inhibition) with respect  to random, whereas where it is above unity reveals an excess of stars, 
either a clustering trend at small spacings (aggregation) or a dispersion trend at large spacings. The $\Psi$ function evaluated for Taurus thus reveals  three spatial regimes  (see Fig.~\ref{Fig:NNDLog_k1Ratio_TauG_Theo}). The stellar clustering regime covers the range of spacings below $\sim 0.1^{\circ}$, where $\Psi (r)> 1$. This  defines the upper clustering  length threshold $r_c$:
\be
r_c = 0.1^{\circ} \sim 0.24 \, \rm{pc} \sim 50\, \rm{kAU}.
\label{Eq:TauG_rc}
\ee
The clustering regime is followed by an <<inhibition zone>> where $\Psi (r)< 1$ associated with a typical length scale  between $\sim 0.1^{\circ}$ ($0.2\, \rm{pc}$) and $\sim 0.3^{\circ}$  ($ 0.7\, \rm{pc}$). Finally, Taurus faces an excess of highly dispersed  stars for typical length scale  above $\sim 0.3^{\circ}$. 
 The <<clustered stars>> are tightly packed together in specific locations while the <<inhibited stars>> are more widely spread between the zones of clustered stars (see  Fig.~\ref{Fig:TauG_Dobashi2005_DHT2001}).

 \subsubsection{Ultra-wide binary regime in Taurus?}

The comparison between the pair correlation and the $\Psi$ functions yields new insight into an extended binary regime. Both of them have the same scale-free trend    at least in the four logarithm bins of separation (see Fig.~\ref{Fig:NNDLog_k1Ratio_TauG_Theo}), from $1.5$  to $5$ kAU. While the correlation function reaches an elbow from that point on, breaking from the binary regime to reach the clustering regime, the $\Psi$ function extends the same power-law trend up to $\sim 100$ kAU.  Thus, although it remains hidden in the correlation function because of mixing between binarity and clustering, the binary regime in Taurus appears to extend to much larger scales than previously realized.
 
 \longtab{
 \onecolumn
 \begin{landscape}
{\scriptsize
\begin{longtable} {|r|rrrrr|rrrrr|rrrrr||}
\caption{\label{Tab:NNSCouples} Catalog of the first nearest neighbor couples of stars}\\
  \hline
 \#  Couple  & Star \#A & & Name (Star A) & $n_{*A}$ & Class (Star A) &Star \#B &  &Name (Star B)& $n_{*B}$ &Class  (Star B)& $\delta_{P}$ [AU]& $n_{*P}$ & HAR\tablefootmark{a}&WP [Conf.]\tablefootmark{b}\\ \hline
  \hline
  \endfirsthead
\caption{continued.}\\
\hline\hline
 \# Couple   & \# Star A & Sp. Bin. & Name (Star A) & $n_{*A}$ & Class (Star A) &\# Star B & Sp. Bin. &Name (Star B)& $n_{*B}$ &Class  (Star B)& $\delta_{P}$ [AU]& $n_{*P}$ & HAR\tablefootmark{a}&WP [Conf.]\tablefootmark{b}
 \\  \hline
\endhead
\hline
1 & 4 &  & HBC360 & 1 & III & 5 &  & HBC361 & 1 & III & 1 018 & 2 & ** & C   \\ 
  2 & 232 &  & IRAS04325+2402C & 1 & II & 233 &  & IRAS04325+2402A+B & 1 & I & 1 144 & 2 & - & C   \\ 
  3 & 244 &  & HPTau/G3 & 2 & III & 245 &  & HPTau/G2 & 1 & III & 1 407 & 3 & ** & C [Y] \\ 
  4 & 186 &  & GGTauBa+Bb & 2 & II & 187 &  & GGTauAa+Ab & 2 & II & 1 454 & 4 & ** & C [Y] \\ 
  5 & 294 &  & LkHa332/G1A+B & 2 & III & 295 &  & V955TauA+B & 2 & II & 1 462 & 4 & ** & C [Y] \\ 
  6 & 121 &  & FVTauA+B & 2 & II & 122 &  & FVTau/cA+B & 2 & II & 1 723 & 4 & ** & C [Y] \\ 
  7 & 290 &  & - & 2 & II & 291 &  & - & 2 & III & 1 725 & 4 & ** & C [Y] \\ 
  8 & 303 &  & HD30171 & 1 & III & 304 &  & IRAS04429+1550 & 1 & II & 1 814 & 2 & * & C   \\
\hline
    \hline
\end{longtable} }
\tablefoot{ 
Amongst the first-nearest-neighbor couples of stars (A,B), the ultra wide  Pairs  (UWPs) are defined as {\it mutual} first nearest neighbors. 
Column 1: Couple reference (this work). Columns 2--5~(Star A): the star reference (this work, see Table~\ref{Tab:stars}), presence of a spectroscopic binary associated to the star/stellar system A ("S" means one spectroscopic binary, "SS" means two spectroscopic binaries), the common Name of the star A, $n_{*A}$ the total number of stars, above the unity if the star is not single (spectroscopic binaries are counted as one component), the class of the (primary star). Columns 6--9~(Star B): the star reference (this work, see Table~\ref{Tab:stars}), presence of a spectroscopic binary associated to the star/stellar system A ("S" means one spectroscopic binary, "SS" means two spectroscopic binaries), the common Name of the star B, $n_{*B}$: the total number of stars, above the unity if the star is not single (spectroscopic binaries are counted as one component), the class of the (primary star). Columns 10--11: the separation of the two stellar components A \& B and their total number of stars as a (hierarchical) couple. Column 12: flag to indicate whether the each star A and B has been observed at high angular resolution (**), if only one component has been observed at HAR (*) or if none of them have been observed at HAR (no mark). Column 13: 
"C" indicates  a  ultra wide pair candidate since the couple are mutual nearest neighbours. The presence of an  [Y], this couple has been confirmed as gravitationally bound binaries using astrometric and spectroscopic techniques as reported in the work of \cite{KrausHillenbrand2009a} who focused on binaries with separation less than $4,200$ AU. The delimitation of separation range in their study is marked by a single horizontal line in the table.
}
 \end{landscape}
 \twocolumn
}
 
\addtocounter{table}{-1}

\subsection{Local neighborhood  of multiples versus single stars}
\label{SubSec:1-NNS-Mult-Sing}

\begin{figure}
\includegraphics[width=\columnwidth]{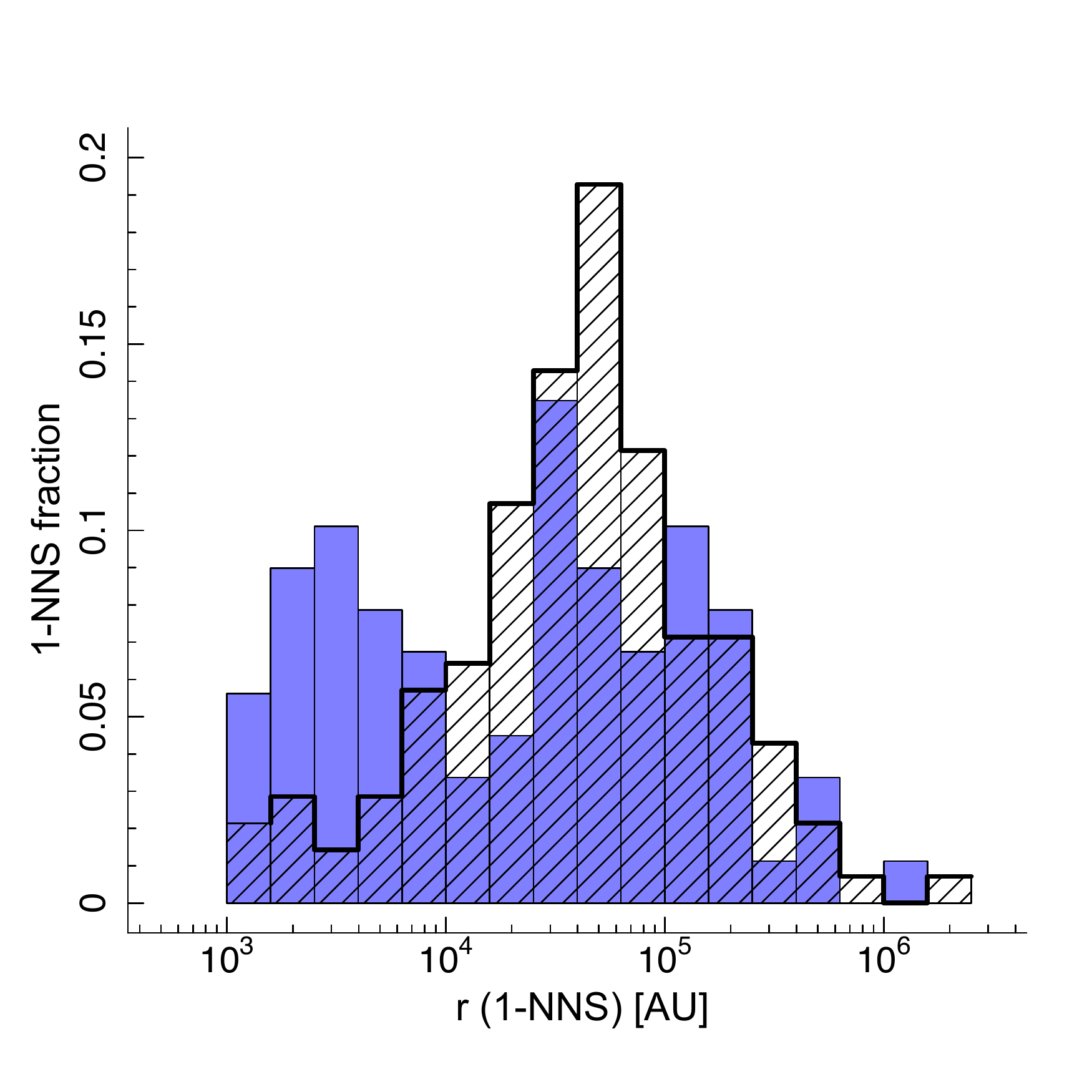}
    \caption{ First nearest neighbor separation fraction  distribution of multiple systems (solid blue histogram) versus single stars (dashed black histogram) for Class II and III objects  observed at high angular resolution (HAR) in Taurus region. Each bin represents the number density of stars per unit logarithmic interval in projected separation, that is, the number density fraction of 1-NNS per ($\Delta \log r = 0.2$) interval. 
There is a marked excess of 1-NNS in the range $1-10$ kAU for the multiple systems with respect to single stars. }
   \label{Fig:TauG_1-NND_Mult_Sing}
\end{figure}

In this subsection, we study the  spatial distribution  of  multiple systems and single stars  based on the comparison of  their 1-NNS statistics.  

The  $S_1$ sample in the whole Taurus contains 96 stars flagged as visual multiple systems (in the range of 10 AU to 1 kAU) and 242 {\it a priori} single stars. To limit biases, we define the  $S_3$ sample composed exclusively of Class II and III stars that have been observed at HAR. From that sub-sample, 
we separately construct the 1-NNS histograms of the 89  multiple systems and the 140 single stars (see Fig.~\ref{Fig:TauG_1-NND_Mult_Sing}) that appear very different one from each other.  Both the   KS and AD statistical tests indicate that we can reject the theory that they come from the same parent population with a confidence level of 99.8\%.   The 1-NNS fraction of  multiple systems having a  projected separation less than $10$ kAU  is noticeably higher than for  the single stars in the same range (see Fig.~\ref{Fig:TauG_1-NND_Mult_Sing}). 

The excess fraction of 1-NNS at close spacings is remarkably high: nearly 40\% of the 1-NNS of all multiple systems are located within the first five bins, that is, between $1$ and $10$ kAU. This is  more than 2.5 times  the fraction found  in the same range of separations for the single stars (15\%). It is also twice as high as
the frequency of visual companions per decade of projected separation found in the $10-2\,000$ AU separation range  (see \citeauthor{DucheneKraus2013} \citeyear{DucheneKraus2013}, Fig.~\ref{Fig:TauG_1-NND_Mult_Sing}).  Moreover, 63\% of the companions of multiple systems in the five first bins are themselves multiple systems (25\% of the total multiple systems), against 43\% for the singles (6\% of total single stars sample).

The local neighborhood of multiple stars is thus more populated  than that of simple stars:   a multiple system has almost three times more chance of having a companion within  $10$ kAU, and that companion in turn has a higher chance of being a close multiple system itself than a single star.

  \subsection{Beyond local neighborhood}
To test whether this striking difference between 
the neighborhood of multiple systems and single stars extends beyond the first neighbor, 
we analyse the second nearest neighbor distributions, again using the bias-free $S_3$ sample.
Although there is still a slight excess of companions for multiple systems within the first bin up to $10$ kAU,   
this difference does not appear statistically significant: the KS and AD tests both give a p-value of 0.15.  We further compare their respective  k-NNS distributions, up to k=8. They too cannot be statistically distinguished (p-value of 1 for both KS and AD tests). This confirms the statistical identity between the neighborhood of the two populations, multiples and singles, beyond their first neighbor. 

Since the 2-NNS distribution is not statistically different for single stars and multiple systems, we can use its median value  ($0.13^\circ$, 0.3 pc or $\sim 60$ kAU)
as a typical length scale at which these two populations cannot be statistically distinguished. This value is  close to the upper clustering threshold $r_c=0.24$ pc derived in the second section. 

Interestingly, we also note that the distribution of the second nearest neighbor of multiple systems cannot be statistically distinguished from the first nearest neighbor of single stars either (p-value of 0.1224 for the KS test). This suggests that, although the neighborhood of multiple systems has been found to be more populated within $10$ kAU, this trend fades away beyond $60$ kAU.  Based on the k-nearest neighbor analysis (k$\geq$2), we therefore define a  scale of $60$ kAU beyond which the stellar environment is statistically identical for single stars and multiple systems.

\section{From multiples to ultra-wide pairs}

\begin{figure}
\includegraphics[width=\columnwidth]{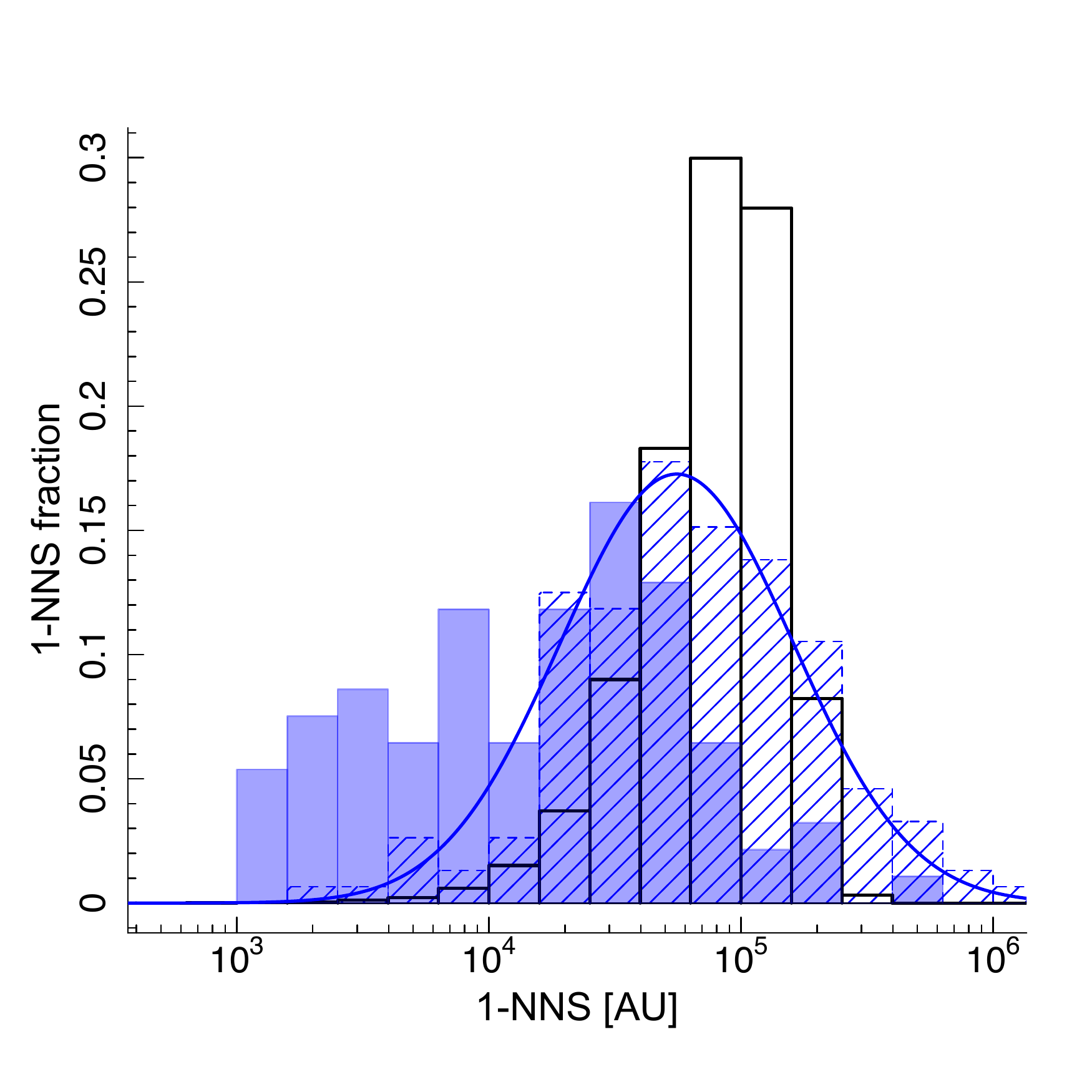}
    \caption{Distribution of ultra-wide pair fraction as a function of separation in the three main Taurus filaments (i.e., for stars enclosed within $W_{in}$, solid blue histogram) versus pair fraction of random mutual pairs (obtained from 10,000 Monte-Carlo samplings, thick black histogram) and  1-NNS fraction of non-mutual pairs (dashed blue histogram). The solid blue line represent a log-normal fit to the distribution of non-mutual pairs in Taurus. The distribution of mutual pairs of Taurus does not result from a random process (p-value of $10^{-16}$), and is statistically different from non-mutual pairs distribution in Taurus  (p-value of $10^{-14}$). Conversely, the  1-NNS distribution of non-mutual pairs in Taurus is reasonably well fitted (p-value of 0.2) by a long-normal function, with a mean value $\mu=4.74\pm0.04$ and a standard deviation $\sigma=0.46 \pm 0.03$.}
   \label{Fig:TauG_1-NNS_Frac_WP_NoWP_RandWP}
\end{figure}

In this section, we further analyze the reasons for such a divergence between neighborhoods of multiple systems and single stars.

\subsection{Mutual nearest neighbors as ultra-wide pairs}
\label{SubSect_UWP}
From the list of all nearest neighbor pairings (see Table~\ref{Tab:NNSCouples}), we selected the subset of all mutual nearest neighbors, that is, pairs in which the nearest neighbors are reciprocal. 
This property serves to probe the most `connected' pairs. It has been used, for instance, as part of the process to identify  physical binary candidates  in simulations \citep{ParkerEtAl2009,KouwenhovenEtAl2010}. 

We find that over  the 338 first nearest neighbor pairings of stars in Taurus, 45\% of pairings are { non-mutual pairs} 
 (within this type of couples, only one star is the nearest neighbor of the other, while the second one has a different nearest neighbor)  
whereas 55\% of the whole pairings  are  mutual pairs. 

The 1-NNS distribution for mutual pairs is markedly
different from that of non-mutual pairs; the KS test returns a p-value of $10^{-14}$ when assuming, as a null hypothesis, that these two populations are drawn from a same parent population (see Fig.~\ref{Fig:TauG_1-NNS_Frac_WP_NoWP_RandWP}). 
The distribution for non-mutual pairs is reasonably well fitted by a log-normal function (p-value = 0.2, showing that two distributions are statistically the same) with a mean $\mu=4.74\pm0.04$ (ie 55 kAU) and a standard deviation $\sigma=0.46 \pm 0.03$ ($\sim 20$ to $160$ kAU range). This suggests that, unlike the mutual pairs, the distribution of 1-NNS for non-mutual pairs in Taurus may be seen as  a statistical realization of a multiplicative process,  which is converted to an additive process in the log domain, for which the central limit theorem may be applied.  
Although it is beyond the scope of this paper to establish the  physical origin of this multiplicative process, we can speculate that relative random motion and gravitational encounters between physically unrelated stars may contribute  to randomization of the spatial location of individual stars. This is not the case for  gravitational binaries or common proper motion pairs, however, as they are kept  together at the same mutual nearest distance over time. 

 The  mutual pairs in Taurus are very unlikely   due to random mutual pairing  (see Fig.~\ref{Fig:TauG_1-NNS_Frac_WP_NoWP_RandWP}). A KS test comparing the distribution of 1-NNS separation of mutual pairs with the $W_{in}$ window to that of the mutual first nearest pairs from an ensemble of random Monte Carlo Poisson samplings within the same window returns a p-value lower than $10^{-16}$.

Moreover, it is worth noting that  the mutual pairs we identified in the separation range 1--5\,kAU, are already known to be true physical binaries, that is, gravitationally bound binaries \citep{KrausHillenbrand2009a}. 
This further validates our selection of mutual nearest neighbors to identify physically related pairs in that range.
We thus expect that a large majority of these mutual pairs to be at least common proper-motion pairs, and plausibly even gravitationally bound. From now on, we refer to these systems as ultra-wide pairs (UWPs).

The present-day distribution of separations for these UWPs should reflect its counterpart at birth, provided that  dynamical encounters are rare enough to keep it unchanged (see discussion section \ref{Sec:Discussion}). These mutual pairs covering  separations from $1$ to $\sim 60$ kAU are the only type of pairs (versus non-mutual pairs) found in the short-separation end of the 1-NNS distribution. Approximately 60\% (resp.  90\%) of 1-NNS have separations below 20 kAU (resp. 60 kAU) with a relatively long-uniform distribution. Thus,   the continuous power-law trend at small and intermediate scales observed in the $\Psi$ function is mainly due to the existence of these mutual pairs, which define the binary regime. They generate a scale-free clustering pattern and constitute a major structural spatial imprint in Taurus.

\subsection{An \"Opik law for the UWPs separation }

\begin{figure}
\includegraphics[width=\columnwidth]{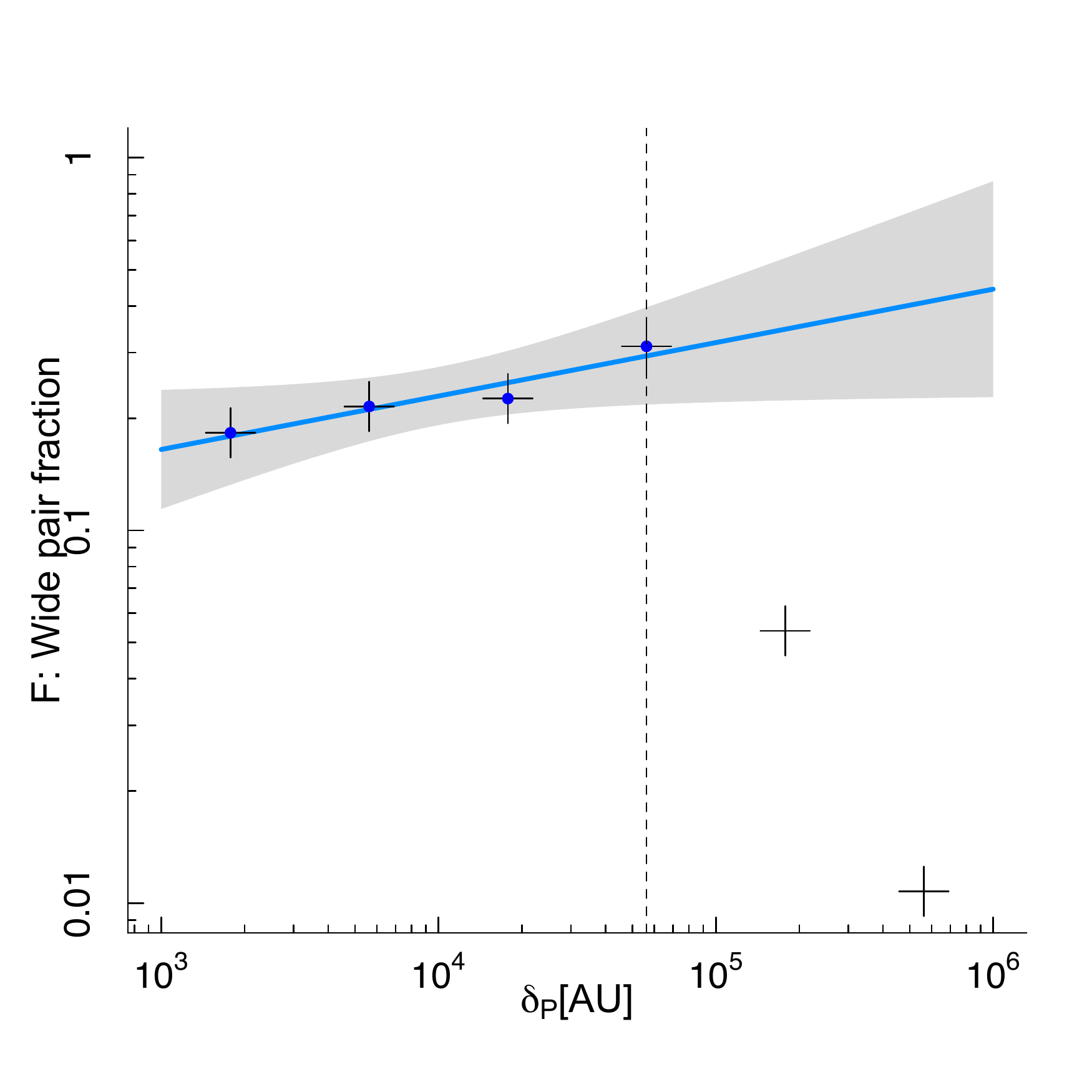}
                 \caption{ 
                 Ultra-wide pair fraction $F$ within the whole Taurus region (sample $S_1$) as a function of separation, plotted using $\Delta \delta_P=0.25$\,dex bins. The solid blue line is a power law fit to the data  ($F=\alpha \,\delta_P^{\beta}$, with $\log \,\alpha=-1.21$ and $\beta=0.14$) between 3 and $\sim$ 60 kAU (vertical black dashed line). The gray area represents the 95\% confidence band. In terms of the power law parameters, the corresponding parameter 95\% confidence intervals are $\log \,\alpha=[-1.78,-0.65]$, $\beta=[0,0.28]$.
                 }
   \label{Fig:TauG_WPairs_sep_v1}   
\end{figure}

In this subsection we discuss the distribution of UWP separation based on the data that we derived  (see Table~\ref{Tab:NNSCouples}).

The UWPs fraction F in Taurus  follows a  slightly increasing power law, 
$ F \propto \delta_P^{0.14}$ 
(see  Fig.~\ref{Fig:TauG_WPairs_sep_v1}), at least up to an upper cut-off of   $\sim 60$ kAU (i.e., 0.27 pc). The robustness of this simple fit is confirmed by performing a power law fit to the empirical cumulative function that explains 97.73\% of observed variations. Beyond this cut-off, there is a sharp decrease  in the two last bins of separations. 
The first two bins  overlap the study conducted by \cite{KrausHillenbrand2009a}, who found that the  separation of binaries below $4.2$ kAU is log-uniform in this range ($F(\log \delta)=Const.$ or, equivalently $F_O(\delta) \propto \frac{1}{\delta}$), i.e., following Opik's law.

\subsection{Class pairing in UWPs}

\begin{figure}
\includegraphics[width=0.45\columnwidth]{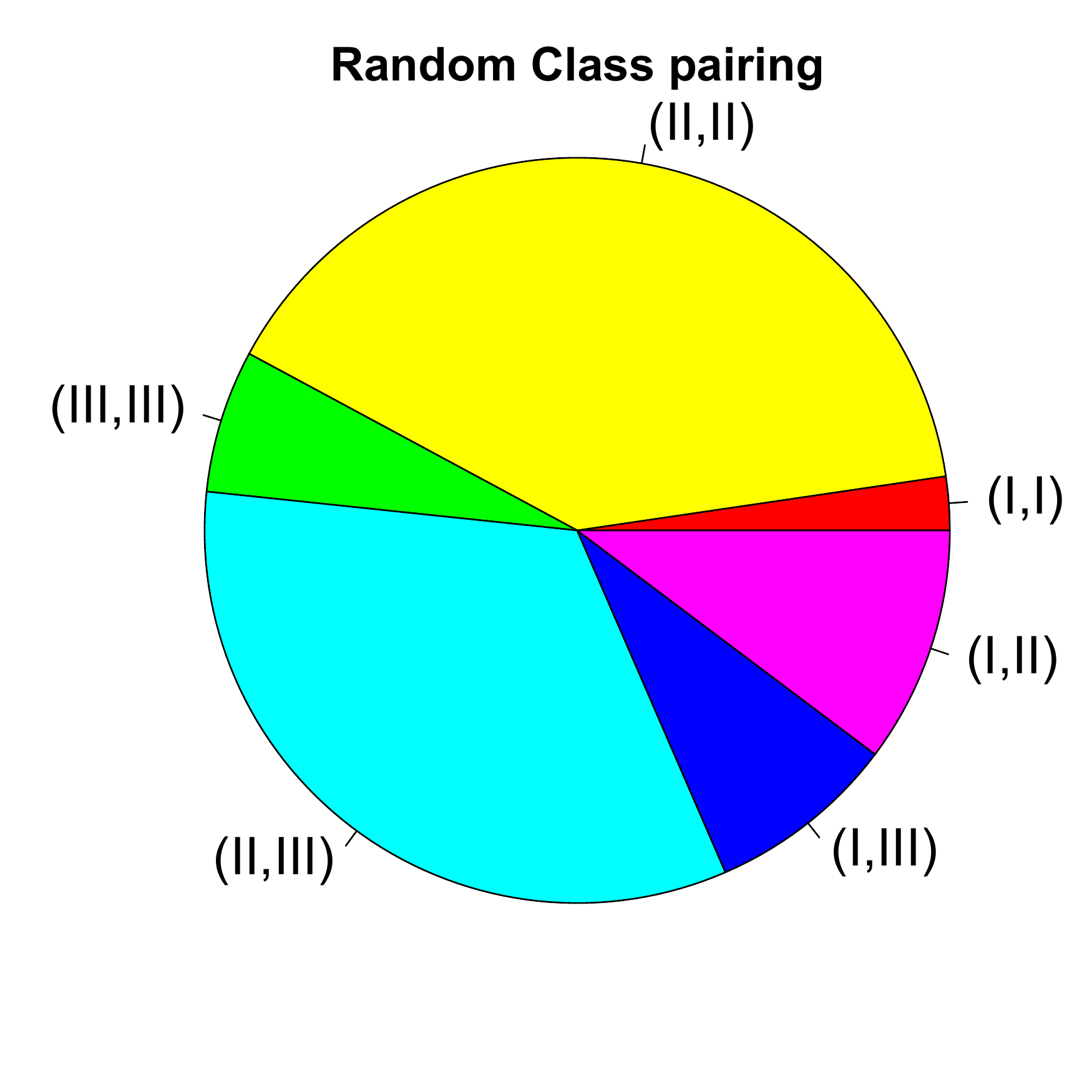}
\includegraphics[width=0.45\columnwidth]{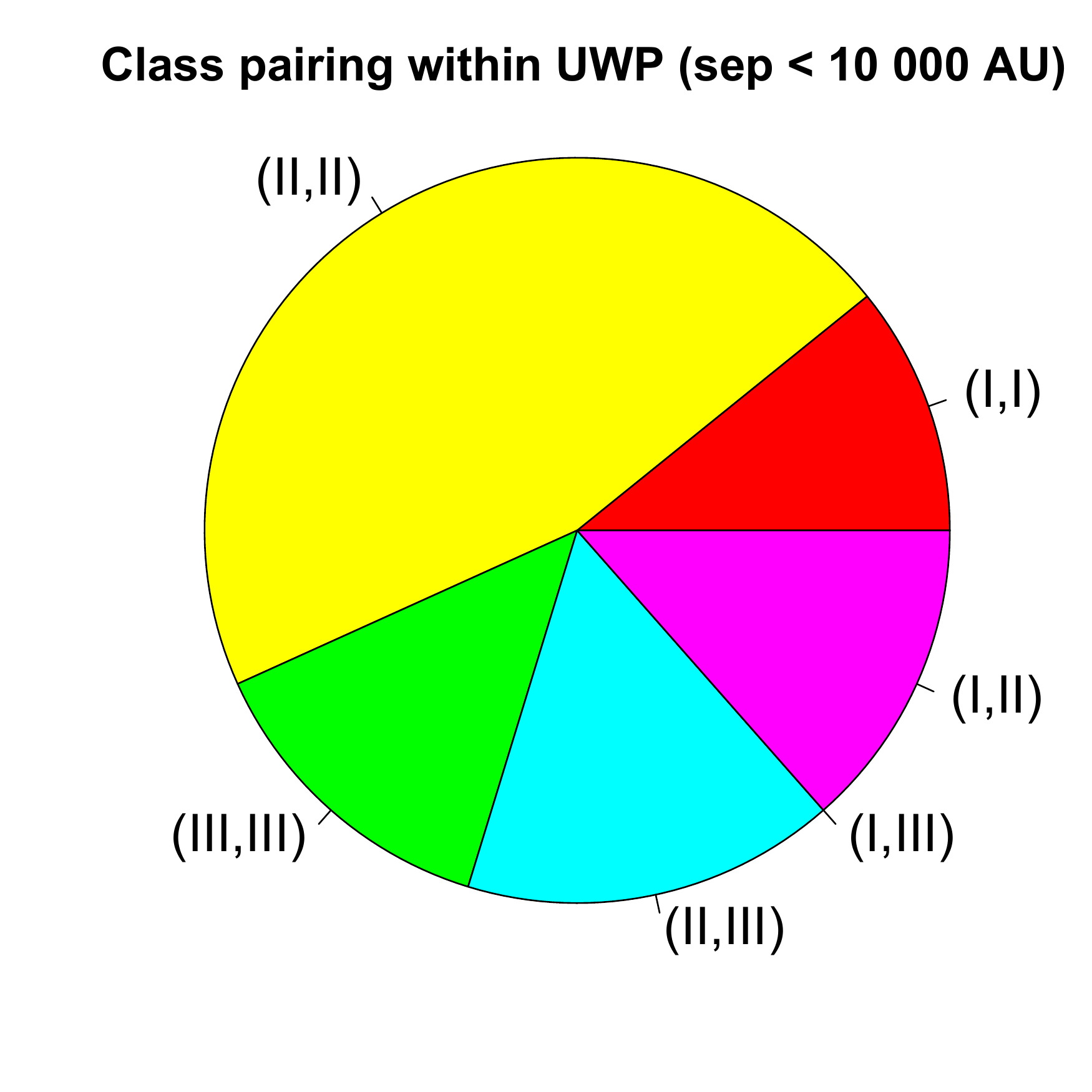}
\caption{Distributions of SED Class pairings expected from a random pairing process based on the global proportions of Class I, II, and III stars in Taurus (left) and observed among Taurus UWPs (right). The SED Class pairing within UWPs deviates significantly from random pairing, suggesting coevality within these systems.}
\label{Fig:TauG_WB_ClassPairs}
\end{figure}

We first focus  on the 37 mutual  pairs with separation less than 10 kAU (from sample $S_1$), a regime where the distinction between multiple systems and single stars strongly suggests that a physical process is at play. 
Within this sub-population, Class I and Class II are over-represented (and, accordingly, Class III under-represented) relative to the overall Taurus population (see Table \ref{Tab:TauG_Class_prob}). 
A chi-squared test confirms that this difference is statistically significant (p-value of 0.01). In other words, the stellar population within UWPs separated by less than 10 kAU is less evolved, i.e., younger, than the rest of the population in the Taurus molecular cloud.

\begin{table}[ht]
\centering
\caption{Class fraction in Taurus}
\label{Tab:TauG_Class_prob}
{\tiny
\begin{tabular}{r|rrr|r}
  \hline
 & Class I &Class II  & Class III & N\\
  \hline
All Taurus stars & 0.11&0.52 &0.37 & 338\\
Stars in UWPs ($\delta \le 10$ kAU) &0.18 &0.61 &0.22 &74\\
Stars in UWPs (all separations) &0.13 &0.56 &0.31 &186\\
  \hline
   \end{tabular}
\tablefoot{ N is the total number of stars in each subsample.}
   }
\end{table}

In the following, we show that the class pairing found in these UWPs deviates from random pairing. 
The probability $P_{ij} $ of getting an unordered pair  $(i,j)$  without replacement is given by: 
\be 
P_{ij} = \gamma \cdot P_{i} \cdot P_{j} = \gamma \cdot \left (N_i/N\right) \cdot \left (N'_j/(N-1) \right)\\
\label{Eq:ProbPairs},
\ee
where    $N$ is the the total number of stars in the sample under consideration,  $\{i,j\}=I, II, III$ is the SED Class of  stars, $P_{i}$ is the fraction of Class $i$ type star, $N_i$ is the number of stars of Class $i$, and $\gamma =1$ and $N'_j= N_i-1$ for $i=j$ whereas  $\gamma =2$ and  $N'_j= N_j$  for ($i\ne j$).

We compute the  random Class pairing probability for each type of pairings from equation \ref{Eq:ProbPairs} and the  number of stars of each Class $N_i$ derived from the first line in Table \ref{Tab:TauG_Class_prob}. These probabilities are quite different from those observed among UWPs in Taurus (see Table \ref{Tab:TauG_WB_prob} and  Fig.~\ref{Fig:TauG_WB_ClassPairs}). The main difference can be summarized as follows: there are more pairs of the same Class, that is, (I,I), (II,II) and (III,III) pairs, than expected from random pairing. Correspondingly, there are fewer pairs from mixed Classes,   (I,II) and (II,III) pairs, and there are no (I,III) pairs at all despite the expectation that there should be $\sim$ 10\% of such pairs based on random pairing. These differences are found to be statistically significant at the 99.74\% level based on a chi-squared test. 

Applied to the larger sample of UWPs  covering the whole range of separation (sample $P3$), we obtain the same result as above with an even higher significance level ($p=10^{-4}$). These results indicate that the UWPs are most likely coeval.

\begin{table}[ht]
\caption{Class pairings  fraction  in UWPs compared to random Class pairing.}
 \label{Tab:TauG_WB_prob}
\centering
{\scriptsize
\begin{tabular}{r|rrrrrrr}
  \hline
WP type & &$p_{(I,I)}$& $p_{(II,II) }$& $p_{(III,III)}$ &$p_{(II,III)}$&$p_{ (I,III)}$ &$p_{ (I,II)}$\\
  \hline
  $\delta_P < 10 $  kAU     & UWPs & .11 & .46 & .14 & .16 &  .00 & .14\\  
&$Rand$& .02 & .36 & .06 & .30 &  .08 & .18\\
     \hline
\end{tabular}
}
\tablefoot{ $p_{(I,I)}$ is either the fraction of Class I stars paired with another Class I star within UWPs (separation $\delta_P$ less than 10 kAU) or the probability to obtain this pairing from a random distribution  ($Rand$).}
\end{table}

\subsection{High multiplicity fraction within UWPs}

\begin{figure}
\includegraphics[width=\columnwidth]{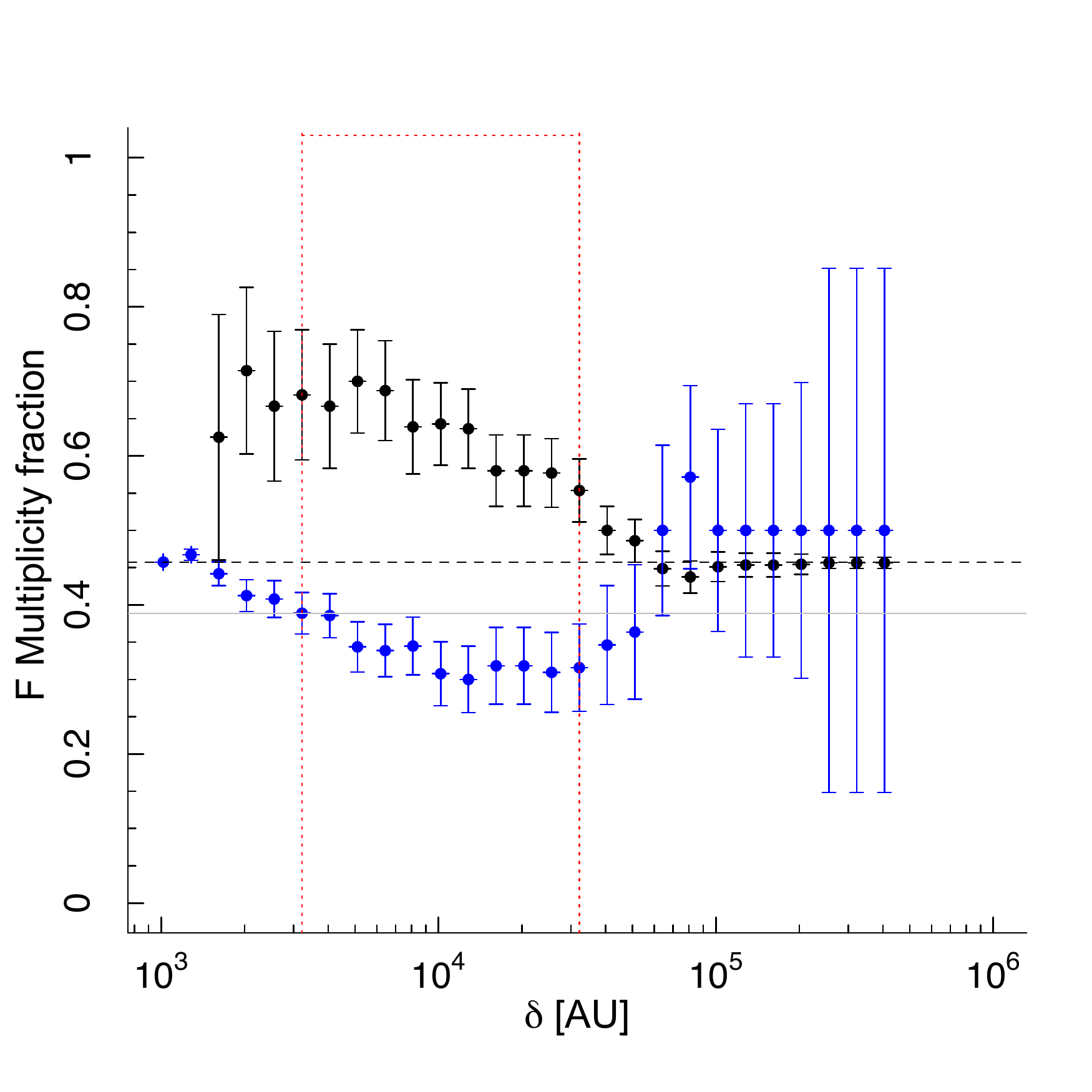}
    \caption{  
      Multiplicity fraction of Class II and III stars observed at HAR and members of UWPs as a function of their separation. 
For a given value of $\delta$, the black (resp. blue) symbol represents the multiplicity fraction computed over all separations smaller (resp. greater)  than $\delta$. Errorbars are standard errors computed from the standard formula: $SE_i=\sqrt{F_i * ( 1 - F_i) / n_i} \cdot \sqrt{ ( N - n_i ) / ( N - 1 ) }$ where  $n_i$ is the sample size, and $N$  is the total population size of  Class II and III stars observed at HAR within UWPs independently of their separation. 
The dashed black line represents the mean multiplicity fraction for stars within UWPs whereas the gray line marks the corresponding mean multiplicity fraction for stars that are not in UWPs. The red dotted lines define the interval within which the multiplicity fraction within UWPs is significantly (p-value of 0.05 or less based on a chi-squared test) higher than the mean value. }
   \label{Fig:TauG_LogNbrePTP_fracMultStar_sep}
\end{figure}
Considering that the stars are part of UWPs, there are almost equal numbers of single stars (50) and multiple systems (44). The resulting multiplicity fraction of 47\% is $\sim 15\%$ higher than that of the sample of Class II and III stars observed at HAR that are not in UWPs, a result that is statistically significant  (p-value of 0.007 for a chi-squared test). In other words, multiple systems are more often within UWPs than not. 
More interestingly, the multiplicity fraction  within UWPs  depends on  the separation range: it declines from a maximum of $\sim 70\%$ at short separations (at approximately $5$ kAU) to the average multiplicity fraction of 47\%  at large separation of approximately $6\times10$ kAU, above which the  multiplicity fraction remains constant  (see Fig.~    \ref{Fig:TauG_LogNbrePTP_fracMultStar_sep}).  Thus, with a confidence level higher than 95\%, the multiplicity fraction within UWPs is higher when their separation is shorter.

\subsection{Multiplicity pairing within UWPs}

In this subsection, we look at the multiplicity nature (single or multiple) of  individual components composing the UWPs. There are 47 UWPs in sample $S_3$ with 12 of them  being composed of two multiple systems (MM pairs), 15 of them of two single stars (SS pairs) and the remaining 20 pairs of one multiple system and one single star (SM pairs). The probabilities associated to random multiplicity pairings are computed from equation  \ref{Eq:ProbPairs} and give 0.28, 0.22, and 0.50  to get a SS, SM or MM pair, respectively. 
Using a Pearson's $\chi^2$ test, we cannot rule out a random multiplicity pairing (p-value 0.59) of the observed UWPs over the whole range of separation. But, the observed probability is not uniform along with the separation. For instance, considering only UWPs whose separation is less than 10 kAU, more than double the expected MM pairs are observed (48\% of the UWPs are of MM pair type, 19\% of SS  and 33\% of SM). The same  $\chi^2$ test indicates that a random multiplicity pairing  can be rejected with a high significance level (p-value 0.018).  Thus the higher multiplicity fraction for tighter UWPs, as shown in the previous subsection, stems primarily from the fact that multiple systems tend to be paired with other multiple systems in that range of separation ($\le 10$ kAU).

\begin{figure}
\includegraphics[width=\columnwidth]{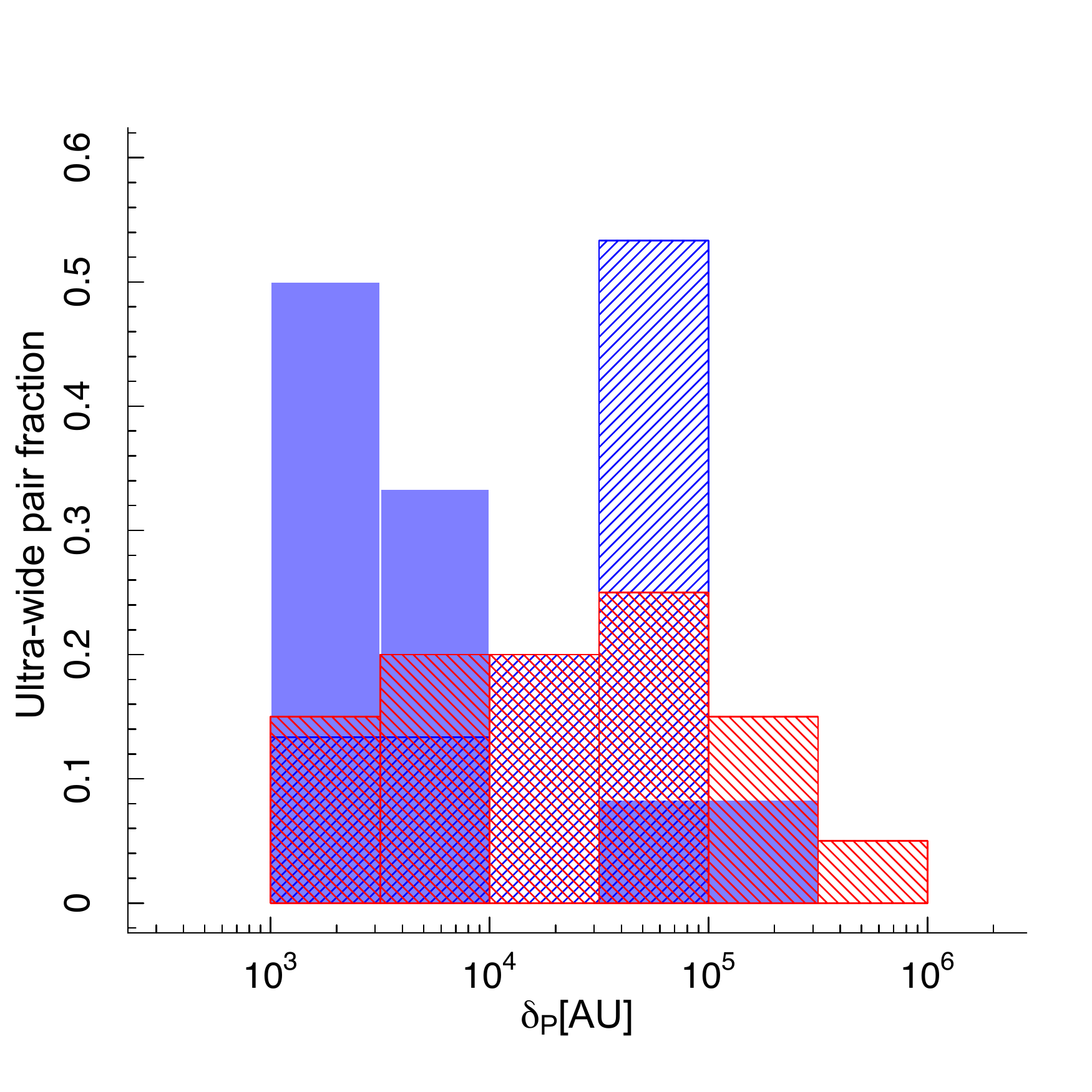}
    \caption{SS (dashed blue), SM (dashed red) and MM (solid blue) pair fraction 
as a function of their separation. MM pairs are tighter (most are found below 10 kAU) whereas SS pairs are far apart (with a peak at approximately 50 kAU); on the other hand, SM pairs are relatively uniformly distributed.}
   \label{Fig:TauG_PTP_P5}
\end{figure}

A comparison of the pair fraction distribution between  MM, SM, and  SS UWPs as a function of their separation  reveals indeed that  nearly 90\% of MM pairs are concentrated below $10$ kAU,  the SM pairs are distributed essentially log-uniformly over the whole range, and SS pairs  are preferentially  separated far apart (median separation at 35 kAU, see Fig.~  \ref{Fig:TauG_PTP_P5}) . As a consequence, MM systems are the main contributor to 
the observed `bump' in the 1-NNS distribution of multiple systems outlined in the previous section. Of the 30 multiple stellar systems  having their  1-NNS less than $10$ kAU, 20 of them are within multiple/multiple UWP (65\%), 7 members (25\%) are  within single/multiple UWPs and 3 (10\%) are non-mutual pairs.  
Therefore, the spatial neighborhood difference between multiple systems versus single stars comes down  to a higher fraction of
`tighter' MM UWPs with respect to the SS UWPs.

\subsection{UWP multiplicity}

In this subsection, we study the properties of UWPs, as a function of their multiplicity  $n_*$, defined as the sum of the total number of stars within each pair ($2\leq n_*\leq5$). 

For this analysis, we continue focusing on sample $S_3$ (half the population of UWPs), which has the most reliable multiplicity data, since each of the components has been observed at HAR. 
Within it, we investigate the distribution of the UWPs as a function of their degree of multiplicity.  The number of UWPs seems to be rather uniform up to a multiplicity of four (see Table  \ref{Tab:TauG_WP_Nbre_Mult}), and exhibits a sharp decrease for a multiplicity of five.  We note that this behavior differs from the geometric progression for pre-main sequence closer binaries in Taurus $N_{n_*} \propto  b^{-n_*}=2.6^{-n_*}$ \citep{DucheneKraus2013}.

 \begin{table}[ht]
 \caption{Number and type of high order multiplicity in UWPs.}
\label{Tab:TauG_WP_Nbre_Mult}
\centering
{\tiny
\begin{tabular}{|r|r|r|r|r|}
  \hline
{\bf $n_*$ }&  2&3  & 4 & 5\\
    \hline
        \hline
$N_P$ &  16 & 14  & 13 & 4\\
    \hline
UWP Type &  (S,S) &  (S,B)  & 8 (B,B)  & 4 (B,T)\\
 &  &   &  5 (S,T) & \\
  \hline
$N_{P_{60}}$ &  13 & 11 & 10 & 4\\ 
    \hline
UWP Type &  (S,S) &  (S,B)  & 6 (B,B)  & 4 (B,T)\\
 &  &   &  4 (S,T) & \\  
    \hline
$N_{P_{10}}$ &  4 & 6 & 7 & 4\\
    \hline
UWP Type &  (S,S) &  (S,B)  & 6 (B,B)  & 4 (B,T)\\
 &  &   &  1 (S,T) & \\  
    \hline
   \end{tabular}
   }
\tablefoot{Number $N_P$ (resp. $N_{P_{60}}$ and $N_{P_{10}}$ ) of UWPs in the whole range of separation (resp. for separation less than 60 kAU and 10 kAU) as a function of their multiplicity $n_*$ and their hierarchical type (S, B,T: single, binary, triple).}
\end{table}

Following  \cite{correia06}, we define the multiplicity fraction per ultra-wide binary ($MF_{uw}$) as the number of higher multiplicity systems over the total number of UWP systems of the sample with projected component separations typically higher than 1 kAU.  
Over the whole range of separation we obtain:
\be
MF_{uw}=\frac{T+Q+\ldots}{B+T+Q+\ldots}  = 65.6 \pm 11.85\%
.\ee
We similarly define the companion frequency per ultra-wide binary ($CF_{uw}$) as:
\be
CF_{uw}=\frac{2 \times T+3 \times Q+\ldots}{B+T+Q+\ldots} = 1.77 \pm 0.19
.\ee
We use the Poisson error to estimate the error made on $MF_{uw}$ (resp. $CF_{uw}$), that is, we use the square root of the number of higher order multiple systems (resp. the total number of companions in higher order systems $2 \times T+3 \times Q+\ldots$) divided by the total number of systems (B, T, \ldots). We also estimate  the multiplicity fraction and companion frequency per ultra-wide binary for UWPs separated by less than 60 kAU (resp. 10 kAU): $MF_{60}=65.79 \pm 13.16 \%$ and  $CF_{60}=1.79 \pm 0.22$ (resp.  $MF_{10}=80.95 \pm 19.63 \%$ and  $CF_{10}=2.33\pm0.33 $).

The values we obtained are far above the values of multiplicity fraction ($26.8 \pm8.1\%$) and companion frequency ($0.68 \pm 0.13$) per wide binary found in young T Tauri wide binaries in the range 14-17 AU up to 1.7-2.3 kAU   \citep{correia06}, highlighting the ubiquity of  high-order ultra-wide binaries in Taurus found in our sample.

\begin{figure}
\includegraphics[width=\columnwidth]{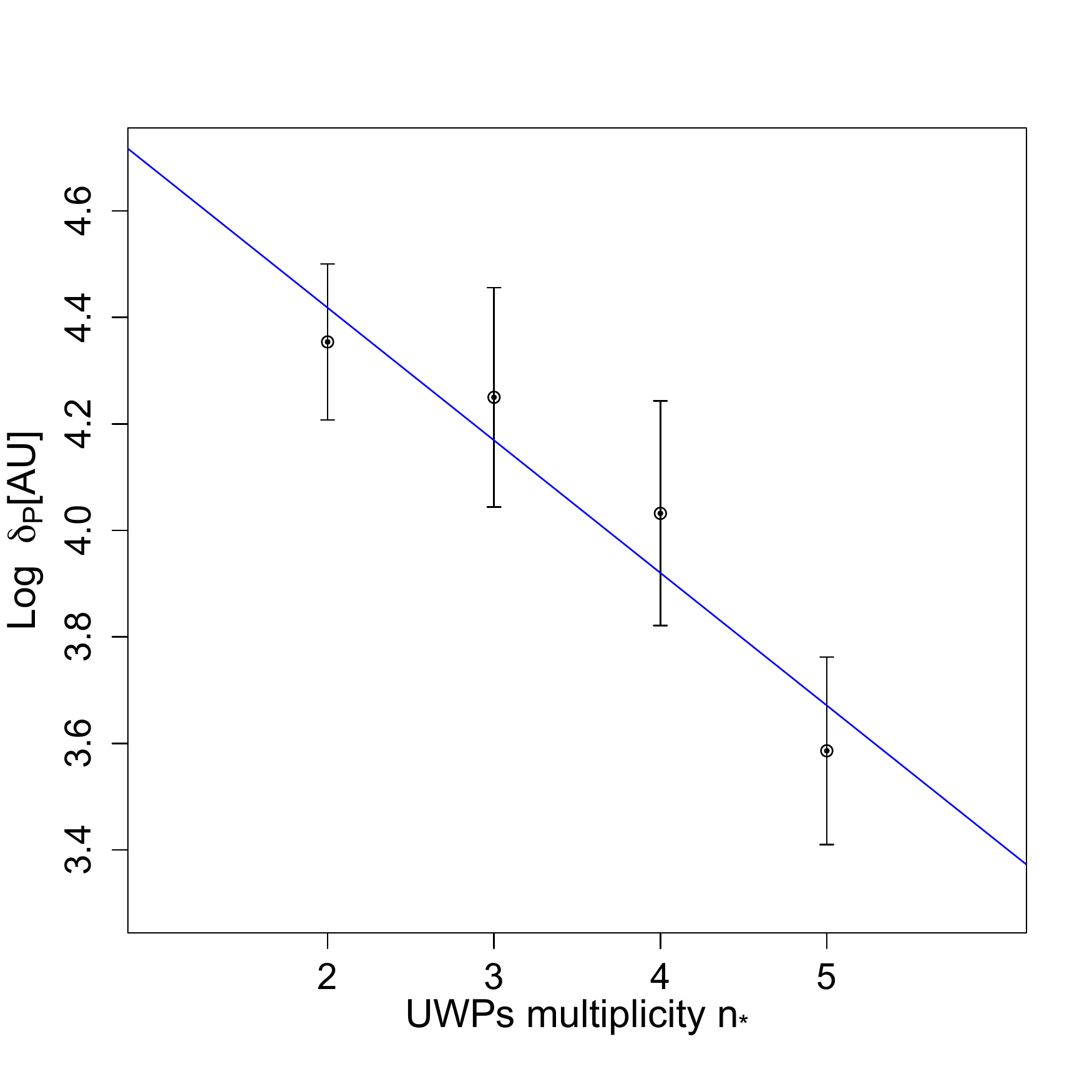}
\caption{ 
Mean separation $\overline{\delta_P}$ of UWPs  (sample $S_3$) and standard errorbars ($\pm \sigma/\sqrt{n_*}$) as a function of their multiplicity $n_*=2, 3,...,$. The separation is consistent with a geometric progression $\delta_P \propto a^{-n_*}$ with  $a\sim 1.8$  (shown as the blue line).}
\label{Fig:TauG_WB_Sep_Nmult_Whole}
\end{figure}

Based on the previous subsection, and since tighter pairs are mostly composed by MM pairs, we expect them to have a higher multiplicity. Indeed, the mean projected separation  $\delta_P$  of UWPs  is negatively correlated with the degree of  multiplicity (see Fig.~\ref{Fig:TauG_WB_Sep_Nmult_Whole}). We obtain the best  geometric progression fit as:
\be 
\overline{\delta_P (n_*)} \propto a^{-n_*} \sim 1.8^{-n_*},
\label{Eq:GeomProDelta}
\ee
 where  $\overline{\delta_P (N_*)}$ is the mean  separation of UWPs per bin of degree of multiplicity $n_*$.  Thus,  UWPs composed of two single stars are wider on average  than  UWPs  composed of  a single star and a binary, which in turn are wider  on average  than UWPs composed of two multiple systems. 
In summary,  tighter UWPs are biased towards higher-order multiplicity.

\subsection{Ultra-wide pairs and mass function}

\begin{figure}
\includegraphics[width=\columnwidth]{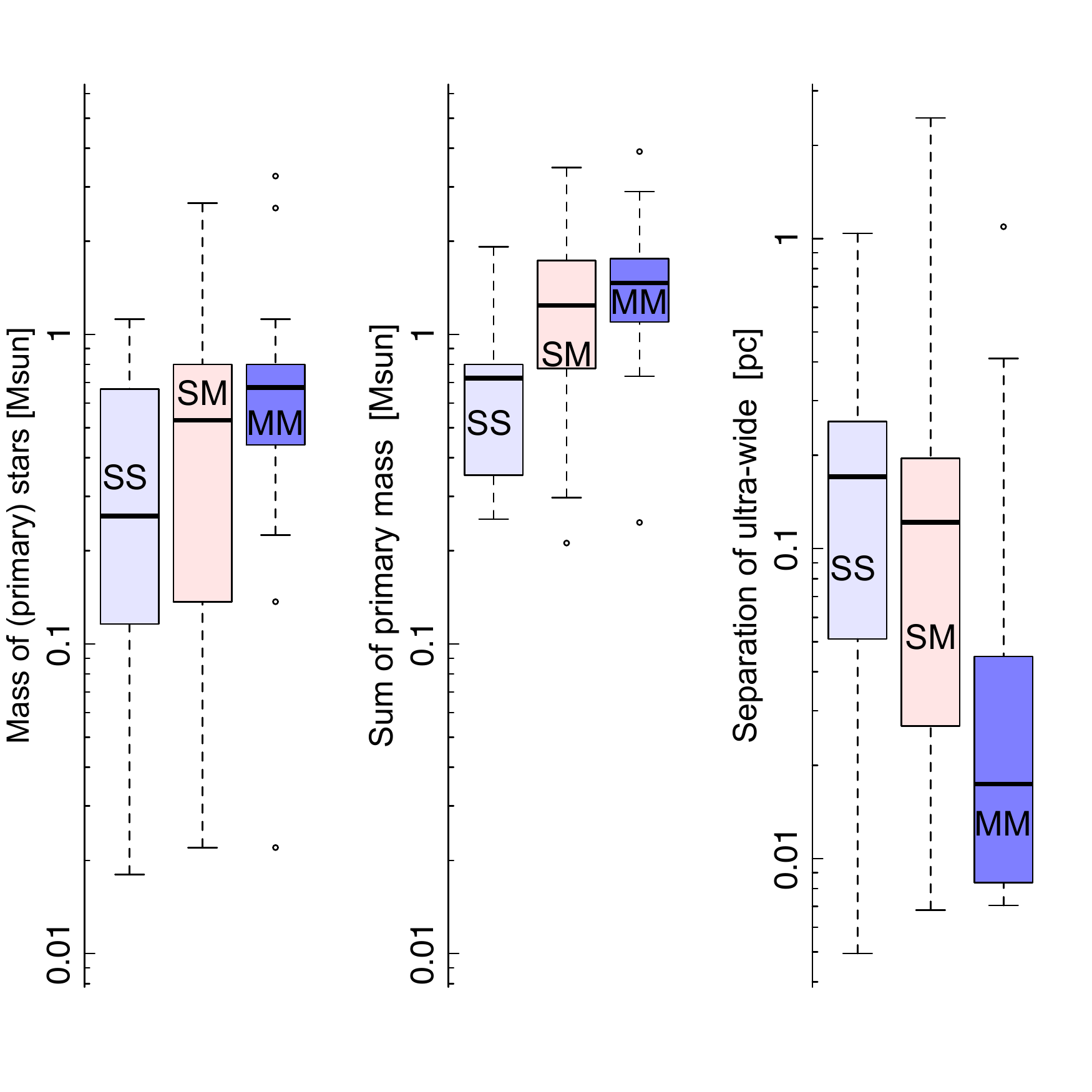}
    \caption{
Boxplot for the distribution of  the (primary) star mass within UWP  (left), the sum of the masses of the two (primary) stars within each UWP (center), and the separation (right) of UWPs. The SS, SM and MM types of systems are shown separately to illustrate that MM pairs are tighter and more massive than SS pairs.
 }
   \label{Fig:TauG_MF_MMpairs}
\end{figure}

The  mass function of single stars and primaries of multiple systems within UWPs depends on their type (see right Fig.~\ref{Fig:TauG_MF_MMpairs}). A KS test indicates that the mass distribution for SM pairs cannot be distinguished from that of either the SS or the MM pairs. 
However, a similar KS test shows that the difference in mass distribution between SS and MM pairs is significant with a confidence level of 97.8\%.

Specifically, compared to SS pairs,  MM pairs exhibit:
\begin{itemize}
\item a  deficit of mass below 0.1 $M_\odot$, with a frequency of only 5\% (against 20\%);
 \item a deficit of stars with mass in the 0.1--0.6 $M_\odot$  range, with a frequency of 25\% for the MM pairs (against 50\%);
 \item an excess of stars with mass between 0.6 and 0.8 $M_\odot$, with a frequency of 45 \% (against 25\%)
 \item a predominance of stars more massive than 0.8 $M_\odot$ (20\% of the most massive stars in Taurus are in MM pairs  but only 5\% are in SS pairs).
\end{itemize} 
Thus, 70\% of stars within SS pairs have a mass of less than 0.6 $M_\odot$ and  $\sim$ 65\% of stars within MM pairs are more massive than 0.6 $M_\odot$. 

As a result of these trends (see Fig.~\ref{Fig:TauG_MF_MMpairs}), the median primary mass of MM pairs is twice as high as the median primary mass in SS pairs ($0.7\, M_\odot$ compared to $0.3\, M_\odot$). Furthermore, they have a much smaller dispersion as indicated by the interquartile range for MM pairs (0.5--0.8\,$ M_\odot$ versus  0.1--0.7\,$ M_\odot$  for SS pairs, see left panel in Fig.~\ref{Fig:TauG_MF_MMpairs}). The  median primary mass of MM pairs is intriguingly close to 
the  unusual peak at $0.8$\,M$_\odot$ in Taurus reported  by \cite{BricenoEtAl2002}.  This  peak was tentatively explained by a fragmentation/ejection hydrodynamical  model   (\citeauthor{GoodwinEtAl2004} \citeyear{GoodwinEtAl2004}) of molecular cores that have unusual properties (extended envelope, a narrow range of core mass function with a peak near 5 M$_\odot$, and a low level of turbulence). The results of simulations show that 50 \% of the low-mass objects that form within the cores are ejected from their cradles to produce an extended population of low-mass stars and brown-dwarfs, whereas the remaining stars in multiple systems accrete the gas reach a mass of up to  0.8-1 M$_\odot$. This type of dynamical ejection model was also advocated by   \cite{ReipurthClarke2001,Boss2001,BateEtAl2003,KroupaBouvier2003b,KroupaEtAl2003a}. 
However,  \cite{BricenoEtAl2002} and \cite{Luhman2006} argued that the absence of a widely dispersed brown-dwarf population is strong enough evidence to reject this type of model. Therefore up to now, there is no clear consensus to explain the peak at 0.8 $M_\odot$ in the mass function of Taurus.

\begin{figure}
\includegraphics[width=\columnwidth]{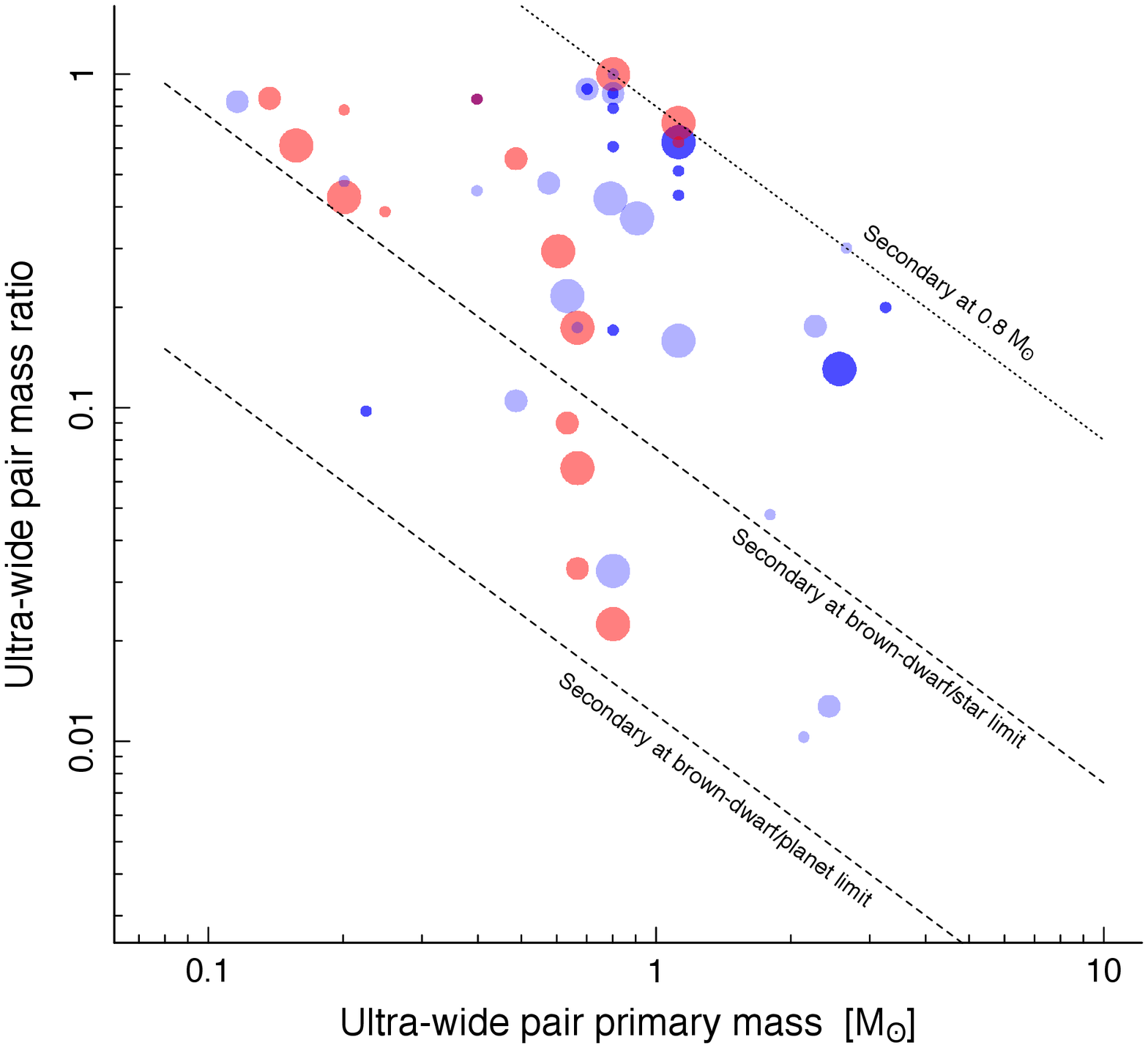}
    \caption{Primary mass versus mass ratio   within MM (filled dark blue circle), SS (filled light red circle), and SM (light blue open circle) UWPs. Intermediate (resp.smallest) size of marks:  UWPs separated by less than 35 kAU (resp. less than 10 kA), the largest size devoted to wider UWPs beyond 35 kAU.  Dashed lines:  brown dwarfs (BD)/planet (0.012 M$_\odot$)  and BD/star (0.075 M$_\odot$) limits.  Dotted line: 0.8 M$_\odot$ star limit. The superpositions are due to the discretized nature of mass determination models from the spectral type and evolutionary tracks.
 }
  \label{Fig:TauG_q_MassP}
\end{figure}

  \begin{figure}
\includegraphics[width=\columnwidth]{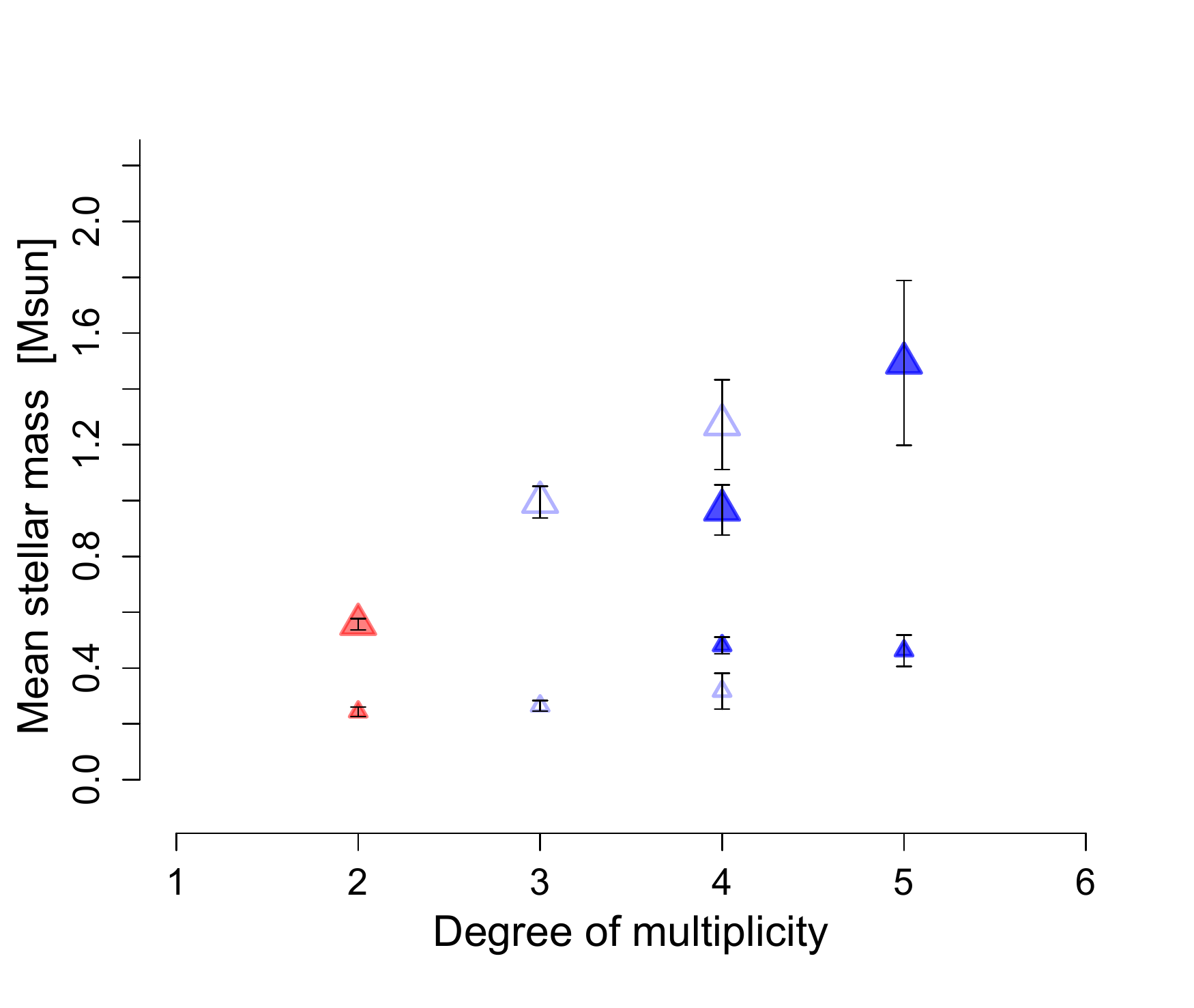}
    \caption{Correlation between multiplicity and primary (large size mark)  \& secondary (small size mark) mass of stars in MM (filled dark blue triangle), SS (filled light red triangle), and SM (light blue open triangle) UWPs.
 }
   \label{Fig:TauG_MaxMinMass_UWP_mult}
\end{figure}

Summing the two (primary) masses within each UWP amplifies the contrast between the lower mass threshold needed to produce MM pairs with respect to SS pairs (see the  middle panel in Fig.~\ref{Fig:TauG_MF_MMpairs}). The median sum of primary masses  within MM pairs  ($1.5\, M_\odot$) is twice as high as that within SS pairs  ($0.7\, M_\odot$),  but the width of the  interquartile ranges are  comparable (1.2--1.7\,$ M_\odot$ and 0.4--0.8\,$M_\odot$ for MM and SS pairs, respectively). 
The difference in mass range between SS and MM pairs would be even more pronounced if the  mass  of inner companions within MM pairs were also included in the sum; unfortunately, the current multiplicity data are insufficient to evaluate this in a consistent manner across the entire sample. 
 In summary, the median of the stellar mass in MM pairs is twice as high as that within SS pairs, and they are also ten times less divergent; the masses of stars within SS pairs spread a larger range unlike the primary masses in MM pairs. The SM pairs have intermediate properties between these two classes.

\subsection{Mass ratio in UWPs}
The study of the mass ratio of the secondary mass over the primary mass in the UWPs as a function of the primary mass  shows several interesting features (see Fig.~\ref{Fig:TauG_q_MassP}). First the secondary mass  covers the whole range of  mass, from equal mass down to brown dwarf type mass (0.075 $M_{\odot}$) suggesting the same continuous formation  scenario from  star to brown dwarf.
Secondly, there is a clumping trend of MM UWPs towards  the upper  part of the diagram. 
 Further, we note that there are no UWPs composed of two brown dwarfs. 

Some other features appear to be spurious. The apparent concentration of secondary brown dwarfs pairing with  0.6--0.8 $M_{\odot}$ stars showing up in Fig.~\ref{Fig:TauG_q_MassP} is in fact due to the preeminence of these 0.6--0.8 $M_{\odot}$ stars in the mass function. By performing 10 000 Monte Carlo mass pairing samplings amongst the mass sample of UWP stars, we obtain indeed a p-value of 0.15 compatible with the hypothesis that this pattern is due to chance pairing. The next intriguing
 pattern that appears at first glance is the empty top right `wedge', where no secondary stars heavier than 0.8 $M_{\odot}$  lie in for primary stars heavier than 0.8 $M_{\odot}$. By performing a similar Monte Carlo sampling, we obtain a p-value of 0.005, which seems to indicate that we  can apparently
reject the hypothesis that this occurs by chance. This result, however, is not based on a large sample size (median expectation: four pairs within that wedge), and the stellar mass determination from theoretical evolutionary tracks, as well as the spectral type classification, artificially
 stretches the  width of this empty wedge. At this stage, we  cannot  therefore  ensure that it is a reliable pattern.

The  increase of the multiplicity goes along with the primary mass of the UWPs (see Fig.~\ref{Fig:TauG_MaxMinMass_UWP_mult}). This trend was also noted for closer binaries \citep{KrausHillenbrand2007}.
The mass of the secondary star  does not  correlate with the multiplicity to the same extent.

\section{Discussion}
\label{Sec:Discussion}
In this section, we discuss the possible nature of the UWPs identified in Taurus.

\subsection{UWPs as physical pairs}

The findings surrounding UWPs obtained in the previous section amount to an array of circumstantial evidence suggesting that these pairs are probably  physical pairs:
\begin{itemize}
 \item non-random  1-NNS distribution of mutual pairs (versus lognormal 1-NNS  distribution for non-mutual pairs);
 \item approximate log-uniform separation distribution (\"Opik's law distribution) for mutual pairs, extending up to $\sim 60$ kAU (i.e., 0.27 pc) at least, observed for binaries at closer separations  \citep{KrausHillenbrand2009a};
 \item non-random Class pairings within mutual pairs suggesting coevality;
 \item  UWPs in the range of separation $< 5$ kAU are already known to be true physical binaries \citep{KrausHillenbrand2009a}.
 \end{itemize}
 
 To be gravitationally bound, the internal energy of UWPs must be negative, that is, $1/2\mu \Delta v^2 -GM_1M_2/r < 0$, with $\mu$ being the reduced mass ($\mu=M_1\,M_2/(M_1+M_2)$, $M_1$ and $M_2$ being  the mass of the stars), $G$ the gravitational constant, $r$ and $\Delta v$ the  separation and the relative velocity between the two stars, respectively. Thus, the condition to meet for a pair to be bound can be rewritten as follows:
 \begin{equation}
\begin{array}{lll}
\Delta v &< & \left (2G (M_1 + M_2)/r\right)^{1/2}\\
&<& 0.3  \left( \left[ \frac{M_1}{0.5 \,M_\odot}\right ]+ \left [ \frac{M_2}{0.5 \,M_\odot}\right ] \right ) \, \left (\frac{r} {10\, kAU} \right)^{-1/2} \, \mathrm{km/s}
\end{array}
\label{Eq:BinLinked}
.\end{equation}
This relative velocity between the stars is low compared to the observed already low velocity dispersion in the Taurus stellar complex; 2-4 km/s in the whole region \citep{FrinkEtAL1997,RiveraEtAl2015} and as low as 1-2 km/s in stellar subgroups \citep{JonesHerbig1979}. However, as we will see in the following subsection, the low stellar density of Taurus and the low stellar velocity dispersion prevent  efficient exchanges of energy during random encounters between wide pairs and other stars, such that the characteristic disruption time $\tau_d$ of  a few $\sim 10^3$ Myr is much longer than the age of the Taurus pre-main sequence population, that is, $1-10$ Myr old 
 \citep{KenyonHartman1995, BertoutEtAl2007, AndrewsEtAl2013}. Therefore, we consider in the following that UWPs are physically linked. This hypothesis will be further tested by results of the Gaia survey that will provide 3D spatial and 2D/3D velocity measurements (proper motions of stars and radial velocity for  part of them).

  \subsection{A short review on wide binaries: models and observations}
  
From the theoretical side, different models and numerical simulations have been performed to produce multiple systems   \citep[see][for a complete review on  recent numerical simulation improvements. ]{ReipurthEtAl2014}. 
One notable result from  simulations of star formation from collapsing gas is that they do not generate wide pairs at birth beyond $1$ kAU because of the strong dynamical interactions that occur in simulations of the collapse of turbulent cores \citep{Bate2012}. 

Nonetheless, wide pairs are present among field stars, albeit in relatively small numbers: approximately 10\% of all solar-type field stars have a companion with separations larger than 1kAU \citep{RaghavanEtAl2010,LepineBongiorno2007,LonghitanoBinggeli2010} and maximum separations as high as 20kAU \citep{ChanameGould2004,ShayaOlling2011,DhitalEtAl2015}.

The low-but-not-negligible fraction of visual wide pairs in the field could not solely be the result of random gravitational capture: the binary creation rate in the field is estimated to be of the order $4\times 10^{-21} \,{\rm{pc}^{-3}} \,{\rm{Gyr}^{-1}}$ \citep{KouwenhovenEtAl2010} showing that random creation of bound pairs in the field, as well as in the star-forming regions,  is an extremely rare event. 

An alternate mechanism must be at play to produce such pairs either at birth or as the result of subsequent dynamical evolution. But wide pairs with separations beyond 0.1 pc appear to be particularly unlikely pristine products of star formation process because their separations exceed the typical size of a collapsing cloud core, precluding a scenario based exclusively on core fragmentation. 

Consequently, two scenarios based on dynamical evolution have been proposed to explain their origin. One possibility is that an initially compact triple system could dynamically unfold, with the tighter pair shrinking while the third component being ejected onto an extremely long orbit \citep{ReipurthMikkola2012}. Alternatively, the gradual dissolution of a star cluster could leave behind wide binaries that are barely bound if two stars happen to be very close in the 6D phase space \citep{KouwenhovenEtAl2010,MoeckelBate2010,MoeckelClarke2011}. For most star clusters, the dissolution time, when this `pairing' could occur, is of the order of a few Gyr \citep{LamersGieles2006}.

The model of  \cite{KouwenhovenEtAl2010} also predicts that these wide pairs are high-order multiples as a reflection of the high multiplicity of individual stars at birth. This matches nicely with our findings on UWPs in Taurus, however their mechanism also induces random pairing with respect to mass and Class because the binding pairs are generated by chance depending on the proximity of the objects once the gas of the cluster is removed. This is not what is observed within  UWPs in Taurus, since we have shown in the previous section that the Class-pairing is not random and that the distributions of mass is not the same in SS and MM pairs. 

Another inadequacy of both the ``pairing during cluster dissolution'' theory and the ``unfolding of triple systems''  theory is that they occur on timescales that are orders of magnitude longer than the age of the Taurus population (0.1--0.5 Myr  for Class I stars and a few Myr for Class II and III stars). 

The  large fraction of UWPs in Taurus is reminiscent of the surprisingly large number of UWPs (10-50 kAU) recently identified in slightly older moving groups (10--100 Myr) such as in the  $\beta$ Pic moving group (\citeauthor{Alonso-FlorianoEtAl2015} \citeyear{Alonso-FlorianoEtAl2015}). A preference has also been identified for high-order multiplicity among these UWPs,  as well as  in AB Doradus, TW Hydrae and Tucanae-Horologium moving groups \citep{TorresEtAl2006,ElliottEtAl2016}. \cite{ElliottBayo2016} suggest that  the binary population from close ($0.1\,$AU) to very wide systems ($100\,$ kAU) in $\beta$ Pic Moving Group can be accounted for by the internal dynamical interaction  of triple systems as proposed by \cite{ReipurthMikkola2012}. But, as an alternative, these very wide pairs may be the survival part of the wide pairs population formed at an earlier time.

Some of these wide pairs may even survive longer since solar-type stars within high-order multiplicity UWPs with periods that peak around  100 yr (separations about 36.5 kAU)  have also been observed in the field \citep{Tokovinin2014} as well as for  late-spectral-type M1-M5 dwarf UWPs \citep{LawEtAl2010}. In the latter case, a  very high value  was found for the multiple fraction per ultra-wide pair, that is, $77^{+9}_{-22}$\% for systems with separations > 4 kAU, very close to what we obtain for the  high-order-multiple fraction derived in Taurus UWPs with separation less than 10 kAU.

However, very wide pairs composed of extremely young ( $\sim 0.1$ Myr) Class 0 objects are relatively common \citep{LooneyEtAl2000,TobinEtAl2010,KempenEtAl2012,ChenEtAl2013,LeeEtAl2015,PinedaEtAl2015,TobinEtAl2016}. We thus conclude that the UWPs in Taurus 
must have been produced by a mechanism that acts extremely early on and, indeed, that these systems may well be the pristine outcome of star formation itself, as is suggested moreover by the coevality trend. 

We contend that the population of UWPs that we identified in Taurus supports the idea that very wide binary systems are almost universally produced in star-forming regions, even though the physical mechanism leading to their formation remains under debate.

\subsection{A pristine origin for these UWPs }
To argue for pristine UWPs, one must ensure that they can survive dynamical interactions. 
The destruction of multiple stars occurs either due to the intrinsic instability of the non-hierarchical multiple stellar system itself or due to the decay of a few Nbody systems driven by hard or soft encounters with other stars of that region. For the former mechanism,  the timescale is very short, that is, a few crossing times, or approximately a few 10 kyr \citep{Reipurth2000}, far less than the Taurus age (few Myr). 

Disruption and dissolution  of wide binary stars in the Galaxy  by gravitational encounters with other stars, molecular clouds, and tidal forces due to gravitational potential of the Galaxy is a long-standing study \citep{Chandrasekhar1944,Heggie1975,Heggie1977,RettererKing1982,BahcallEtAl1985,JiangTremaine2010}. It was estimated that in the solar neighborhood, the half-life of a binary composed of two solar-type stars  separated by $31$ kAU is 10 Gyr \citep{JiangTremaine2010}.

We expect that in a low-stellar-density region such as Taurus  (i.e., 1-10 pc$^{-2}$),  dynamical interactions between stellar systems  are rare and soft. We thus expect minor dynamical destruction of the wide pairs in Taurus, such that most probably wide binaries in Taurus reveal their pristine configuration. The following arguments on characteristic times associated with the destruction of binaries due to dynamical encounters will give some support to these expectations.

\cite{BinneyTremaine2008} have estimated the effects of dynamical encounters of binaries  with  random stars, allowing us to derive  two destruction timescales for wide binaries. First, we consider catastrophic single encounters, characterized by a timescale
 \begin{equation}
\begin{array}{lll}
\tau_D&=&  \sqrt{3/14}\, \pi^{-1} \, M_{tot}^{1/2} \,(G^{1/2} \,\rho_c \,M_c \,a^{3/2})^{-1}  ;\\
\tau_D & \sim & 7.2 \,{\rm {Gyr}} \, \left( \frac{M_{tot}}{2\,\rm{M_\odot}} \right)^{1/2} \frac{0.05\, {\rm{pc^{-3}}}   }{\rho_c} \frac{1\,\rm{M_\odot}}{M_c} \left( \frac{10^4  {\rm{AU}}}{a} \right)^{3/2}.
\end{array}
\label{Eq:TpsDestCat}
\end{equation}
Secondly, we evaluate the destruction timescale in the diffusive regime:
 \begin{equation}
\begin{array}{lll}
\tau_d & \sim& 0.085 \, (\sigma_{rel} M_{tot} b_{min}^2)(G M_c^2 \rho_c a^3)^{-1} ;\\
\tau_d & \sim & 3.5\,{\rm{Gyr}} \frac{k_{diff}}{0.085} \frac{\sigma_{rel}}{4 \,{\rm{km/s}} } \frac{M_{tot}} {2 {\rm{M_\odot}}}  \left ( \frac{1 {\rm{M_\odot}}} {M_{c}} \right )^2 \frac{0.05\, {\rm{pc^{-3}}}}{\rho_c}\frac{10^4\, {\rm{AU}}}{a}.
\end{array}
\label{Eq:TpsDestDif}
\end{equation}
In both cases, $a$ is the semimajor axis, $M_{tot}$ is the total mass of the pair, $\rho_c$ is the stellar volumic density, $M_c$ is the mass of the random encounter star,  $\sigma_{rel}$ is the stellar dispersion velocity, and    $k_{diff} = 0.085 (b_{min}/a)^2$ is a diffusive parameter with $b_{min}\sim a$ being the impact parameter cut off. 

While the semi-major axis $a$ cannot be estimated directly for these long period binaries, it is statistically correlated with the projected separation $\delta$, as $\delta \approx a$ \citep{leinert93,TokovininLepine2012}. The closeness  of the two values depends on the chosen eccentricity distribution type, either  flat ($f(e)=Cste$), as observed in solar-type  stars from the {\it Hipparcos} catalog by  \cite{RaghavanEtAl2010} for wide binaries, or thermalized ($ f(e )=2e$),  as first theoretically proposed by \cite{Jeans1919}, generalized to broader types of distribution by \cite{Ambartsumian1937} and observed  in field stars by \cite{DM91}.

We evaluate the volume density through $\rho_c=\rho_w/\Delta \sim1/25\sim 0.04 \, {\rm{pc^{-3}}}$, with $\rho_w$  being the mean projected surface density (equation \ref{Eq:P_meanRho_wind_taurus})  and $\Delta \sim 20\, {\rm{pc}}$  the depth of Taurus  \citep{TorresEtAl2007}.

Both destruction times are much larger than the age of the young Taurus star-forming region. Therefore, in such an environment, it is extremely unlikely that wide binaries, even with separation as large as $100$ kAU, can be destroyed by dynamical encounters.
Numerical simulations  have indeed confirmed that loose associations such as Taurus provide an environment that is inefficient at binary disruption even at large separation \citep{KroupaBouvier2003,MarksKroupa2012}. 
Even if they use an initial  distribution of binary separation up to $10$ kAU, we expect that a same result would be obtained with an extended binary separation up to $100$ kAU. 

We may thus feel confident that the very wide pairs populations in Taurus may trace the initial spatial distribution at birth without being destroyed by dynamical encounters up to now. 

\begin{figure}
\includegraphics[width=0.8\columnwidth]{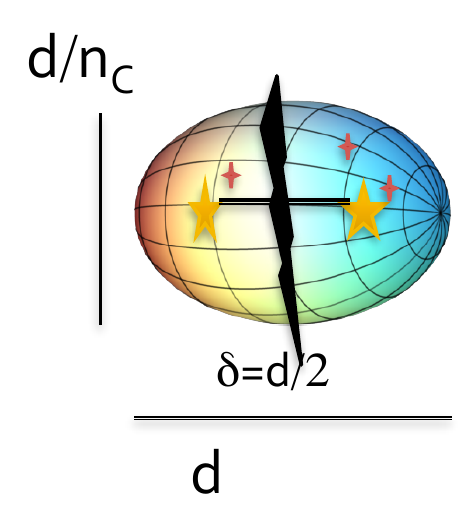}
   \caption{Schematics for a prolate core with a length $d$ in the main direction and a width of $d/n_c$ giving birth to two multiple systems whose primaries are separated by $\delta$.}
      \label{Fig:TauG_cartoon_coreWP}
\end{figure}

 \subsection{Core properties for pristine UWPs}

 In the context of a pristine configuration for the wide pairs, we look for the initial conditions that could lead to the formation of such pairs in a core fragmentation model (see Fig.~\ref{Fig:TauG_cartoon_coreWP}). 
 
 We thus make the hypotheses that; (1) these UWPs  are pristine remnants of the star-formation process, and (2) they form within a single core with a core-to-star  efficiency factor $\alpha_C \sim 40 \%$ \citep{KonyvesEtAl2015}. The total mass of parental clump can be evaluated as $M_{tot}(1+\alpha_B)/\alpha_C$, where $M_{tot}$ is the sum of (primary) masses and $\alpha_B \, M_{tot}$  is the mass of inner companion stars ($\alpha_B=0$ for SS pairs). We also introduce a geometrical shape factor $n_c$ to quantify the propension of clumps to be elongated and  prolate  rather than roundish or oblate \citep[$n_c\sim2$--4,][]{MyersEtAl1991,Ryden1996,LomaxEtAl2013}. From those assumptions,  we get the mean particle density $\rho$ of an initial core as:
 \be
     \begin{array}{lll}
               \rho & = & M_{tot}/(\mu\,m_H )\, (4/3\,\pi \, n_C^2 \delta^3)^{-1}\,(1+\alpha_B)/ \alpha_C  \\
       \end{array}
 ,\ee
 where $\mu=2.33$ is the mean molecular weight and $m_H$ is the atomic hydrogen mass.
 
 We proceed to compute the median density  estimate of initial cores that would be required to produce each type of pair $ \rho_{SS},  \rho_{SM},  \rho_{MM}$:
 \begin{subequations}
  \begin{equation}
  \begin{array}{lll}
       \rho_{SS}   &\sim   10^{4} \,{\rm cm^{-3}}\, \left [\frac{M_{tot}}{0.72 \,M_\odot} \right ] \left [\frac{0.17 \,{\rm pc}}{\delta} \right ]^{3}  \frac{[n_C/4]^2}{[\alpha_C/0.5]},
    \end{array}
\end{equation}

  \begin{equation}
  \begin{array}{lll}
         \rho_{SM} &  \sim   10^{5}  \,{\rm cm^{-3}}\, \left [ \frac{M_{tot}}{1.24 \,M_\odot} \right ]   \left [ \frac{0.12 \,{\rm pc}}{\delta} \right ]^{3}  \frac{[n_C/4]^2\,(1+[\alpha_B/0.2])}{[\alpha_C/0.5]},\\
    \end{array}
\end{equation}  \begin{equation}
    \begin{array}{lll}
        \rho_{MM}  & \sim   3\times 10^{7} \,{\rm cm^{-3}}\, \left [ \frac{M_{tot}}{1.47 \,M_\odot} \right ] \left [\frac{0.017 \,{\rm pc}}{\delta} \right ]^{3} \frac{[n_C/4]^2\,(1+[\alpha_B/0.2]}{[\alpha_C/0.5]}.\\
          \end{array}
\end{equation}
 \end{subequations}

We then question whether those kinds of cores would be instable or stable against any perturbation. As the Jeans instability length is given by:
 \be
  \lambda =  c_s \left ( \frac{\pi}{G \rho} \right )^{1/2} 
    \label{Eq_JeanLength}
  ,\ee
  where $c_s = 0.2$ km/s is the sound speed at T=10K, we  derive the following Jeans length estimates for the three types of pairs:
 \begin{subequations}
  \begin{equation}
  \lambda_{SS}=1.2 \left [\frac{\delta_{SS}}{ 0.17} \right ]\, {\rm pc}
  \label{Eq_JeanLength_SS}
   ,\end{equation}
  \begin{equation}
      \lambda_{SM}=0.7 \, \left [\frac{\delta_{SM}}{ 0.12} \right ] {\rm pc}
        \label{Eq_JeanLength_SM}
,\end{equation}
  \begin{equation}
   \lambda_{MM}=0.2 \, \left [ \frac{\delta_{MM}}{ 0.017} \right ] {\rm pc }
     \label{Eq_JeanLength_MM}
   ,\end{equation}
  \end{subequations}
  where $\delta_{SS},\delta_{SM},\delta_{MM}$ are the median separations of the three types of UWPs.

The Jeans length is therefore slightly larger than the characteristic  length of the SS  fictitious clumps (Equation   \ref{Eq_JeanLength_SS}), while it is smaller for the MM and SM ones (Equations   \ref{Eq_JeanLength_SM} and \ref{Eq_JeanLength_MM}). In other words, the latter two appear  to be instable against gravitational (Jeans) instability, while the former appears to be marginally stable.

Therefore, it is plausible that SM and MM pairs formed within the same core, while the SS pairs may have formed through an alternative scenario. We may still consider a pristine origin for them. The two stars may be born  in two distinct but proximate cores.  Or alternatively, the two stars may be born within the same low-density core that becomes supercritical due to the external potential of the Galaxy \citep{BallesterosEtAl2009a,BallesterosEtAl2009b} or due to the tidal potential of the parent molecular cloud  \citep{HortonEtAl2001}, both scenarios triggering enhanced fragmentation. 

\subsection{Towards a fragmentation cascade scenario}

 We note a negative correlation between the multiplicity within the UWPs and their separation  (see Fig.~\ref{Fig:TauG_WB_Sep_Nmult_Whole}): the  multiplicity decreases with wider separation. The fragmentation of dense molecular cores into UWPs with at least one  high-mass ($\gtrsim0.5\,M_\odot$) component seems to have   an impact
on the probability that either one component, and most probably both, will further fragment. The extent of this fragmentation cascade would depend on the separation of the first two fragments. It suggest a competitive scenario between collapse and an efficient fragmentation process of the gas core/clump to create multiple systems rather than one single object. The densest and most massive molecular cores in Taurus would produce high multiplicity hierarchical MM UWPs. Based on the work of \cite{KrausEtAl2011}, this fragmentation cascade scenario  may be inhibited at lower spatial scales, since they found  that there is no relation between a 1-5 kAU-wide binary and innermost high-order multiplicity distribution.

The probable primordial nature of UWPs suggests a scenario in which the turbulent fragmentation of a molecular filament forms over-densities \citep{PinedaEtAl2015} that undergo a fragmentation cascade, producing wide pairs that are themselves closer multiple systems. These systems may be stable enough to survive 1-30 Myr in a low-density environment. This link  between multiple systems and wide pairs has also been outlined by millimeter observations that show that fragmentation of clumps can proceed through separate envelopes or inside a common envelope \citep{LooneyEtAl2000}.

\subsection{Distributed and clustered modes of formation clues in Taurus}
Taurus is considered as the archetypical star-forming region that gives rise to isolated prestellar cores, with a typical density of $10^5\, {\rm cm}^{-3}$ and a size of 0.1 pc. The consensus model for star formation therefore predicts that this cloud will produce  a  distributed star population, 
in contrast with the clustered star-formation mode that is associated with denser regions such as $\rho$\,Ophiuchus,  with typical core densities of $10^7\, {\rm cm}^{-3}$ and sizes of 0.02-0.03 pc \citep{Ward-ThompsonEtAl2007}. 
In this context, it is worth noting the significant (more than two orders of magnitude) difference between the typical densities derived for the hypothesized cores that give birth to the SM pairs and MM pairs. These results suggest that, despite its overall low core densities, the Taurus molecular cloud may harbor the two modes of star formation, isolated and clustered.

\section{Conclusion}
In the Taurus star-forming complex, we have identified an extended binary regime  up to $\sim 60$ kAU using the  one point correlation function, $\Psi$,
which further allowed us to  distinguish three spatial regimes. 
The clustering regime has an upper clustering threshold of 0.1$^\circ$ (0.24 pc, i.e., 50 kAU, at the distance of Taurus). The  $\Psi$ function  appears to be scale-free  ($\psi= r^{-1.53}$) over three decades and extends  the usual binary regime, as defined by the two-point correlation function, to a very
wide pair regime. 
 We note that the value of the upper clustering length coincides with the local maximum of Taurus NH$^{\mathrm{2+}}$ molecular cores correlation function  reported by \cite{HacarEtAl2013}. As the latter indicates a typical separation between the cores, we expect to see a clustering imprint of stars starting at this value, as is observed.
This length is more than twice the universal typical width (0.1 pc)  of filamentary structures highlighted by Herschel in Taurus  and more generally in Gould Belt star-forming regions  \citep{AndreEtAl2014}.
The clustering regime is  followed by  a regime of stellar inhibition between 0.1$^\circ$ and 0.5$^\circ$ (0.24 pc up to 1.2 pc), and  ends beyond this with a  third regime associated to highly isolated stars. 

We have highlighted a major structural pattern in the spatial distribution of stars in Taurus based on their multiplicity status.
Distinguishing single stars from multiple systems, defined as having at least one physical companion within 1 kAU,  we highlight a major and unexpected difference in their spatial distribution based on 1-NNS study: the stellar neighborhood in the range of $1-10$ kAU of multiple stars is more `crowded' than the single star neighborhood: 40\% of multiple systems have a companion within that range compared to 15\% for single stars.  The probability that a multiple system has a wide companion is threefold that of a single star. 
We have shown that this excess  within 10 kAU is due to the very wide pairs candidates that are moreover generally composed themselves of two multiple systems.

We have identified a potential very wide pairs population in Taurus.
Our work shows that 55\% of our stellar catalog in Taurus are UWPs in the range of $1-60$ kAU. These UWPs are composed preferentially of stars of the same Class, departing from random Class pairing, and, thus, pointing towards coeval pairs. The pair fraction of UWPs follows an almost flat \"Opik function ($r^{0.14}$), extending what has been found for the young binaries at lower range. 
This coevality clue and \"Opik law behavior coupled with the fact that their 1-NNS distribution differs from  random mutual pairs suggests that these may be true physical pairs; indeed, UWPs with separation less than $5$ kAU have been found to be wide binaries \citep{KrausHillenbrand2009a}. 

Amongst the UWP population in Taurus, we distinguish three different types of UWPs, depending on whether they are composed either of two multiple systems (MM pairs), a single star and a multiple system (SM pairs), or two single stars (SS pairs). Their properties differ in terms of (primary) mass and separation range. The MM pairs are composed of more massive stars (median mass: 0.65 $M_\odot$) and tighter  (median value $3.5$ kAU, 75\% of them having separation below $9$ kAU). The median mass of stars within SS pairs is 0.25 $M_\odot$ and their median separation is $35$ kAU (75\% of them have a separation  above $12$ kAU). SM pairs have intermediate properties   (median mass of primaries of 0.5 $M_\odot$ and a uniform separation distribution covering the whole range $1-60$ kAU).  Multiplicity in these UWPs  is a decreasing function of stellar primary mass. We also show that the multiplicity  within UWPs   increases as the separation of (primary) stars decreases, showing an  increased high-order-multiple fraction for the tightest UWP  targets in Taurus. Class

These differences suggest a different scenario for the formation of UWPs in Taurus. The `massive' MM pairs most probably reflect the imprints of the star-formation process within the same molecular core. Based on estimates of their hypothesized natal core properties, Taurus may generate both isolated and clustered star formation. The `low-mass' SS pairs could point to a different formation scenario or may result from a dynamical process, such as low-mass star ejection from multiple systems. 

 Obtaining parallaxes and proper motions for all Taurus members would help tremendously in identifying which of the UWPs are physical binaries, common proper motion pairs and those which may form from a dynamical process. This will ultimately be achieved thanks to the {\it Gaia} survey that will determine the  astrometric quantities (angular position, proper motion, and parallax)  of stars with an accuracy of 20 $\mu$ and at a brightness of 15 mag,  and determine the radial velocity of stars brighter than 16 mag.

 Unfortunately, the first {\it Gaia} release \citep{GaiaDr1-2016} does not provide enough data on the whole sample of stars in Taurus (only $\sim$ 5\%). We expect that future releases of {\it Gaia}, perhaps as early as the second release, will allow us to firmly establish the physical status of the UWPs, in turn providing a means to discriminate between different scenarios for their formation, and more generally will provide invaluable information on the structure and dynamics of Taurus as a young association \citep{Moraux2016}. 
 
In summary, we identified a new category of ultra-wide pairs (UWPs) in Taurus outlining a high order of multiplicity for tighter ones (separation less than 10 kAU). We suggest that part of these UWPs may be pristine imprints of their spatial configuration at  birth and put forward a cascade fragmentation scenario for their formation.
Furthmore, UWPs may constitute  a potential link between the wide pairs recently  identified in very young Class 0 type objects ($\lesssim$0.5 Myr) and the somewhat older but still young moving groups ($\sim 20-30$ Myr).

\begin{acknowledgements} 
The authors would like to thank Lee Mundy for helpful discussions and Marc Pound for reading the first draft of this article. We also want to  sincerely thank the anonymous referee who helped us to improve both the scientific and language aspects of this paper.
This work was funded  by the French national research agency through ANR 2010 JCJC 0501-1 DESC (PI E. Moraux).
\end{acknowledgements}



\begin{thebibliography}{156}
\expandafter\ifx\csname natexlab\endcsname\relax\def\natexlab#1{#1}\fi

\bibitem[{Ahn \& Fessler(2003)}]{AhnEtAl2003}
Ahn, S. \& Fessler, A. 2003, Standard Errors of Mean, Variance, and Standard
  Deviation Estimators, Tech. rep., Technical Report, Ann Arbor, MI, USA

\bibitem[{{Alonso-Floriano} {et~al.}(2015){Alonso-Floriano}, {Caballero},
  {Cort{\'e}s-Contreras}, {Solano}, \& {Montes}}]{Alonso-FlorianoEtAl2015}
{Alonso-Floriano}, F.~J., {Caballero}, J.~A., {Cort{\'e}s-Contreras}, M.,
  {Solano}, E., \& {Montes}, D. 2015, \aap, 583, A85

\bibitem[{{Alves} {et~al.}(2007){Alves}, {Lombardi}, \& {Lada}}]{AlvesEtAl2007}
{Alves}, J., {Lombardi}, M., \& {Lada}, C.~J. 2007, \aap, 462, L17

\bibitem[{{Ambartsumian}(1937)}]{Ambartsumian1937}
{Ambartsumian}, V.~A. 1937, Astronomicheskii Zhurnal, 14, 207

\bibitem[{{Andr{\'e}} {et~al.}(2014){Andr{\'e}}, {Di Francesco},
  {Ward-Thompson}, {Inutsuka}, {Pudritz}, \& {Pineda}}]{AndreEtAl2014}
{Andr{\'e}}, P., {Di Francesco}, J., {Ward-Thompson}, D., {et~al.} 2014,
  Protostars and Planets VI, 27

\bibitem[{{Andre} {et~al.}(1993){Andre}, {Ward-Thompson}, \&
  {Barsony}}]{AndreEtAl1993}
{Andre}, P., {Ward-Thompson}, D., \& {Barsony}, M. 1993, \apj, 406, 122

\bibitem[{{Andrews} {et~al.}(2013){Andrews}, {Rosenfeld}, {Kraus}, \&
  {Wilner}}]{AndrewsEtAl2013}
{Andrews}, S.~M., {Rosenfeld}, K.~A., {Kraus}, A.~L., \& {Wilner}, D.~J. 2013,
  \apj, 771, 129

\bibitem[{Baddeley \& Turner(2005)}]{R-Spatstat2005}
Baddeley, A. \& Turner, R. 2005, Journal of Statistical Software, 12, 1

\bibitem[{{Bahcall} {et~al.}(1985){Bahcall}, {Hut}, \&
  {Tremaine}}]{BahcallEtAl1985}
{Bahcall}, J.~N., {Hut}, P., \& {Tremaine}, S. 1985, \apj, 290, 15

\bibitem[{{Ballesteros-Paredes}
  {et~al.}(2009{\natexlab{a}}){Ballesteros-Paredes}, {G{\'o}mez}, {Loinard},
  {Torres}, \& {Pichardo}}]{BallesterosEtAl2009b}
{Ballesteros-Paredes}, J., {G{\'o}mez}, G.~C., {Loinard}, L., {Torres}, R.~M.,
  \& {Pichardo}, B. 2009{\natexlab{a}}, \mnras, 395, L81

\bibitem[{{Ballesteros-Paredes}
  {et~al.}(2009{\natexlab{b}}){Ballesteros-Paredes}, {G{\'o}mez}, {Pichardo},
  \& {V{\'a}zquez-Semadeni}}]{BallesterosEtAl2009a}
{Ballesteros-Paredes}, J., {G{\'o}mez}, G.~C., {Pichardo}, B., \&
  {V{\'a}zquez-Semadeni}, E. 2009{\natexlab{b}}, \mnras, 393, 1563

\bibitem[{{Bate}(2012)}]{Bate2012}
{Bate}, M.~R. 2012, \mnras, 419, 3115

\bibitem[{{Bate} {et~al.}(2003){Bate}, {Bonnell}, \& {Bromm}}]{BateEtAl2003}
{Bate}, M.~R., {Bonnell}, I.~A., \& {Bromm}, V. 2003, \mnras, 339, 577

\bibitem[{Benaglia {et~al.}(2009)Benaglia, Chauveau, Hunter, \&
  Young}]{R-mixtools2009}
Benaglia, T., Chauveau, D., Hunter, D.~R., \& Young, D. 2009, Journal of
  Statistical Software, 32, 1

\bibitem[{{Bertout} {et~al.}(2007){Bertout}, {Siess}, \&
  {Cabrit}}]{BertoutEtAl2007}
{Bertout}, C., {Siess}, L., \& {Cabrit}, S. 2007, \aap, 473, L21

\bibitem[{{Binney} \& {Tremaine}(2008)}]{BinneyTremaine2008}
{Binney}, J. \& {Tremaine}, S. 2008, {Galactic Dynamics: Second Edition}
  (Princeton University Press)

\bibitem[{{Boss}(2001)}]{Boss2001}
{Boss}, A.~P. 2001, \apjl, 551, L167

\bibitem[{{Brice{\~n}o} {et~al.}(2002){Brice{\~n}o}, {Luhman}, {Hartmann},
  {Stauffer}, \& {Kirkpatrick}}]{BricenoEtAl2002}
{Brice{\~n}o}, C., {Luhman}, K.~L., {Hartmann}, L., {Stauffer}, J.~R., \&
  {Kirkpatrick}, J.~D. 2002, \apj, 580, 317

\bibitem[{{Burrows} {et~al.}(1996){Burrows}, {Stapelfeldt}, {Watson}, {Krist},
  {Ballester}, {Clarke}, {Crisp}, {Gallagher}, {Griffiths}, {Hester},
  {Hoessel}, {Holtzman}, {Mould}, {Scowen}, {Trauger}, \&
  {Westphal}}]{burrows96}
{Burrows}, C.~J., {Stapelfeldt}, K.~R., {Watson}, A.~M., {et~al.} 1996, \apj,
  473, 437

\bibitem[{{Carney} {et~al.}(2016){Carney}, {Y{\i}ld{\i}z}, {Mottram}, {van
  Dishoeck}, {Ramchandani}, \& {J{\o}rgensen}}]{CarneyEtAl2016}
{Carney}, M.~T., {Y{\i}ld{\i}z}, U.~A., {Mottram}, J.~C., {et~al.} 2016, \aap,
  586, A44

\bibitem[{Chakraborty {et~al.}(2014)Chakraborty, Feigelson, \&
  Babu}]{R-astrolab2014}
Chakraborty, A., Feigelson, E.~D., \& Babu, G.~J. 2014, {astrolabe}: Astronomy
  Users Library for R, r package version 0.1

\bibitem[{{Chakraborty} \& {Ge}(2004)}]{chakraborty04}
{Chakraborty}, A. \& {Ge}, J. 2004, \aj, 127, 2898

\bibitem[{{Chanam{\'e}} \& {Gould}(2004)}]{ChanameGould2004}
{Chanam{\'e}}, J. \& {Gould}, A. 2004, \apj, 601, 289

\bibitem[{{Chandrasekhar}(1943)}]{Chandrasekhar1943}
{Chandrasekhar}, S. 1943, Reviews of Modern Physics, 15, 1

\bibitem[{{Chandrasekhar}(1944)}]{Chandrasekhar1944}
{Chandrasekhar}, S. 1944, \apj, 99, 54

\bibitem[{{Chen} {et~al.}(2013){Chen}, {Arce}, {Zhang}, {Bourke}, {Launhardt},
  {J{\o}rgensen}, {Lee}, {Foster}, {Dunham}, {Pineda}, \&
  {Henning}}]{ChenEtAl2013}
{Chen}, X., {Arce}, H.~G., {Zhang}, Q., {et~al.} 2013, \apj, 768, 110

\bibitem[{{Connelley} {et~al.}(2008){Connelley}, {Reipurth}, \&
  {Tokunaga}}]{connelley08}
{Connelley}, M.~S., {Reipurth}, B., \& {Tokunaga}, A.~T. 2008, \aj, 135, 2496

\bibitem[{{Connelley} {et~al.}(2009){Connelley}, {Reipurth}, \&
  {Tokunaga}}]{connelley09}
{Connelley}, M.~S., {Reipurth}, B., \& {Tokunaga}, A.~T. 2009, \aj, 138, 1193

\bibitem[{{Correia} {et~al.}(2006){Correia}, {Zinnecker}, {Ratzka}, \&
  {Sterzik}}]{correia06}
{Correia}, S., {Zinnecker}, H., {Ratzka}, T., \& {Sterzik}, M.~F. 2006, \aap,
  459, 909

\bibitem[{{Covey} {et~al.}(2006){Covey}, {Greene}, {Doppmann}, \&
  {Lada}}]{CoveyEtAl2006}
{Covey}, K.~R., {Greene}, T.~P., {Doppmann}, G.~W., \& {Lada}, C.~J. 2006, \aj,
  131, 512

\bibitem[{{Daemgen} {et~al.}(2015){Daemgen}, {Bonavita}, {Jayawardhana},
  {Lafreni{\`e}re}, \& {Janson}}]{daemgen15}
{Daemgen}, S., {Bonavita}, M., {Jayawardhana}, R., {Lafreni{\`e}re}, D., \&
  {Janson}, M. 2015, \apj, 799, 155

\bibitem[{Dahl(2014)}]{R-xtable2014}
Dahl, D.~B. 2014, xtable: Export tables to LaTeX or HTML, r package version
  1.7-4

\bibitem[{{Dhital} {et~al.}(2015){Dhital}, {West}, {Stassun}, {Schluns}, \&
  {Massey}}]{DhitalEtAl2015}
{Dhital}, S., {West}, A.~A., {Stassun}, K.~G., {Schluns}, K.~J., \& {Massey},
  A.~P. 2015, \aj, 150, 57

\bibitem[{{Di Folco} {et~al.}(2014){Di Folco}, {Dutrey}, {Le Bouquin},
  {Lacour}, {Berger}, {K{\"o}hler}, {Guilloteau}, {Pi{\'e}tu}, {Bary}, {Beck},
  {Beust}, \& {Pantin}}]{difolco14}
{Di Folco}, E., {Dutrey}, A., {Le Bouquin}, J.-B., {et~al.} 2014, \aap, 565, L2

\bibitem[{{Dobbs} {et~al.}(2014){Dobbs}, {Krumholz}, {Ballesteros-Paredes},
  {Bolatto}, {Fukui}, {Heyer}, {Low}, {Ostriker}, \&
  {V{\'a}zquez-Semadeni}}]{DobbsEtAl2014}
{Dobbs}, C.~L., {Krumholz}, M.~R., {Ballesteros-Paredes}, J., {et~al.} 2014,
  Protostars and Planets VI, 3

\bibitem[{{Duch{\^e}ne}(1999)}]{duchene99}
{Duch{\^e}ne}, G. 1999, \aap, 341, 547

\bibitem[{{Duch{\^e}ne} {et~al.}(2007){Duch{\^e}ne}, {Bontemps}, {Bouvier},
  {Andr{\'e}}, {Djupvik}, \& {Ghez}}]{duchene07}
{Duch{\^e}ne}, G., {Bontemps}, S., {Bouvier}, J., {et~al.} 2007, \aap, 476, 229

\bibitem[{{Duch{\^e}ne} {et~al.}(2003){Duch{\^e}ne}, {Ghez}, {McCabe}, \&
  {Weinberger}}]{duchene03}
{Duch{\^e}ne}, G., {Ghez}, A.~M., {McCabe}, C., \& {Weinberger}, A.~J. 2003,
  \apj, 592, 288

\bibitem[{{Duch{\^e}ne} \& {Kraus}(2013)}]{DucheneKraus2013}
{Duch{\^e}ne}, G. \& {Kraus}, A. 2013, \araa, 51, 269

\bibitem[{{Duquennoy} \& {Mayor}(1991)}]{DM91}
{Duquennoy}, A. \& {Mayor}, M. 1991, \aap, 248, 485

\bibitem[{{Elliott} \& {Bayo}(2016)}]{ElliottBayo2016}
{Elliott}, P. \& {Bayo}, A. 2016, \mnras, 459, 4499

\bibitem[{{Elliott} {et~al.}(2016){Elliott}, {Bayo}, {Melo}, {Torres},
  {Sterzik}, {Quast}, {Montes}, \& {Brahm}}]{ElliottEtAl2016}
{Elliott}, P., {Bayo}, A., {Melo}, C.~H.~F., {et~al.} 2016, \aap, 590, A13

\bibitem[{{Esplin} {et~al.}(2014){Esplin}, {Luhman}, \&
  {Mamajek}}]{EsplinEtAl2014}
{Esplin}, T.~L., {Luhman}, K.~L., \& {Mamajek}, E.~E. 2014, \apj, 784, 126

\bibitem[{{Frink} {et~al.}(1997){Frink}, {R{\"o}ser}, {Neuh{\"a}user}, \&
  {Sterzik}}]{FrinkEtAL1997}
{Frink}, S., {R{\"o}ser}, S., {Neuh{\"a}user}, R., \& {Sterzik}, M.~F. 1997,
  \aap, 325, 613

\bibitem[{{Gaia Collaboration} {et~al.}(2016){Gaia Collaboration}, {Brown},
  {Vallenari}, {Prusti}, {de Bruijne}, {Mignard}, {Drimmel}, \&
  {co-authors}}]{GaiaDr1-2016}
{Gaia Collaboration}, {Brown}, A.~G.~A., {Vallenari}, A., {et~al.} 2016, ArXiv
  e-prints [\eprint[arXiv]{1609.04172}]

\bibitem[{{Ghez} {et~al.}(1993){Ghez}, {Neugebauer}, \& {Matthews}}]{ghez93}
{Ghez}, A.~M., {Neugebauer}, G., \& {Matthews}, K. 1993, \aj, 106, 2005

\bibitem[{{Ghez} {et~al.}(1997){Ghez}, {White}, \& {Simon}}]{ghez97}
{Ghez}, A.~M., {White}, R.~J., \& {Simon}, M. 1997, \apj, 490, 353

\bibitem[{{Gladwin} {et~al.}(1999){Gladwin}, {Kitsionas}, {Boffin}, \&
  {Whitworth}}]{GladwinEtAl1999}
{Gladwin}, P.~P., {Kitsionas}, S., {Boffin}, H.~M.~J., \& {Whitworth}, A.~P.
  1999, \mnras, 302, 305

\bibitem[{{Gomez} {et~al.}(1993){Gomez}, {Hartmann}, {Kenyon}, \&
  {Hewett}}]{GomezEtAl1993}
{Gomez}, M., {Hartmann}, L., {Kenyon}, S.~J., \& {Hewett}, R. 1993, \aj, 105,
  1927

\bibitem[{{Goodwin} {et~al.}(2008){Goodwin}, {Nutter}, {Kroupa},
  {Ward-Thompson}, \& {Whitworth}}]{GoodwinEtAl2008}
{Goodwin}, S.~P., {Nutter}, D., {Kroupa}, P., {Ward-Thompson}, D., \&
  {Whitworth}, A.~P. 2008, \aap, 477, 823

\bibitem[{{Goodwin} {et~al.}(2004){Goodwin}, {Whitworth}, \&
  {Ward-Thompson}}]{GoodwinEtAl2004}
{Goodwin}, S.~P., {Whitworth}, A.~P., \& {Ward-Thompson}, D. 2004, \aap, 419,
  543

\bibitem[{Graffelman(2013)}]{R-calibrate2013}
Graffelman, J. 2013, calibrate: Calibration of Scatterplot and Biplot Axes, r
  package version 1.7.2

\bibitem[{{Greene} {et~al.}(1994){Greene}, {Wilking}, {Andre}, {Young}, \&
  {Lada}}]{GreeneEtAl1994}
{Greene}, T.~P., {Wilking}, B.~A., {Andre}, P., {Young}, E.~T., \& {Lada},
  C.~J. 1994, \apj, 434, 614

\bibitem[{{Hacar} {et~al.}(2013){Hacar}, {Tafalla}, {Kauffmann}, \&
  {Kov{\'a}cs}}]{HacarEtAl2013}
{Hacar}, A., {Tafalla}, M., {Kauffmann}, J., \& {Kov{\'a}cs}, A. 2013, \aap,
  554, A55

\bibitem[{{Harrell} {et~al.}(2015){Harrell}, {Dupont}, \&
  {others}}]{R-Hmisc2015}
{Harrell}, F., E., {Dupont}, C., \& {others}. 2015, Hmisc: Harrell
  Miscellaneous, r package version 3.16-0

\bibitem[{Harris(2013)}]{R-FITSio2013}
Harris, A. 2013, FITSio: FITS (Flexible Image Transport System) utilities, r
  package version 2.0-0

\bibitem[{{Hartmann}(2002)}]{Hartmann2002}
{Hartmann}, L. 2002, \apj, 578, 914

\bibitem[{{Heggie}(1975)}]{Heggie1975}
{Heggie}, D.~C. 1975, \mnras, 173, 729

\bibitem[{{Heggie}(1977)}]{Heggie1977}
{Heggie}, D.~C. 1977, \rmxaa, 3

\bibitem[{{Herczeg} \& {Hillenbrand}(2014)}]{HerczegHillenbrand2014}
{Herczeg}, G.~J. \& {Hillenbrand}, L.~A. 2014, \apj, 786, 97

\bibitem[{{Holman} {et~al.}(2013){Holman}, {Walch}, {Goodwin}, \&
  {Whitworth}}]{HolmanEtAl2013}
{Holman}, K., {Walch}, S.~K., {Goodwin}, S.~P., \& {Whitworth}, A.~P. 2013,
  \mnras, 432, 3534

\bibitem[{{Horton} {et~al.}(2001){Horton}, {Bate}, \&
  {Bonnell}}]{HortonEtAl2001}
{Horton}, A.~J., {Bate}, M.~R., \& {Bonnell}, I.~A. 2001, \mnras, 321, 585

\bibitem[{{Ireland} \& {Kraus}(2008)}]{ireland08}
{Ireland}, M.~J. \& {Kraus}, A.~L. 2008, \apjl, 678, L59

\bibitem[{{Itoh} {et~al.}(2005){Itoh}, {Hayashi}, {Tamura}, {Tsuji}, {Oasa},
  {Fukagawa}, {Hayashi}, {Naoi}, {Ishii}, {Mayama}, {Morino}, {Yamashita},
  {Pyo}, {Nishikawa}, {Usuda}, {Murakawa}, {Suto}, {Oya}, {Takato}, {Ando},
  {Miyama}, {Kobayashi}, \& {Kaifu}}]{itoh05}
{Itoh}, Y., {Hayashi}, M., {Tamura}, M., {et~al.} 2005, \apj, 620, 984

\bibitem[{{Jeans}(1919)}]{Jeans1919}
{Jeans}, J.~H. 1919, \mnras, 79, 408

\bibitem[{{Jiang} \& {Tremaine}(2010)}]{JiangTremaine2010}
{Jiang}, Y.-F. \& {Tremaine}, S. 2010, \mnras, 401, 977

\bibitem[{{Jones} \& {Herbig}(1979)}]{JonesHerbig1979}
{Jones}, B.~F. \& {Herbig}, G.~H. 1979, \aj, 84, 1872

\bibitem[{{Juvela} \& {Montillaud}(2016)}]{JuvelaEtAl2016}
{Juvela}, M. \& {Montillaud}, J. 2016, \aap, 585, A78

\bibitem[{Kendall \& Stuart(1977)}]{KendallStuart1977}
Kendall, M. \& Stuart, A. 1977, The advanced theory of statistics, 4th edn.,
  Vol. 1: Distribution theory (New York, NY: Macmillan)

\bibitem[{{Kenyon} \& {Hartmann}(1995)}]{KenyonHartman1995}
{Kenyon}, S.~J. \& {Hartmann}, L. 1995, \apjs, 101, 117

\bibitem[{{Kerscher} {et~al.}(2000){Kerscher}, {Szapudi}, \&
  {Szalay}}]{KerscherEtAl2000}
{Kerscher}, M., {Szapudi}, I., \& {Szalay}, A.~S. 2000, \apjl, 535, L13

\bibitem[{{Kirk} \& {Myers}(2011)}]{KirkMyers2011}
{Kirk}, H. \& {Myers}, P.~C. 2011, \apj, 727, 64

\bibitem[{{Kohler} \& {Leinert}(1998)}]{koehler98}
{Kohler}, R. \& {Leinert}, C. 1998, \aap, 331, 977

\bibitem[{{Konopacky} {et~al.}(2007){Konopacky}, {Ghez}, {Rice}, \&
  {Duch{\^e}ne}}]{konopacky07}
{Konopacky}, Q.~M., {Ghez}, A.~M., {Rice}, E.~L., \& {Duch{\^e}ne}, G. 2007,
  \apj, 663, 394

\bibitem[{{K{\"o}nyves} {et~al.}(2015){K{\"o}nyves}, {Andr{\'e}},
  {Men'shchikov}, {Palmeirim}, {Arzoumanian}, {Schneider}, {Roy}, {Didelon},
  {Maury}, {Shimajiri}, {Di Francesco}, {Bontemps}, {Peretto}, {Benedettini},
  {Bernard}, {Elia}, {Griffin}, {Hill}, {Kirk}, {Ladjelate}, {Marsh}, {Martin},
  {Motte}, {Nguy{\^e}n Luong}, {Pezzuto}, {Roussel}, {Rygl}, {Sadavoy},
  {Schisano}, {Spinoglio}, {Ward-Thompson}, \& {White}}]{KonyvesEtAl2015}
{K{\"o}nyves}, V., {Andr{\'e}}, P., {Men'shchikov}, A., {et~al.} 2015, \aap,
  584, A91

\bibitem[{{Koresko}(2000)}]{koresko00}
{Koresko}, C.~D. 2000, \apjl, 531, L147

\bibitem[{{Kouwenhoven} {et~al.}(2010){Kouwenhoven}, {Goodwin}, {Parker},
  {Davies}, {Malmberg}, \& {Kroupa}}]{KouwenhovenEtAl2010}
{Kouwenhoven}, M.~B.~N., {Goodwin}, S.~P., {Parker}, R.~J., {et~al.} 2010,
  \mnras, 404, 1835

\bibitem[{{Kraus} \& {Hillenbrand}(2007)}]{KrausHillenbrand2007}
{Kraus}, A.~L. \& {Hillenbrand}, L.~A. 2007, \apj, 662, 413

\bibitem[{{Kraus} \& {Hillenbrand}(2008)}]{KrausHillenbrand2008}
{Kraus}, A.~L. \& {Hillenbrand}, L.~A. 2008, \apjl, 686, L111

\bibitem[{{Kraus} \& {Hillenbrand}(2009{\natexlab{a}})}]{KrausHillenbrand2009b}
{Kraus}, A.~L. \& {Hillenbrand}, L.~A. 2009{\natexlab{a}}, \apj, 704, 531

\bibitem[{{Kraus} \& {Hillenbrand}(2009{\natexlab{b}})}]{KrausHillenbrand2009a}
{Kraus}, A.~L. \& {Hillenbrand}, L.~A. 2009{\natexlab{b}}, \apj, 703, 1511

\bibitem[{{Kraus} \& {Hillenbrand}(2012)}]{kraus12}
{Kraus}, A.~L. \& {Hillenbrand}, L.~A. 2012, \apj, 757, 141

\bibitem[{{Kraus} {et~al.}(2011{\natexlab{a}}){Kraus}, {Ireland}, {Martinache},
  \& {Hillenbrand}}]{KrausEtAl2011}
{Kraus}, A.~L., {Ireland}, M.~J., {Martinache}, F., \& {Hillenbrand}, L.~A.
  2011{\natexlab{a}}, \apj, 731, 8

\bibitem[{{Kraus} {et~al.}(2011{\natexlab{b}}){Kraus}, {Ireland}, {Martinache},
  \& {Hillenbrand}}]{kraus11}
{Kraus}, A.~L., {Ireland}, M.~J., {Martinache}, F., \& {Hillenbrand}, L.~A.
  2011{\natexlab{b}}, \apj, 731, 8

\bibitem[{{Kraus} {et~al.}(2006){Kraus}, {White}, \& {Hillenbrand}}]{kraus06}
{Kraus}, A.~L., {White}, R.~J., \& {Hillenbrand}, L.~A. 2006, \apj, 649, 306

\bibitem[{{Kroupa} \& {Bouvier}(2003{\natexlab{a}})}]{KroupaBouvier2003b}
{Kroupa}, P. \& {Bouvier}, J. 2003{\natexlab{a}}, \mnras, 346, 369

\bibitem[{{Kroupa} \& {Bouvier}(2003{\natexlab{b}})}]{KroupaBouvier2003}
{Kroupa}, P. \& {Bouvier}, J. 2003{\natexlab{b}}, \mnras, 346, 343

\bibitem[{{Kroupa} {et~al.}(2003){Kroupa}, {Bouvier}, {Duch{\^e}ne}, \&
  {Moraux}}]{KroupaEtAl2003a}
{Kroupa}, P., {Bouvier}, J., {Duch{\^e}ne}, G., \& {Moraux}, E. 2003, \mnras,
  346, 354

\bibitem[{{Lada}(1987)}]{Lada1987}
{Lada}, C.~J. 1987, in IAU Symposium, Vol. 115, Star Forming Regions, ed.
  M.~{Peimbert} \& J.~{Jugaku}, 1--17

\bibitem[{{Lada} \& {Lada}(2003)}]{LadaEtAl2003}
{Lada}, C.~J. \& {Lada}, E.~A. 2003, \araa, 41, 57

\bibitem[{{Lada} {et~al.}(2008){Lada}, {Muench}, {Rathborne}, {Alves}, \&
  {Lombardi}}]{LadaEtAl2008}
{Lada}, C.~J., {Muench}, A.~A., {Rathborne}, J., {Alves}, J.~F., \& {Lombardi},
  M. 2008, \apj, 672, 410

\bibitem[{{Lada} \& {Wilking}(1984)}]{LadaWilking1984}
{Lada}, C.~J. \& {Wilking}, B.~A. 1984, \apj, 287, 610

\bibitem[{{Lamers} \& {Gieles}(2006)}]{LamersGieles2006}
{Lamers}, H.~J.~G.~L.~M. \& {Gieles}, M. 2006, \aap, 455, L17

\bibitem[{{Landy} \& {Szalay}(1993)}]{LandyEtAl1993}
{Landy}, S.~D. \& {Szalay}, A.~S. 1993, \apj, 412, 64

\bibitem[{{Larson}(1995)}]{Larson1995}
{Larson}, R.~B. 1995, \mnras, 272, 213

\bibitem[{{Law} {et~al.}(2010){Law}, {Dhital}, {Kraus}, {Stassun}, \&
  {West}}]{LawEtAl2010}
{Law}, N.~M., {Dhital}, S., {Kraus}, A., {Stassun}, K.~G., \& {West}, A.~A.
  2010, \apj, 720, 1727

\bibitem[{{Lee} {et~al.}(2015){Lee}, {Dunham}, {Myers}, {Tobin}, {Kristensen},
  {Pineda}, {Vorobyov}, {Offner}, {Arce}, {Li}, {Bourke}, {J{\o}rgensen},
  {Goodman}, {Sadavoy}, {Chandler}, {Harris}, {Kratter}, {Looney}, {Melis},
  {Perez}, \& {Segura-Cox}}]{LeeEtAl2015}
{Lee}, K.~I., {Dunham}, M.~M., {Myers}, P.~C., {et~al.} 2015, \apj, 814, 114

\bibitem[{{Leinert} {et~al.}(1997){Leinert}, {Richichi}, \& {Haas}}]{leinert97}
{Leinert}, C., {Richichi}, A., \& {Haas}, M. 1997, \aap, 318, 472

\bibitem[{{Leinert} {et~al.}(1993){Leinert}, {Zinnecker}, {Weitzel},
  {Christou}, {Ridgway}, {Jameson}, {Haas}, \& {Lenzen}}]{leinert93}
{Leinert}, C., {Zinnecker}, H., {Weitzel}, N., {et~al.} 1993, \aap, 278, 129

\bibitem[{{L{\'e}pine} \& {Bongiorno}(2007)}]{LepineBongiorno2007}
{L{\'e}pine}, S. \& {Bongiorno}, B. 2007, \aj, 133, 889

\bibitem[{{Lomax} {et~al.}(2013){Lomax}, {Whitworth}, \&
  {Cartwright}}]{LomaxEtAl2013}
{Lomax}, O., {Whitworth}, A.~P., \& {Cartwright}, A. 2013, \mnras, 436, 2680

\bibitem[{{Longhitano} \& {Binggeli}(2010)}]{LonghitanoBinggeli2010}
{Longhitano}, M. \& {Binggeli}, B. 2010, \aap, 509, A46

\bibitem[{{Looney} {et~al.}(2000){Looney}, {Mundy}, \&
  {Welch}}]{LooneyEtAl2000}
{Looney}, L.~W., {Mundy}, L.~G., \& {Welch}, W.~J. 2000, \apj, 529, 477

\bibitem[{{Luhman}(2004)}]{Luhman2004b}
{Luhman}, K.~L. 2004, \apj, 617, 1216

\bibitem[{{Luhman}(2006)}]{Luhman2006}
{Luhman}, K.~L. 2006, \apj, 645, 676

\bibitem[{{Luhman} {et~al.}(2010){Luhman}, {Allen}, {Espaillat}, {Hartmann}, \&
  {Calvet}}]{LuhmanEtAl2010}
{Luhman}, K.~L., {Allen}, P.~R., {Espaillat}, C., {Hartmann}, L., \& {Calvet},
  N. 2010, \apjs, 186, 111

\bibitem[{{Luhman} {et~al.}(2016){Luhman}, {Mamajek}, {Shukla}, \&
  {Loutrel}}]{LuhmanEtAl2016Arxiv}
{Luhman}, K.~L., {Mamajek}, E.~E., {Shukla}, S.~J., \& {Loutrel}, N.~P. 2016,
  ArXiv e-prints [\eprint[arXiv]{1610.09412}]

\bibitem[{{Marks} \& {Kroupa}(2012)}]{MarksKroupa2012}
{Marks}, M. \& {Kroupa}, P. 2012, \aap, 543, A8

\bibitem[{{Marsh} {et~al.}(2016){Marsh}, {Kirk}, {Andr{\'e}}, {Griffin},
  {K{\"o}nyves}, {Palmeirim}, {Men'shchikov}, {Ward-Thompson}, {Benedettini},
  {Bresnahan}, {Francesco}, {Elia}, {Motte}, {Peretto}, {Pezzuto}, {Roy},
  {Sadavoy}, {Schneider}, {Spinoglio}, \& {White}}]{MarshEtAl2016}
{Marsh}, K.~A., {Kirk}, J.~M., {Andr{\'e}}, P., {et~al.} 2016, \mnras, 459, 342

\bibitem[{{Mathieu}(1994)}]{Mathieu1994}
{Mathieu}, R.~D. 1994, \araa, 32, 465

\bibitem[{{Moeckel} \& {Bate}(2010)}]{MoeckelBate2010}
{Moeckel}, N. \& {Bate}, M.~R. 2010, \mnras, 404, 721

\bibitem[{{Moeckel} \& {Clarke}(2011)}]{MoeckelClarke2011}
{Moeckel}, N. \& {Clarke}, C.~J. 2011, \mnras, 415, 1179

\bibitem[{{Monnier} {et~al.}(2008){Monnier}, {Tannirkulam}, {Tuthill},
  {Ireland}, {Cohen}, {Danchi}, \& {Baron}}]{monnier08}
{Monnier}, J.~D., {Tannirkulam}, A., {Tuthill}, P.~G., {et~al.} 2008, \apjl,
  681, L97

\bibitem[{{Moraux}(2016)}]{Moraux2016}
{Moraux}, E. 2016, ArXiv e-prints [\eprint[arXiv]{1607.00027}]

\bibitem[{{Myers} {et~al.}(1991){Myers}, {Fuller}, {Goodman}, \&
  {Benson}}]{MyersEtAl1991}
{Myers}, P.~C., {Fuller}, G.~A., {Goodman}, A.~A., \& {Benson}, P.~J. 1991,
  \apj, 376, 561

\bibitem[{Nychka {et~al.}(2015)Nychka, Furrer, \& Sain}]{R-fields2015}
Nychka, D., Furrer, R., \& Sain, S. 2015, R-fields: Tools for Spatial Data, r
  package version 8.2-1

\bibitem[{{Offner} {et~al.}(2014){Offner}, {Clark}, {Hennebelle}, {Bastian},
  {Bate}, {Hopkins}, {Moraux}, \& {Whitworth}}]{OffnerEtAl2014}
{Offner}, S.~S.~R., {Clark}, P.~C., {Hennebelle}, P., {et~al.} 2014, Protostars
  and Planets VI, 53

\bibitem[{{Padgett} {et~al.}(1999){Padgett}, {Brandner}, {Stapelfeldt},
  {Strom}, {Terebey}, \& {Koerner}}]{padgett99}
{Padgett}, D.~L., {Brandner}, W., {Stapelfeldt}, K.~R., {et~al.} 1999, \aj,
  117, 1490

\bibitem[{{Parker} {et~al.}(2009){Parker}, {Goodwin}, {Kroupa}, \&
  {Kouwenhoven}}]{ParkerEtAl2009}
{Parker}, R.~J., {Goodwin}, S.~P., {Kroupa}, P., \& {Kouwenhoven}, M.~B.~N.
  2009, \mnras, 397, 1577

\bibitem[{{Peebles}(1980)}]{Peebles1980}
{Peebles}, P.~J.~E. 1980, in Annals of the New York Academy of Sciences, Vol.
  336, Ninth Texas Symposium on Relativistic Astrophysics, ed. J.~{Ehlers},
  J.~J. {Perry}, \& M.~{Walker}, 161--171

\bibitem[{{Pineda} {et~al.}(2015){Pineda}, {Offner}, {Parker}, {Arce},
  {Goodman}, {Caselli}, {Fuller}, {Bourke}, \& {Corder}}]{PinedaEtAl2015}
{Pineda}, J.~E., {Offner}, S.~S.~R., {Parker}, R.~J., {et~al.} 2015, \nat, 518,
  213

\bibitem[{{R Core Team}(2015)}]{R-ManualR2015}
{R Core Team}. 2015, R: A Language and Environment for Statistical Computing, R
  Foundation for Statistical Computing, Vienna, Austria

\bibitem[{{Raghavan} {et~al.}(2010){Raghavan}, {McAlister}, {Henry}, {Latham},
  {Marcy}, {Mason}, {Gies}, {White}, \& {ten Brummelaar}}]{RaghavanEtAl2010}
{Raghavan}, D., {McAlister}, H.~A., {Henry}, T.~J., {et~al.} 2010, \apjs, 190,
  1

\bibitem[{{Rebull} {et~al.}(2010){Rebull}, {Padgett}, {McCabe}, {Hillenbrand},
  {Stapelfeldt}, {Noriega-Crespo}, {Carey}, {Brooke}, {Huard}, {Terebey},
  {Audard}, {Monin}, {Fukagawa}, {G{\"u}del}, {Knapp}, {Menard}, {Allen},
  {Angione}, {Baldovin-Saavedra}, {Bouvier}, {Briggs}, {Dougados}, {Evans},
  {Flagey}, {Guieu}, {Grosso}, {Glauser}, {Harvey}, {Hines}, {Latter},
  {Skinner}, {Strom}, {Tromp}, \& {Wolf}}]{RebullEtAl2010}
{Rebull}, L.~M., {Padgett}, D.~L., {McCabe}, C.-E., {et~al.} 2010, \apjs, 186,
  259

\bibitem[{{Reipurth}(2000)}]{Reipurth2000}
{Reipurth}, B. 2000, \aj, 120, 3177

\bibitem[{{Reipurth} \& {Clarke}(2001)}]{ReipurthClarke2001}
{Reipurth}, B. \& {Clarke}, C. 2001, \aj, 122, 432

\bibitem[{{Reipurth} {et~al.}(2014){Reipurth}, {Clarke}, {Boss}, {Goodwin},
  {Rodr{\'{\i}}guez}, {Stassun}, {Tokovinin}, \&
  {Zinnecker}}]{ReipurthEtAl2014}
{Reipurth}, B., {Clarke}, C.~J., {Boss}, A.~P., {et~al.} 2014, Protostars and
  Planets VI, 267

\bibitem[{{Reipurth} \& {Mikkola}(2012)}]{ReipurthMikkola2012}
{Reipurth}, B. \& {Mikkola}, S. 2012, \nat, 492, 221

\bibitem[{{Reipurth} \& {Zinnecker}(1993)}]{reipurth93}
{Reipurth}, B. \& {Zinnecker}, H. 1993, \aap, 278, 81

\bibitem[{{Retterer} \& {King}(1982)}]{RettererKing1982}
{Retterer}, J.~M. \& {King}, I.~R. 1982, \apj, 254, 214

\bibitem[{{Rivera} {et~al.}(2015){Rivera}, {Loinard}, {Dzib}, {Ortiz-Le{\'o}n},
  {Rodr{\'{\i}}guez}, \& {Torres}}]{RiveraEtAl2015}
{Rivera}, J.~L., {Loinard}, L., {Dzib}, S.~A., {et~al.} 2015, \apj, 807, 119

\bibitem[{Robotham(2015)}]{R-magicaxis}
Robotham, A. 2015, magicaxis: Pretty Scientific Plotting with Minor-Tick and
  log Minor-Tick Support, r package version 1.9.4

\bibitem[{{Ryden}(1996)}]{Ryden1996}
{Ryden}, B.~S. 1996, \apj, 471, 822

\bibitem[{{Sakhr} \& {Nieminen}(2006)}]{SakhrNieminen2006}
{Sakhr}, J. \& {Nieminen}, J.~M. 2006, \pre, 73, 036201

\bibitem[{{Sartoretti} {et~al.}(1998){Sartoretti}, {Brown}, {Latham}, \&
  {Torres}}]{sartoretti98}
{Sartoretti}, P., {Brown}, R.~A., {Latham}, D.~W., \& {Torres}, G. 1998, \aap,
  334, 592

\bibitem[{{Shaya} \& {Olling}(2011)}]{ShayaOlling2011}
{Shaya}, E.~J. \& {Olling}, R.~P. 2011, \apjs, 192, 2

\bibitem[{{Shu} {et~al.}(1987){Shu}, {Adams}, \& {Lizano}}]{ShuEtAl1987}
{Shu}, F.~H., {Adams}, F.~C., \& {Lizano}, S. 1987, \araa, 25, 23

\bibitem[{{Simon}(1997)}]{Simon1997}
{Simon}, M. 1997, \apjl, 482, L81

\bibitem[{{Simon} {et~al.}(1999){Simon}, {Beck}, {Greene}, {Howell}, {Lumsden},
  \& {Prato}}]{simon99}
{Simon}, M., {Beck}, T.~L., {Greene}, T.~P., {et~al.} 1999, \aj, 117, 1594

\bibitem[{{Simon} {et~al.}(1995){Simon}, {Ghez}, {Leinert}, {Cassar}, {Chen},
  {Howell}, {Jameson}, {Matthews}, {Neugebauer}, \& {Richichi}}]{simon95}
{Simon}, M., {Ghez}, A.~M., {Leinert}, C., {et~al.} 1995, \apj, 443, 625

\bibitem[{{Smith} {et~al.}(2005){Smith}, {Balega}, {Duschl}, {Hofmann},
  {Lachaume}, {Preibisch}, {Schertl}, \& {Weigelt}}]{smith05}
{Smith}, K.~W., {Balega}, Y.~Y., {Duschl}, W.~J., {et~al.} 2005, \aap, 431, 307

\bibitem[{{Tafalla} \& {Hacar}(2015)}]{TafallaHacar2015}
{Tafalla}, M. \& {Hacar}, A. 2015, \aap, 574, A104

\bibitem[{Therneau(2014)}]{R-deming}
Therneau, T. 2014, deming: Deming, Thiel-Sen and Passing-Bablock Regression, r
  package version 1.0-1

\bibitem[{{Tobin} {et~al.}(2010){Tobin}, {Hartmann}, {Looney}, \&
  {Chiang}}]{TobinEtAl2010}
{Tobin}, J.~J., {Hartmann}, L., {Looney}, L.~W., \& {Chiang}, H.-F. 2010, \apj,
  712, 1010

\bibitem[{{Tobin} {et~al.}(2016){Tobin}, {Looney}, {Li}, {Chandler}, {Dunham},
  {Segura-Cox}, {Sadavoy}, {Melis}, {Harris}, {Kratter}, \&
  {Perez}}]{TobinEtAl2016}
{Tobin}, J.~J., {Looney}, L.~W., {Li}, Z.-Y., {et~al.} 2016, \apj, 818, 73

\bibitem[{{Todorov} {et~al.}(2010){Todorov}, {Luhman}, \& {McLeod}}]{todorov10}
{Todorov}, K., {Luhman}, K.~L., \& {McLeod}, K.~K. 2010, \apjl, 714, L84

\bibitem[{{Todorov} {et~al.}(2014){Todorov}, {Luhman}, {Konopacky}, {McLeod},
  {Apai}, {Ghez}, {Pascucci}, \& {Robberto}}]{todorov14}
{Todorov}, K.~O., {Luhman}, K.~L., {Konopacky}, Q.~M., {et~al.} 2014, \apj,
  788, 40

\bibitem[{{Tokovinin}(2014)}]{Tokovinin2014}
{Tokovinin}, A. 2014, \aj, 147, 87

\bibitem[{{Tokovinin} \& {L{\'e}pine}(2012)}]{TokovininLepine2012}
{Tokovinin}, A. \& {L{\'e}pine}, S. 2012, \aj, 144, 102

\bibitem[{{Torres} {et~al.}(2006){Torres}, {Quast}, {da Silva}, {de La Reza},
  {Melo}, \& {Sterzik}}]{TorresEtAl2006}
{Torres}, C.~A.~O., {Quast}, G.~R., {da Silva}, L., {et~al.} 2006, \aap, 460,
  695

\bibitem[{{Torres} {et~al.}(2007){Torres}, {Loinard}, {Mioduszewski}, \&
  {Rodr{\'{\i}}guez}}]{TorresEtAl2007}
{Torres}, R.~M., {Loinard}, L., {Mioduszewski}, A.~J., \& {Rodr{\'{\i}}guez},
  L.~F. 2007, \apj, 671, 1813

\bibitem[{{van Kempen} {et~al.}(2012){van Kempen}, {Longmore}, {Johnstone},
  {Pillai}, \& {Fuente}}]{KempenEtAl2012}
{van Kempen}, T.~A., {Longmore}, S.~N., {Johnstone}, D., {Pillai}, T., \&
  {Fuente}, A. 2012, \apj, 751, 137

\bibitem[{Vincenty(1975)}]{Vincenty75}
Vincenty, T. 1975, Survey Review, 22, 88

\bibitem[{{Ward-Thompson} {et~al.}(2007){Ward-Thompson}, {Andr{\'e}},
  {Crutcher}, {Johnstone}, {Onishi}, \& {Wilson}}]{Ward-ThompsonEtAl2007}
{Ward-Thompson}, D., {Andr{\'e}}, P., {Crutcher}, R., {et~al.} 2007, Protostars
  and Planets V, 33

\bibitem[{{White} \& {Ghez}(2001)}]{white01}
{White}, R.~J. \& {Ghez}, A.~M. 2001, \apj, 556, 265

\bibitem[{{Ysard} {et~al.}(2013){Ysard}, {Abergel}, {Ristorcelli}, {Juvela},
  {Pagani}, {K{\"o}nyves}, {Spencer}, {White}, \& {Zavagno}}]{YsardEtAl2013}
{Ysard}, N., {Abergel}, A., {Ristorcelli}, I., {et~al.} 2013, \aap, 559, A133

\end{thebibliography}


\appendix

\section{Spherical geometry}
\label{Appendix_SpheGeo}
As a reminder, this section gives the formulae to obtain the angular distance $\Delta \sigma$ between two stars on the celestial sphere, given by the spherical law of cosines:
\begin{equation}
\Delta\sigma=\arccos\bigl(\sin\alpha_1\sin\alpha_2+\cos\alpha_1\cos\alpha_2\cos\Delta\delta \bigr)
,\end{equation}
where  $\phi_i,\lambda_i$ ($i=1,2$)  are, respectively, the  declination  and right ascension of the star $i$, and $\Delta\delta= |\delta_2-\delta_1|$ their absolute difference in declination.
Since this above equation is ill-conditioned for small $\Delta \sigma$, it is better to use the following Vincenty formula \citep{Vincenty75} applied to a sphere 

{\small
\begin{equation}
\Delta\sigma=\arctan\left(\frac{\sqrt{\left(\cos\alpha_2\sin\Delta\delta\right)^2+\left(\cos\alpha_1\sin\alpha_2-\sin\phi_1\cos\alpha_2\cos\Delta\delta\right)^2}}{\sin\alpha_1\sin\alpha_2+\cos\alpha_1\cos\alpha_2\cos\Delta\delta}\right).
\end{equation}
}

\section{Analytics for k-Nearest Neighborhood statistics}
\label{Appendix_k-NND}
\subsection{Theoretical probability density function (PDF) and cumulative distribution of 1-nearest neighbor separation (1-NNS) for spatial random distribution}
In this section, we derive the theoretical distribution of  1-NNS for a  random point process in an infinite medium, 
following the work done in 3D by \cite{Chandrasekhar1943}. The probability $w( r ) dr $ that the first neighbor of a star is located  at a distance between $r$ and $r+dr$ in 2D,  is given by the product of the probability that there is no stars in a disk of radius $r$ multiplied by the probability  of having stars in a  shell area between $r$ and $r+dr$:
\begin{equation}
w ( r ) dr =\left (1-\int_0^r w( r ) {\rm d}r \right )\times 2 \pi r \;{\rm d}r\; \rho 
\label{Eq:P_nnd_k1}
,\end{equation}
where $\rho$ is the mean intensity of the process in an infinite medium, that is, the average  number of stars per unit of surface. Equation \ref{Eq:P_nnd_k1} may be written as:
\begin{equation}
\frac{{\rm d}}{{\rm d} r} \left ( \frac{w ( r )}{2 \pi r \rho} \right ) = -2 \pi r \rho \frac{w ( r )}{2 \pi r \rho}
.\end{equation}
Once integrated, from the equation above, we get the probability density function (PDF) of the first neighbor distance for a random Poissonnian process:
\begin{equation}
w( r )=  2 \pi \rho r \exp(- \pi \rho r^2)
\label{Eq:NND_k1_pdf_theo}
.\end{equation}
The PDF integral is unity (i.e., we have $\int_0^{+\infty} w(r ) \rm{d}r = \left [ -\exp(-\pi \rho r^2) \right ]_0^{+\infty}=1$). 

Furthermore, from the 1-NNS PDF (Eq. \ref{Eq:NND_k1_pdf_theo}), we get the related cumulative distribution function $W( r )$ as:
\begin{equation}
\barr{lll}
W( r ) &= &\int_0^{r} 2 \pi \rho r \exp(- \pi \rho r^2) {\rm d}r\\
&=& 1 -\exp(-\pi \rho r^2) 
\earr
\label{Eq:NND_k1_cdf_theo}
,\end{equation}
that allows us to define the characteristic value $r_q$ associated to the quantile $q$ from: $q = W(r_q) = 1 -\exp(-\pi \rho r^2_q)$, to get:
\be
\barr{lll}
  r_q&= & \sqrt{\ln(1/(1-q))/ (\pi \rho)}\\
\earr 
.\ee

\subsection{Moments of first nearest neighbor distribution}

\subsubsection{Mean}
The theoretical mean $\bar{r_t} $ of the first nearest neighbor distance in an infinite medium is computed from the PDF as:
\begin{equation}
\begin{array}{lll} 
\bar{r_t} & = \mathbb{E}(r) = \int_0^{+\infty} r \, w( r ) dr \\
&  =\left [  \frac{\erf(\sqrt{\pi \rho} r)}{2\sqrt{\rho}} - r \exp(-\pi\rho r^2) \right ]_0^{+\infty} &
\end{array}
\label{Eq:NND_k1_mean_theo_full}
,\end{equation}
where $= \mathbb{E}(r)$ is the expected value of $r$, and $\erf(x)$ is the error function. Taking the value at $0$ and infinity, we end up  with a simple analytical result  of first nearest neighbor distance theoretical mean $\bar{r_t} $ in a random process of intensity $\rho$ in an infinite medium,  i.e., $\bar{r_t} $ is strictly equal to half  of the inverse square root of the density:
\begin{equation}
\bar{r_t} \simeq \frac{1}{2 \sqrt{\rho}}
\label{Eq:NND_k1_mean_theo}
.\end{equation}

\subsubsection{Variance}
From the PDF $w(r )$, we now compute the theoretical  variance $\sigma_t^2$ ($\sigma_t$  being the standard deviation) of the first nearest neighbor distance for a random Poisson process, 
\begin{equation}
\begin{array}{lll}
\sigma_t^2 & = & \int_0^{+\infty} (r-\bar{r})^2 w(r ) \,  \rm{d}r \\
&= & 2 \,  \pi  \,  \rho  \,  \int_0^{+\infty} r  \, \left (r-\frac{1}{2 \sqrt{\rho}}\right )^2  \exp(- \pi \rho r^2) \, \rm{d}r\\
\end{array}
.\end{equation}
The integration gives:

 \begin{equation}
\begin{array}{lll}
\sigma_t^2  &= &\frac{-1}{4 \pi \rho}   \Bigg[ \exp ( - \pi \rho r^2) \Big( 2 \pi \exp(\pi \rho \,r^2) \erf(\sqrt{\pi\rho}\, r) \Big) \\
&+ & \pi (1-2\sqrt{a} \; r)^2 +4 \Bigg]_0^{+\infty}
\end{array}
.\end{equation}

We then compute series expansions of the integral at $x=0$:
 \begin{equation}
\begin{array}{lll}
\lim_{r \to 0} \sigma_t^2 & = & -\frac{4+\pi}{4\pi\rho}+\frac{\pi r^2}{4}-\frac{2}{3} \pi \sqrt{\rho} r^3 \\
& -& \frac{1}{8}(\pi-4)\pi \rho r^4 +O(r^5)
\end{array}
,\end{equation}

and then at $x=+\infty$:
 \begin{equation}
\begin{array}{lll}
\lim_{r \to +\infty} \sigma_t^2  &= & -1/(2\rho) \Bigg( \exp(-\pi r^2) \big[ 2 \rho r^2-2 \sqrt{\rho}\\
&+ &(1/2+2/\pi) -(\sqrt{\rho}\, \pi r)^{-1}\\
& + & (2\rho^{2/3} \pi^2 r3)^{-1} + O(r^{-4}) \big] +1 \Bigg)
\end{array}
,\end{equation}
to finally get:
\begin{equation}
\begin{array}{lll}
\sigma_t^2 & = &- (2 \rho)^{-1}+(4+ \pi)(4 \pi \rho)^{-1} \\
& = & (4- \pi) (4 \pi \rho)^{-1}\\
\end{array}
\label{Eq:NND_k1_sigma_theo}
.\end{equation}

\subsubsection{Skewness}
The skewness of a distribution quantifies its asymmetry, and is defined from the third moment of the distribution. When it is positive, the peak of the distribution is shifted to the left, that is, there is a tail towards the right. When it is negative, it is the other way round.
The skewness may be defined from the moments of the distribution:
\be
\begin{array}{lll} 
\gamma_t & = &  \left (\mathbb{E}(r^3) -3 \bar{r_t}  \mathbb{E}(r^2) +2 \bar{r_t}^3 \right ) / (\sigma_t)^{3} \\
&=& \left (3/(4 \pi \rho^{3/2}) -3  \bar{r_t} / (\pi \rho) + 2 \bar{r_t}^3\right )  / (\sigma_t)^{3}\\
&=& 2 \sqrt{\pi} (\pi-3)/(4-\pi)^{3/2} \\
&\approx&0.63 \\
\end{array} 
.\ee
We note that the skewness of the first nearest neighbor distribution of a random spatial distribution is  constant, that is, independent of the density/intensity of the random process.

\subsection{Log PDF}
Based on transformation rules, we then derive the theoretical PDF for the Log(1-NNS) in a random process. Let  $y= Log( r )$, where $r$ is the 1-NNS variable, and thus we have:
\begin{equation}
w( r ) {\rm d}r=  w( y ) {\rm d}y
,\end{equation}
where $w( r ) $ is the PDF of the 1-NNS variable (Eq. \ref{Eq:NND_k1_pdf_theo}) and $w(y)$ is the PDF of the logarithm variable. We get the following moments for the random k1-NNS theoretical distributions:
\begin{equation}
w( y ) =  w( r ) \left |  \frac{\partial r}{ \partial y} \right | =  w( r ) \frac{{\rm d}r}{ {\rm d}y}
.\end{equation}
From this relation, we get the PDF of the first neighbor distance logarithm for a random process as:
\begin{equation}
w( y ) = 2 \, \ln(10)  \, \pi  \,  \rho  \,  r^2  \,  \exp(- \pi \rho r^2)
\label{Eq:LogNND_k1_pdf_theo}
,\end{equation}
where it reaches its maximun value (mode) at $r_{mod}= \sqrt{\pi \rho}$
. From Taylor expansion for $x\ll1$, we get:
\begin{equation}
w( y ) \sim 2 \, \ln(10)  \, \pi  \,  \rho  \,  r^2  \, +  {O}(x^4)
\label{Eq:LogNND_k1_pdf_theo_dl0}
,\end{equation}
as we expect in 2D for a random distribution, the  first nearest neighbor 1-NNS distribution function increases as a squared power law  for $r \ll1/\sqrt{\pi \rho}$. From equation \ref{Eq:LogNND_k1_pdf_theo} above, we derive its logarithmic expression:
\begin{equation}
\log w( y ) = \log(2 \, \ln(10)  \, \pi  \,  \rho  \,)+2 \, r   \,- (\pi \rho/\ln 10\,) r^2
\label{Eq:NNDLog_k1_LogTheo}
.\end{equation}

In a Poisson point process in a $d$-dimensional space with intensity $\rho$, the distance $r$ between a
point and its $k^{th}$ neighbor is distributed according to the generalized Gamma distribution:

\begin{equation}
  \begin{array}{ll}
  { \mathcal P}_k(r )          \approx \frac{d}{r  }  \, \frac{\left  [ \rho V_d (r ) \right ]^{k}}{\Gamma(k)} \cdot \exp(-\rho V_d (r )) \\
  \label{eq:KNN_End}
\end{array}
.\end{equation}
where $V_d (r ) $ is the volume ball or radius $r$ in $d$-space.
\begin{equation}
\begin{split}
V_d ( r )  =c_d r^d\\
\end{split}
\label{eq:VolBall}
,\end{equation}
and   $c_d$ is the unit volume ball in $d$-space given by:
\begin{equation}
c_d  = \pi^{d/2}  /\Gamma (d/2 +1)
\label{eq:PreFactorDspace}
,\end{equation}
and $\Gamma $ is the Gamma function. So, in a planar (2D) distribution,  we get:
\begin{equation}
    \mathcal{P}_k(r ) =  \frac{2 (\pi \rho)^k}{\Gamma(k)} \cdot r^ {2k-1}\cdot \mathrm{exp}(-\rho \pi r^2)
\label{Eq:RandDistr_knd_k}
.\end{equation}
The cumulative distribution is then:
\begin{equation}
\mathcal{W}_k(r) =  \int_0^r \mathcal{P}_k(r ) {\rm d}r
.\end{equation}
So for the second nearest neighbor cumulative distribution we get:
\begin{equation}
\mathcal{W}_2(r) =  \int_0^r \mathcal{P}_2(r ) {\rm d}r = 1- (\pi \rho r^2+1)\, exp(-\pi r^2)
\label{Eq:RandCumu_knd_k2}
,\end{equation}
and for the third nearest neighbor cumulative distribution, we have:
\begin{equation}
\mathcal{W}_3(r) =  \int_0^r \mathcal{P}_3(r ) {\rm d}r = 1/2 \left[ exp(-\pi r^2) \left( -\pi \rho r^2(\pi \rho r^2 +2)-2 \right)+2 \right]
\label{Eq:RandCumu_knd_k3}
.\end{equation}

\section{Catalogs}
This sections gather the catalogs described respectively in Sects. \ref{Sect_Data} and \ref{SubSect_UWP} of the paper.

\end{document}